\documentclass[a4paper,12pt]{article}
\usepackage{amsmath,amsthm,amsfonts,amssymb,natbib,graphicx,booktabs,float,caption,subcaption,multirow,dsfont,setspace,chngcntr,afterpage,lscape,etoolbox,geometry}
\usepackage{mathrsfs}
\usepackage{enumerate}

\usepackage[utf8]{inputenc}
\usepackage[table]{xcolor}
\usepackage{enumitem}
\usepackage{multicol}
\usepackage{longtable}
\usepackage{fancyhdr}
\usepackage{tikz}
\usepackage[sectionbib]{chapterbib}
\usepackage{etoc}
\usepackage[colorlinks=true, linkcolor=blue, citecolor=blue, urlcolor=blue]{hyperref}

\newcommand{\spacingset}[1]{%
  \renewcommand{\baselinestretch}{#1}%
  \small\normalsize
}

\newtheorem{definition}{Definition}

\newtheorem{theorem}{Theorem}

\newtheorem{corollary}[theorem]{Corollary}

\newtheorem{lemma}[theorem]{Lemma}

\newtheorem{assumption}{Assumption}
\newtheorem*{assumption*}{Assumption}
\newtheorem{proposition}[theorem]{Proposition}

\newtheorem*{start*}{Starting Point}
\newtheorem*{proof*}{Proof}
\newtheorem*{note*}{Side Remark}

\AfterEndEnvironment{proposition}{\noindent\ignorespaces}
\AfterEndEnvironment{assumption}{\noindent\ignorespaces}
\AfterEndEnvironment{proof}{\noindent\ignorespaces}
\AfterEndEnvironment{corollary}{\noindent\ignorespaces}
\AfterEndEnvironment{remark}{\noindent\ignorespaces}

\newcommand\fnote[1]{\captionsetup{font=small}\caption*{#1}}

\newcommand{\ManuscriptMainSpacing}{1.5}

\newcommand{\ManuscriptMainFrontMatter}{%
  \spacingset{1}
  \title{Lost in the Bins: Treatment Effects with Discretized Covariates\thanks{I am indebted to Daniel Gutknecht for many helpful discussions and advice. I am also grateful to Timo Dimitriadis, Uwe Hassler, Thomas Otter, Mengshan Xu, Sirui Zhou, and seminar participants at Goethe University Frankfurt for helpful comments and discussions.}}
  \author{Cenchen Liu\thanks{%
  Faculty of Economics and Business, Goethe University Frankfurt, 60629
  Frankfurt am Main, Germany. E-mail: \texttt{liu@wiwi.uni-frankfurt.de}.%
  }}
  \date{November 9, 2026}
  \maketitle

\begin{abstract}
\noindent
Mean independence restrictions are widely used to identify average treatment
effects, but the covariates used to formulate these restrictions are often
observed or used in coarsened form. This paper shows that mean
independence conditional on a finely measured covariate need not hold after the
covariate is replaced by a coarsened version. Coarsening can therefore change
the identifying content of the comparison rather than only the precision of
estimation. I characterize the resulting coarsening bias for the ATT and ATE.
For the ATT, the bias is driven by within-cell composition in untreated
potential outcomes. For the ATE, an additional treatment effect heterogeneity
term appears. I then introduce $c$-mean dependence, a nonparametric relaxation
of exact mean independence at the coarsened level, and derive sharp bounds for
average treatment effects. The parameter $c$ determines how far comparisons may
extend across neighboring values of the coarsened covariate. I propose sample
analogue estimators and bootstrap inference procedures for the resulting bound
functions. In an application to voluntary military service using data from \citet{angrist_estimating_1998}, positive effects on early earnings, which primarily reflect military pay, hold even under the weakest form of $c$-mean dependence, whereas the evidence on later effects on civilian earnings is inconclusive once mean independence is relaxed.

\bigskip\noindent\textbf{JEL Classification:} C14, C18, C21

\bigskip \noindent \textbf{Keywords:} Coarsening Bias, Mean Independence, Identification, Sensitivity Analysis, Treatment Effects, Discretized Covariates
\end{abstract}
}

\newcommand{\ManuscriptSupplementSetup}{%
  \spacingset{1.5}
  \setcounter{page}{1}
  \setcounter{section}{0}
  \setcounter{equation}{0}
  \setcounter{theorem}{0}
  \setcounter{definition}{0}
  \setcounter{assumption}{0}
  \renewcommand{\thetheorem}{S.\arabic{theorem}}
  \renewcommand{\thedefinition}{S.\arabic{definition}}
  \renewcommand{\theassumption}{S.\arabic{assumption}}
  \renewcommand{\thepage}{S.\arabic{page}}
  \renewcommand{\thesection}{S.\arabic{section}}
  \renewcommand{\theequation}{S.\arabic{equation}}
  \renewcommand{\theHsection}{supp.\arabic{section}}
  \renewcommand{\theHtheorem}{supp.\arabic{theorem}}
  \renewcommand{\theHdefinition}{supp.\arabic{definition}}
  \renewcommand{\theHassumption}{supp.\arabic{assumption}}
  \renewcommand{\theHequation}{supp.\arabic{equation}}
}

\newcommand{\ManuscriptSupplementFrontMatter}{%
  \begin{center}
  {\LARGE\bf Supplementary Material to \textit{Lost in the Bins: Treatment Effects with Discretized Covariates}}
  \end{center}

  \bigskip
  \begingroup
    \expandafter\def\csname @pnumwidth\endcsname{2.7em}
    \expandafter\def\csname @tocrmarg\endcsname{3.7em}
    \renewcommand{\numberline}[1]{\makebox[\dimexpr\csname @tempdima\endcsname+0.5em\relax][l]{##1}}
    \etocsettagdepth{main}{none}
    \etocsettagdepth{supplement}{subsection}
    \etocsetnexttocdepth{all}
    \tableofcontents
  \endgroup
  \newpage
}

\begin{document}
\ManuscriptMainFrontMatter
\newpage
\etocdepthtag.toc{main}
\spacingset{\ManuscriptMainSpacing}

\section{Introduction}
\label{sec:coarsened_covariates_reorganized}

Mean independence assumptions are widely used in treatment effect models. These
assumptions state that, after conditioning on observed covariates, treatment
assignment may be related to the distribution of potential outcomes, but not to
their conditional means. They are weaker than statistical independence while
still identifying the average treatment effect
(ATE) and the average treatment effect on the treated (ATT)
\citep[see, e.g.,][]{heckman_matching_1998,frolich_propensity_2007}. The same logic also
appears in difference-in-differences (DiD) designs, where conditional parallel
trends can be written as a mean independence restriction on untreated potential
outcome differences.

In empirical applications, however, the covariates used to formulate such
restrictions are often collected, released, or used in coarsened form. Work
experience may be measured in years rather than months and then further grouped
into intervals, income may be discretized into equal-width or quantile bins, and
education may be collapsed into broader levels such as ``low'', ``medium'', and
``high''. When a covariate is a curve, such as the past path of an interest rate, researchers often condition on its values at a finite number of dates, which form a coarsened version of the whole path \citep[see the discussion in][]{kurisu_nonparametric_2026}. Researchers may also discretize continuous covariates to simplify the analysis, avoid small cells, or mitigate the curse of dimensionality in nonparametric estimation \citep[see, e.g.,][]{iacus_causal_2012,lee_weidner_bounding_2026}. For example, \citet{finkelstein_gentzkow_li_2025} group age into five-year intervals to study how location affects risky prescription opioid use, and \citet{adams_prassl_huttunen_nix_zhang_2024} match workers on broad education categories rather than specific degrees to study the effects of workplace violence. This raises a natural question: if mean independence
is credible given a finely measured covariate, does it remain plausible after
that covariate has been coarsened?

The first result of the paper characterizes the consequences of covariate coarsening for the identification of average treatment effects. Let
$X$ denote a finely measured covariate and $W=g(X)$ a coarsened
version of $X$. Mean independence that holds conditional on $X$ need not hold
conditional on $W$. I derive a decomposition of the bias induced by replacing $X$ with $W$ in the ATT and ATE comparisons.
For the ATT, the decomposition contains a composition bias term that reflects within-cell differences in mean outcomes without treatment between treated and untreated units. For the ATE, an additional treatment effect heterogeneity term arises from within-cell differences in their average treatment effects. I also provide sufficient conditions for these bias terms to vanish and characterize their signs under monotone selection.

Second, I study what can be learned about average treatment effects when exact
mean independence is relaxed at the coarsened level. I introduce a new
nonparametric class of assumptions, which I call $c$-mean dependence. For units with a given treatment status, the potential outcome under the other treatment is not observed. Mean independence given $W$ requires the average of this unobserved potential outcome at each value of $W$ to equal the corresponding observed mean at that value. The $c$-mean-dependence assumption allows the average unobserved potential outcome to lie between the lowest and highest corresponding observed means at nearby values of $W$. The parameter $c$ determines the size of this neighborhood. This class nests exact mean independence given $W$ as the case $c=0$ and becomes weaker as $c$ increases. Under $c$-mean dependence, I derive sharp bounds for the ATT and ATE. For ordered cells, such as income brackets, I motivate $c=1$ as a natural benchmark. Outcomes measured before treatment provide an additional, empirical benchmark. Since the treatment cannot affect these outcomes, the bounds on its effect should contain zero, and the smallest value of $c$ at which the data do not reject this necessary condition can serve as this benchmark. Beyond these benchmarks, researchers can vary $c$ to assess sensitivity to departures from mean independence, in the spirit of the $c$-dependence analysis of \citet{masten_identification_2018} and the breakdown frontier approach of \citet{masten_inference_2020}. Substantive knowledge about which values of $W$ are comparable can also guide the choice of $c$. For estimation and inference, I develop sample analogue estimators and confidence bands that are asymptotically valid uniformly over the sensitivity values under consideration. Since the bound functionals are Hadamard directionally differentiable but need not be fully Hadamard differentiable, I follow \citet{fang_inference_2019} and use bootstrap inference based on the numerical directional derivative of \citet{hongNumericalDeltaMethod2018}. I also extend the inference procedure to grouped data, in which the outcome is reported only through cell-level summaries.

I apply the framework to the earnings effects of voluntary military service using the grouped data studied by \citet{angrist_estimating_1998}. The data report Armed Forces Qualification Test (AFQT) scores only in groups, which can contain applicants with different underlying scores. Extending Angrist's specification check, I first use earnings before application, which military service cannot affect, as a placebo outcome. For pre-application earnings, the placebo tests reject mean independence given the reported covariates, the restriction behind Angrist's matching estimates, in various comparisons for both white and nonwhite applicants, but they reject $1$-mean dependence in none. The finding that military service raised earnings in 1982, when veterans' earnings largely reflect military pay, is robust to all values of $c$ for both groups. For later years, however, the matching estimates, which suggest an earnings loss for white applicants in 1986 and earnings gains for nonwhite applicants in 1986 and 1991, are not robust. Their signs rely on mean independence, and at $c=1$ the evidence on the sign of the effect is inconclusive for both groups.

The paper is related to earlier work on adjustment for coarsened covariates. \citet{cochran1968effectiveness} studies residual bias after subclassification in a regression framework. \citet{iacus_multivariate_2011} bound the error of matching within coarsened cells when the outcome without treatment has bounded derivatives in the finely measured covariates. \citet{ogburn_bias_2013} establish conditions under which adjustment for coarsened confounders reduces bias relative to no adjustment. \citet{manski_inference_2002} study identification and inference for regressions with interval data on a regressor or outcome. \citet{fankhauser_sharp_2026} characterize sharp identified sets in conditional moment models when covariates are observed only as intervals or are missing. In practice, researchers often compare treated and untreated units within broad groups, such as income categories or education levels, rather than at exact covariate values. Whether the identifying assumption still holds once units are grouped in this way has received little attention in economics. This paper shows that it can fail even when it holds at the exact values, characterizes the resulting bias for the ATT and ATE, and develops a sensitivity analysis that remains informative when only the groups are observed. Since coarsening can invalidate mean independence, the analysis is also related to the literature on partial identification under weakened mean independence restrictions. \citet{manski_anatomy_1989,manski_nonparametric_1990} derives bounds when potential outcomes are only restricted to lie in known bounded intervals. \citet{manski_monotone_2000,manski_more_2009} replace mean independence with monotonicity restrictions on average potential outcomes. \citet{hotz_bounding_1997} analyze a contaminated instrument that satisfies the exclusion restriction in the conditional mean function only for a subpopulation. Related work in DiD relaxes parallel trends, the corresponding mean independence restriction on untreated potential outcome differences. \citet{manski_how_2018} bound violations of parallel trends by those observed in pre-treatment periods, while \citet{rambachan_more_2023} and \citet{ban_robust_2023} develop related bounding approaches. The paper is also related to the broader literature on relaxing independence assumptions between treatment and potential outcomes. \citet{masten_identification_2018} relax statistical independence, which is stronger than mean independence, and \citet{masten_choosing_2023} study mean and quantile independence and propose a weaker form of quantile independence.

The rest of the paper is organized as follows.
Section~\ref{sec:coarsening_setup} formalizes discretization and derives the
coarsening bias decomposition for the ATT and ATE.
Section~\ref{sec:c_mean_dependence} introduces $c$-mean dependence, derives
sharp bounds for average treatment effects, and discusses how to choose the
sensitivity parameter. Section~\ref{sec:estimation_c_mean_dependence} develops
estimation and inference for these bounds. Section~\ref{sec:application} revisits the matching estimates of \citet{angrist_estimating_1998} for the earnings effects of voluntary military service. Section~\ref{sec:conclusion} concludes. The Supplementary Material contains the proofs and additional details.

\section{Pitfalls of Discretization}
\label{sec:coarsening_setup}

\subsection{Modeling Discretization}

 Let $X$ denote the covariate and $\mathcal X$ its support.
The distribution of $X$ may be discrete, continuous, or mixed. For ease of
exposition, suppose $\mathcal X\subseteq\mathbb{R}$. The extension to vector-valued
covariates is straightforward.

Coarsening groups values of \(X\) into categories. It includes discretizing a continuous covariate and combining categories of a discrete covariate. Represent this grouping by a measurable function \(g:\mathcal X\to I\), where \(I\) is an at most countable set of categories. The coarsened covariate and its support are
\[
W:=g(X),\qquad
\mathcal W:=\{w\in I\mid\Pr(W=w)>0\}.
\]
The following examples illustrate common choices of \(g\).

\emph{Equal-width binning.}
Equal-width binning groups values of \(X\) into intervals of a common width. For an origin \(a\in\mathbb R\) and a bin width \(b>0\), define
\[
g(x):=k\quad\text{if }a+kb\leq x<a+(k+1)b,
\qquad k\in\mathbb Z.
\]
The origin \(a\) determines the location of the cutoffs, while \(b\) determines the width of each bin.
For example, \citet{finkelstein_gentzkow_li_2025} group age into five-year intervals to study how geographic location affects risky prescription opioid use. These age groups, together with race and gender, define the covariate cells used to compare movers with non-movers in a conditional difference-in-differences design. Similarly, in a propensity score matching specification,
\citet{baylis_boomhower_2026} match homes in code-required
jurisdictions to homes in no-code jurisdictions using a propensity score based
on effective year built, ground slope, and fuel model. Before matching, they
coarsen effective year built for older homes: pre-1935 values are rounded to
five-year bins, and pre-1900 values are grouped at 1900.

\emph{Quantile binning.}
Quantile binning uses quantiles of the distribution of \(X\) as category cutoffs. For a chosen \(K\in\mathbb N\), let \(q_j:=F_X^{-1}(j/K)\), \(j=1,\ldots,K-1\), where \(F_X^{-1}\) is the generalized inverse distribution function. With \(q_0:=-\infty\) and \(q_K:=\infty\), define
\[
g(x):=k\quad\text{if }q_k\leq x<q_{k+1},
\qquad k=0,\ldots,K-1.
\]
For example, \citet{cingano_palomba_pinotti_rettore_2025} estimate the effects of investment subsidies for firms away from the assignment cutoff. Their analysis relies on mean independence of potential outcomes from the running variable and common support conditional on firm covariates. Their conditioning vector includes deciles of lagged employment growth, deciles of average wages, and deciles of employment growth among similar firms.

\emph{Categorical grouping.}
Categorical grouping combines detailed categories into broader groups. In their matched difference-in-differences analysis of workplace violence, \citet{adams_prassl_huttunen_nix_zhang_2024} match workers on education level, together with other characteristics and pre-incident outcomes. For example, workers whose highest degree is vocational upper secondary and those whose highest degree is academic upper secondary are both counted as having a high school education. Here, \(X\) is the detailed degree and \(W=g(X)\) is the education level. Mean independence that holds given the degree need not hold given the level.

Coarsening can also arise from restrictions on data access. For example, \citet{angrist_estimating_1998} studies the labor market effects of voluntary military service using Social Security earnings data that report average earnings for cells rather than for individuals. These cells are defined by Armed Forces Qualification Test (AFQT) score groups, schooling levels, and other characteristics.

\subsection{When Discretization Goes Wrong}

This subsection studies how discretizing a covariate can change treatment effect
comparisons. Discretization is not inherently problematic. For descriptive or predictive
targets, replacing \(X\) by \(W\) only changes the information set with respect
to which the object is defined. Let
\(m(X):=\mathrm{E}\left[Y\mid X\right]\) and \(u:=Y-m(X)\), so that
\(\mathrm{E}\left[u\mid X\right]=0\) a.s. Since \(W\) is a measurable function
of \(X\), the law of iterated expectations gives
$
\mathrm{E}\left[u\mid W\right]
=
\mathrm{E}\left[
\mathrm{E}\left[u\mid X\right]
\mid W
\right]
=
0.
$
The coarsened conditional expectation function and its residual are
\begin{align*}
\overline m(W)
&:=
\mathrm{E}\left[Y\mid W\right]
=
\mathrm{E}\left[m(X)\mid W\right],\\
v
&:=
Y-\overline m(W)
=
u+m(X)-\mathrm{E}\left[m(X)\mid W\right],
\end{align*}
with \(\mathrm{E}\left[v\mid W\right]=0\). Thus,
\(Y=\overline m(W)+v\) remains a valid conditional expectation function
decomposition of \(Y\) with respect to
the coarsened covariate \(W\).

In treatment effect models, mean independence is an identifying assumption, rather than a property that follows from defining a conditional mean. This assumption may hold given \(X\) but fail given its coarsened version \(W\). Let $Y\left(1\right)$ and
$Y\left(0\right)$ denote the potential outcomes, and
$D\in\left\{0,1\right\}$ denote treatment status. The observed outcome satisfies
\begin{align}
Y=D Y\left(1\right)+\left(1-D\right)Y\left(0\right).
\label{eq:observed_outcome}
\end{align}
Two common causal parameters of interest are
\begin{align*}
\tau_{\mathrm{ATT}}
&:=
\mathrm{E}\left[
Y\left(1\right)-Y\left(0\right)\mid D=1
\right],
\\
\tau_{\mathrm{ATE}}
&:=
\mathrm{E}\left[
Y\left(1\right)-Y\left(0\right)
\right].
\end{align*}
The identifying conditions are stated first for the original covariate $X$.

\begin{assumption}[Mean independence for \(Y\left(0\right)\)]
\label{ass:mean_independence_y0}
\begin{align*}
\mathrm{E}\left[Y\left(0\right)\mid X,D\right]
=
\mathrm{E}\left[Y\left(0\right)\mid X\right]\qquad\text{a.s.}
\end{align*}
\end{assumption}

\begin{assumption}[Mean independence for \(Y\left(1\right)\)]
\label{ass:mean_independence_y1}
\begin{align*}
\mathrm{E}\left[Y\left(1\right)\mid X,D\right]
=
\mathrm{E}\left[Y\left(1\right)\mid X\right]\qquad\text{a.s.}
\end{align*}
\end{assumption}
The conditional independence restriction
\(\left(Y\left(1\right),Y\left(0\right)\right)\perp D\mid X\) is sufficient
for Assumptions~\ref{ass:mean_independence_y0}--\ref{ass:mean_independence_y1}, but is stronger than needed for average
treatment effects \citep[see, e.g.,][]{frolich_propensity_2007}.

Write
\(\mu_d\left(x\right):=\mathrm{E}\left[Y\left(d\right)\mid X=x\right]\)
for $d\in\left\{0,1\right\}$ and $x\in\mathcal{X}$, and define the fine
propensity score and the conditional ATE given $X$ by
\begin{align*}
e^*\left(x\right)
&:=
\mathrm{Pr}\left(D=1\mid X=x\right),
\\
\tau\left(x\right)
&:=
\mu_1\left(x\right)-\mu_0\left(x\right).
\end{align*}
The standard overlap condition is also imposed for the covariate $X$.
\begin{assumption}[Overlap]
\label{ass:overlap}
There exists $\eta>0$ such that
\begin{align*}
\eta
\leq
e^*\left(X\right)
\leq
1-\eta
\qquad
\text{a.s.}
\end{align*}
\end{assumption}
Under the observed outcome equation \eqref{eq:observed_outcome} and
Assumption~\ref{ass:overlap},
Assumption~\ref{ass:mean_independence_y0} identifies the ATT, since \(Y\left(1\right)\) is observed for treated units:
\begin{align*}
\tau_{\mathrm{ATT}}
&=
\mathrm{E}\left[
Y
-
\mathrm{E}\left[Y\mid D=0,X\right]
\mid D=1
\right].
\end{align*}
The ATE also requires Assumption~\ref{ass:mean_independence_y1}, because mean outcomes under
both treatment states must be recovered for the full population. It is then identified by
\begin{align*}
\tau_{\mathrm{ATE}}
&=
\mathrm{E}\left[
\mathrm{E}\left[Y\mid D=1,X\right]
-
\mathrm{E}\left[Y\mid D=0,X\right]
\right].
\end{align*}

The coarsened analogue of the \(X\)-conditional treatment-control comparison
replaces conditioning on \(X\) by conditioning on \(W\). For
$w\in\mathcal{W}$, define the coarsened propensity score and the
coarsened treatment-control difference in mean outcomes by
\begin{align*}
e\left(w\right)
&:=
\mathrm{Pr}\left(D=1\mid W=w\right)
=
\mathrm{E}\left[e^*\left(X\right)\mid W=w\right],
\\
\theta\left(w\right)
&:=
\mathrm{E}\left[Y\mid D=1,W=w\right]
-
\mathrm{E}\left[Y\mid D=0,W=w\right].
\end{align*}
Here and below, \(e^*\) denotes the propensity score conditional on the fine
covariate \(X\), while \(e\) denotes the propensity score conditional on the coarsened
covariate \(W\).
Assumption~\ref{ass:overlap} implies
$\eta\leq e\left(W\right)\leq 1-\eta$ a.s. The corresponding coarsened estimands are
\begin{align*}
\theta_{\mathrm{ATT}}
&:=
\mathrm{E}\left[\theta\left(W\right)\mid D=1\right],
\\
\theta_{\mathrm{ATE}}
&:=
\mathrm{E}\left[\theta\left(W\right)\right].
\end{align*}

The next proposition characterizes the bias induced by replacing the original covariate \(X\) with the coarsened covariate \(W\). For a generic random variable \(Z\), write
\[
\Delta_Z(W):=\mathrm E[Z\mid D=1,W]-\mathrm E[Z\mid D=0,W]
\]
for the treatment-control difference in its mean within a coarsened cell. In particular,
\[
\Delta_{\mu_0(X)}(W)
=
\mathrm E[\mu_0(X)\mid D=1,W]
-
\mathrm E[\mu_0(X)\mid D=0,W].
\]
This difference compares the average values of the same conditional mean function \(\mu_0(X)\) across treated and comparison units within the cell. Similarly, \(\Delta_{\tau(X)}(W)\) compares their average treatment effects.

Under Assumptions~\ref{ass:mean_independence_y0} and~\ref{ass:overlap}, since \(W\) is a function of \(X\), the law of iterated expectations gives
\[
\mathrm E[Y(0)\mid D=d,W]=\mathrm E[\mu_0(X)\mid D=d,W],
\qquad d\in\{0,1\}.
\]
Mean independence for \(Y(0)\) given \(W\) therefore holds if and only if \(\Delta_{\mu_0(X)}(W)=0\) a.s. This difference need not be zero, because treated and comparison units in the same cell can have different distributions of \(X\). The same argument applies to \(Y(1)\) under Assumption~\ref{ass:mean_independence_y1}, with \(\mu_1\) in place of \(\mu_0\).

\begin{proposition}[Coarsening bias decomposition]
\label{prop:coarsen_bias_decomposition}
Suppose Assumption~\ref{ass:overlap} holds and \(\mathrm{E}\left[\left|Y\left(d\right)\right|\right]<\infty\) for all
\(d\in\left\{0,1\right\}\).
\begin{enumerate}[label=\textup{(\roman*)}]
\item If
Assumption~\ref{ass:mean_independence_y0} holds, then
\begin{align*}
\theta_{\mathrm{ATT}}
=
\tau_{\mathrm{ATT}}
+
\widetilde B_0,
\end{align*}
where
\begin{align*}
\widetilde B_0
:=\mathrm{E}\left[
\Delta_{\mu_0\left(X\right)}\left(W\right)
\mid D=1
\right].
\end{align*}
\item If Assumption~\ref{ass:mean_independence_y1} also holds, then
\begin{align*}
\theta_{\mathrm{ATE}}
=
\tau_{\mathrm{ATE}}+B_{\tau}+B_0,
\end{align*}
where
\begin{align*}
B_{\tau}
&:=\mathrm{E}\left[
\left(1-e\left(W\right)\right)
\Delta_{\tau\left(X\right)}\left(W\right)
\right],\\
B_0
&:=\mathrm{E}\left[
\Delta_{\mu_0\left(X\right)}\left(W\right)
\right].
\end{align*}
\end{enumerate}
\end{proposition}

The ATE decomposition in Proposition~\ref{prop:coarsen_bias_decomposition} contains two coarsening bias components. The treatment effect heterogeneity component \(B_\tau\) reflects differences in average treatment effects between treated and comparison units within the same coarsened cell. The composition bias component \(B_0\) reflects differences in their mean untreated potential outcomes. These differences can arise because treated and comparison units with the same observed category \(W\) may have different distributions of the underlying covariate \(X\), on which \(\mu_0(X)\) and \(\tau(X)\) depend. For the ATT, the treatment effect heterogeneity component drops out. The remaining composition bias \(\widetilde B_0\) averages the same within-cell differences in mean untreated potential outcomes over the treated population rather than the full population.

\begin{figure}[htbp]
\centering
\includegraphics[width=\textwidth]{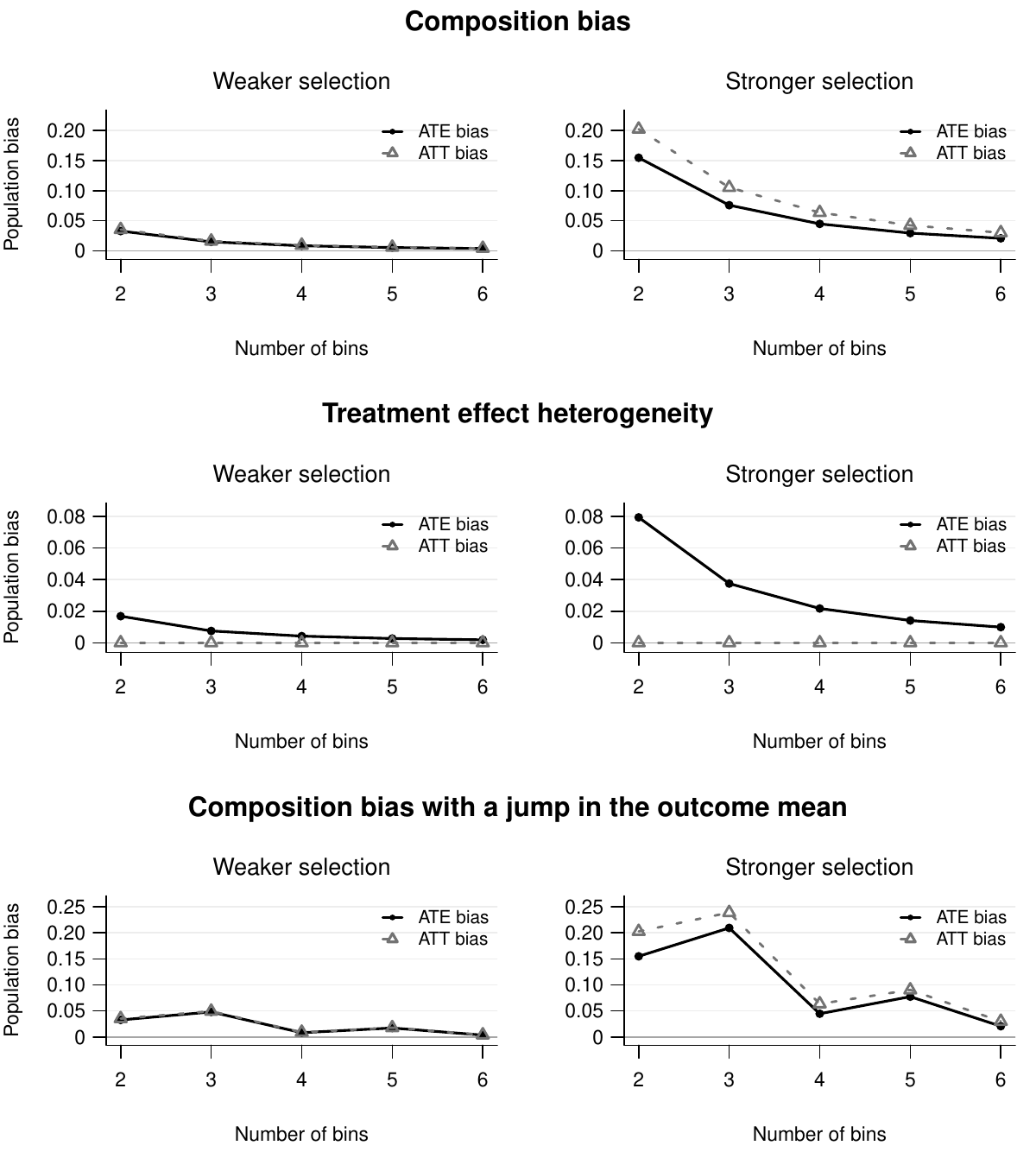}
\caption{A Numerical Illustration of Coarsening Bias}
\label{fig:coarsening_bias_illustration}
\fnote{\textit{Notes:} Each panel plots the coarsening bias for the ATE and the ATT against the number of equal-width bins \(K\) into which the support of \(X\sim U[0,1]\) is divided. Treatment is assigned with probability \(0.5+s(x-0.5)\), with \(s=0.2\) under weaker selection and \(s=0.8\) under stronger selection. The top panels set \(\mu_0(x)=2x^3\) and \(\tau(x)=1\), the middle panels set \(\mu_0(x)=0\) and \(\tau(x)=1+2x\), and the bottom panels set \(\mu_0(x)=2x^3+3\cdot\mathbb{I}\{x\geq 1/2\}\) and \(\tau(x)=1\).}
\end{figure}

Figure~\ref{fig:coarsening_bias_illustration} gives a simple numerical illustration of these bias terms. The first two designs vary the potential outcome mean functions to separate the composition channel from the treatment effect heterogeneity channel, and the third shows that more bins do not always reduce the bias. For each design, I compare weaker and stronger treatment selection. Let $X\sim U[0,1]$, and divide its support into $K$ equal-width bins, where $K\in\{2,\ldots,6\}$. Treatment is assigned with probability
\[
\Pr(D=1\mid X=x)=0.5+s(x-0.5),
\qquad s\in\{0.2,0.8\}.
\]
A larger value of $s$ makes treatment probabilities vary more with $X$. Potential outcomes are $Y(0)=\mu_0(X)+\varepsilon_0$ and $Y(1)=\mu_0(X)+\tau(X)+\varepsilon_1$, where $\varepsilon_0$ and $\varepsilon_1$ are independent standard normal shocks, independent of $(D,X)$. The top panels isolate composition bias by setting $\mu_0(x)=2x^3$ and $\tau(x)=1$, and the middle panels isolate the treatment effect heterogeneity component by setting $\mu_0(x)=0$ and $\tau(x)=1+2x$. The bottom panels add a jump at $x=1/2$ to the design of the top panels, with $\mu_0(x)=2x^3+3\cdot\mathbb{I}\{x\geq 1/2\}$ and $\tau(x)=1$. Within each row, treatment selection is weaker on the left ($s=0.2$) and stronger on the right ($s=0.8$).

The top panels show positive composition bias for both the ATE and ATT. The middle panels show positive ATE bias from treatment effect heterogeneity, while the ATT bias is zero by construction. In these two designs, bias decreases as the number of bins increases, and stronger treatment selection results in a larger ATE bias at each $K$. In the bottom panels, by contrast, the bias does not decrease monotonically, because the jump at $x=1/2$ coincides with a bin boundary when $K$ is even and lies strictly inside a bin when $K$ is odd.

\begin{figure}[htbp]
\centering
\definecolor{groupzero}{HTML}{0072B2}
\definecolor{groupone}{HTML}{C04F00}
\begin{tikzpicture}[x=0.8cm,y=0.78cm]
\draw[black!25] (0,0) -- (0,5.3);
\draw[black!25] (8,0) -- (8,5.3);
\draw[->] (-0.3,0) -- (8.8,0);
\draw (0,-0.12) -- (0,0.12);
\draw (8,-0.12) -- (8,0.12);
\node[below] at (0,-0.12) {\footnotesize $50{,}000$};
\node[below] at (8,-0.12) {\footnotesize $75{,}000$};
\node at (4,-1.05) {\footnotesize exact family income $X$ inside one reported category $W=k$};
\draw[thick,groupzero] plot[domain=0:8,samples=101]
  (\x,{0.55*(0.2+24*(\x/8)*(1-\x/8)^4)});
\draw[thick,densely dashed,groupone] plot[domain=0:8,samples=101]
  (\x,{0.55*(0.2+24*(\x/8)^4*(1-\x/8))});
\node[align=center,groupzero] at (0.8,2.05) {\footnotesize public school\\[-2pt]\footnotesize $D=0$};
\node[align=center,groupone] at (7.2,2.05) {\footnotesize private school\\[-2pt]\footnotesize $D=1$};
\draw[thick] plot[domain=0:8,samples=81] (\x,{3.3+1.75*(2*\x/8-(\x/8)^2)});
\node[anchor=east] at (-0.15,3.3) {\footnotesize $\mu_0(x)$};
\draw[groupzero] (0,4.1833) -- (8.6,4.1833);
\draw[densely dashed,groupone] (0,4.7833) -- (8.6,4.7833);
\node[anchor=south west,groupone,inner sep=1pt] at (0.1,4.83) {\footnotesize private school average};
\node[anchor=north east,groupzero,inner sep=1pt] at (7.9,4.14) {\footnotesize public school average};
\draw[<->] (8.6,4.1833) -- (8.6,4.7833);
\node[right,align=left] at (8.75,4.4833) {\footnotesize $\Delta_{\mu_0(X)}(k)$\\[-2pt]\footnotesize $>0$};
\end{tikzpicture}
\caption{Within-Cell Sorting and Positive Composition Bias}
\label{fig:within_cell_sorting_composition}
\fnote{\textit{Notes:} The solid and dashed curves at the bottom show the densities of exact family income \(X\) for public and private school students within one reported income category \(W=k\). The upward-sloping curve shows the mean public school outcome \(\mu_0(x)\). The solid and dashed horizontal lines show the averages of \(\mu_0(X)\) over these two densities, and the vertical gap between them is \(\Delta_{\mu_0(X)}(k)\).}
\end{figure}
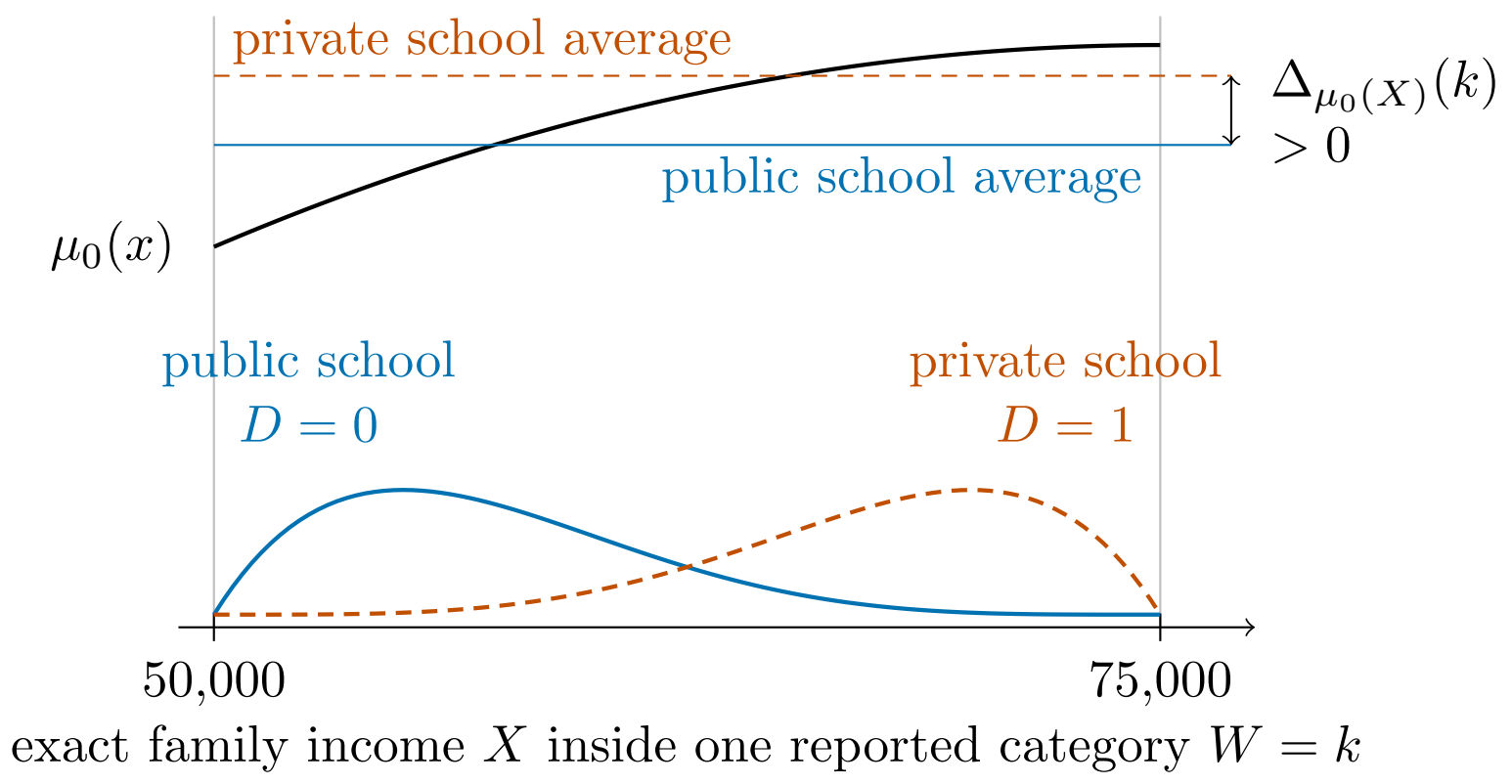

Consider the effect of private school attendance on children's test scores. Let $X$ denote exact family income and $Y(0)$ the test score a child would attain in public school. Suppose mean independence for $Y(0)$ holds conditional on $X$, but the researcher observes only income categories $W$. Figure~\ref{fig:within_cell_sorting_composition} illustrates one such category, covering incomes from $50{,}000$ to $75{,}000$. In this illustration, children attending private school are concentrated at higher family incomes within the category, and the mean public school outcome $\mu_0(x)$ increases with income.

The vertical gap in Figure~\ref{fig:within_cell_sorting_composition} represents the positive difference in mean public school potential outcomes between private and public school students within the same income category, $\Delta_{\mu_0(X)}(k)$. Averaging these gaps over the full population yields $B_0$, while averaging over the treated population yields $\widetilde B_0$. If the effects of private school attendance also vary with family income, the same sorting may generate differences in average treatment effects within a category and contribute to $B_\tau$ in the ATE decomposition.

The same channels can also be written at the level of a coarsened cell.
Section~\ref{sec:supp_conditional_coarsening_bias} of the supplement gives the
conditional analogues.

The next result gives three cases in which some or all coarsening bias terms
vanish.

\begin{corollary}[Sufficient conditions for vanishing coarsening channels]
\label{cor:coarsen_harmless}
Under the assumptions of Proposition~\ref{prop:coarsen_bias_decomposition}:
\begin{enumerate}[label=\textup{(\roman*)}]
\item If $\tau\left(X\right)=r_\tau\left(W\right)$ a.s.\ for some measurable
function $r_\tau$, then \(B_{\tau}=0\).
\item If $\mu_0(X)=r_0\left(W\right)$ a.s.\ for some measurable function
$r_0$, then \(B_0=\widetilde B_0=0\).
\item If \(e^*\left(X\right)=e\left(W\right)\) a.s., then
\(B_{\tau}=B_0=\widetilde B_0=0\).
\end{enumerate}
\end{corollary}

Corollary~\ref{cor:coarsen_harmless} gives useful sufficient conditions under
which specific coarsening bias channels vanish. In the school choice example,
the first condition says that average effects of private school do not vary
within each family income category. Treated and comparison children may still be
sorted differently across exact family income values inside the category, but
this sorting cannot change the cell average of \(\tau\left(X\right)\). Constant
average treatment effects are a special case: if the average effect of private
school is the same at all exact family income values, the treatment effect
heterogeneity channel is absent under any coarsening. The second condition gives
the same logic for average test scores in the absence of private school: if
\(\mu_0\left(x\right)\) does not vary within each family income category,
differences in exact family income composition do not translate into composition
bias. If
\(e^*\left(X\right)=e\left(W\right)\) a.s., then private school attendance is
not associated with exact family income after conditioning on the family income
category. In that case, for each \(w\in\mathcal{W}\), one can show that \(X\mid D=1,W=w\sim X\mid D=0,W=w\), that is, treated and comparison children have the same conditional distribution of exact family income within the category. All within-cell sorting disappears, and
all three coarsening bias terms vanish.

The next corollary considers selection toward larger average treatment effects or larger average outcomes without treatment within each coarsened cell. Under monotone selection, these forms of sorting yield positive coarsening bias terms whenever selection varies within at least one cell.

\begin{corollary}[Coarsening bias under monotone selection]
\label{cor:positive_coarsening_bias}
Suppose the assumptions of Proposition~\ref{prop:coarsen_bias_decomposition} hold. For each $w\in\mathcal{W}$, define
\begin{align*}
e_{\tau}\left(y,w\right)
&:=\mathrm{Pr}\left(D=1\mid \tau\left(X\right)=y,W=w\right),\\
e_0\left(y,w\right)
&:=\mathrm{Pr}\left(D=1\mid \mu_0\left(X\right)=y,W=w\right).
\end{align*}
\begin{enumerate}[label=\textup{(\roman*)}]
\item Suppose $y\mapsto e_{\tau}\left(y,w\right)$ is weakly increasing on the conditional support of $\tau\left(X\right)$ given $W=w$ for all $w\in\mathcal{W}$. If $e_{\tau}\left(\tau\left(X\right),w\right)$ is not a.s.\ constant conditional on $W=w$ for some $w\in\mathcal{W}$, then
\begin{align*}
B_{\tau}>0.
\end{align*}
\item Suppose $y\mapsto e_0\left(y,w\right)$ is weakly increasing on the conditional support of $\mu_0\left(X\right)$ given $W=w$ for all $w\in\mathcal{W}$. If $e_0\left(\mu_0\left(X\right),w\right)$ is not a.s.\ constant conditional on $W=w$ for some $w\in\mathcal{W}$, then
\begin{align*}
B_0>0,\qquad \widetilde B_0>0.
\end{align*}
\end{enumerate}
\end{corollary}

The monotonicity conditions can follow from restrictions on the fine covariate. If \(e^*(x)\) and \(\tau(x)\) are both weakly increasing on the conditional support of \(X\) given \(W=w\), then \(e_\tau(y,w)\) is weakly increasing in \(y\). To see the implication, note that \(\tau(X)\) and \(W\) are functions of \(X\). Hence, iterated expectations yield
\begin{align*}
e_\tau(y,w)
&=\mathrm E\!\left[\mathrm E[D\mid X,\tau(X),W]\mid\tau(X)=y,W=w\right]\\
&=\mathrm E[e^*(X)\mid\tau(X)=y,W=w].
\end{align*}
Intuitively, within a coarsened cell, units with larger average gains have higher values of \(X\) because \(\tau(x)\) is weakly increasing. Since \(e^*(x)\) is also weakly increasing, these units are at least as likely to receive treatment. Thus, \(e_\tau(y,w)\) is weakly increasing in \(y\). The same argument applies to \(e_0(y,w)\) when \(\mu_0(x)\) replaces \(\tau(x)\).

Corollary~\ref{cor:positive_coarsening_bias} translates monotone within-cell
sorting into sign restrictions on the bias terms. Monotone selection also appears in Proposition~3.1 of \citet{masten_choosing_2023}, where mean independence fails if the latent propensity score, the probability of treatment given the potential outcome, is weakly monotone and nonconstant on the interior of its domain, under regularity conditions. In Corollary~\ref{cor:positive_coarsening_bias}, by contrast, mean independence holds given the fine covariate \(X\), while monotone sorting within cells of \(W\) determines the sign of the coarsening bias. In the school choice example, \(\tau(x)\) is the average effect of private school on test scores for children with exact family income \(x\), and \(\mu_0(x)\) is their average test score in the absence of private school. Suppose that, within each family income category, both the probability of private school attendance and the average gains from attending are weakly increasing in income. These restrictions imply the monotonicity condition on \(e_\tau\). Treated children then have weakly larger average gains than comparison children within the same category. If this selection on average gains is nontrivial in at least one category, the treatment effect heterogeneity component \(B_\tau\) is positive.

The same reasoning applies to composition bias. If the probability of private school attendance and average test scores in the absence of private school are both weakly increasing in income within each category, then the monotonicity condition on \(e_0\) holds. Consequently, \(B_0\) and \(\widetilde B_0\) are nonnegative, and both are positive if selection on average test scores without private school is nontrivial in at least one category. If attendance is weakly increasing in income but average test scores without private school are weakly decreasing, the composition terms are instead nonpositive, and both are negative if this selection is nontrivial in at least one category.
At the cell level, the same monotonicity argument
can be applied to the conditional formulas in
Section~\ref{sec:supp_conditional_coarsening_bias} of the supplement.

The coarsening problem also affects quantile treatment effects. Section~\ref{sec:supp_quantile_coarsening_bias} of the supplement compares quantiles of potential outcome distributions with those obtained by averaging observed conditional distribution functions over the coarsened covariate and then inverting. Since generalized inversion is nonlinear, the resulting quantile biases need not admit the additive channel decomposition in Proposition~\ref{prop:coarsen_bias_decomposition}.

The same issue arises in DiD designs when conditional parallel trends holds given \(X\) but not given \(W\). Section~\ref{sec:supp_did_coarsening_bias} of the supplement derives the resulting ATT bias and shows how restrictions relating it to pre-treatment comparisons can be used to correct or bound it.

Taken together, the results in this section show that a valid comparison based on \(X\) can become biased when it is formed using \(W\). The decomposition identifies the channels of this failure, but it does not by itself restore identification when only \(W\) is available. Section~\ref{sec:c_mean_dependence} addresses this problem by restricting average unobserved potential outcomes using observed means in neighboring cells.

\section{\texorpdfstring{\(c\)}{c}-Mean Dependence}
\label{sec:c_mean_dependence}

\subsection{Identification under \texorpdfstring{\(c\)}{c}-Mean Dependence}

The preceding section studies how coarsening can invalidate mean independence that holds given the finely measured covariate \(X\). In many applications, however, the finer covariate is unavailable, or researchers choose to form comparisons using the coarsened covariate \(W\). To study identification of the ATT and ATE in this setting, I introduce \(c\)-mean dependence, a nonparametric class of assumptions that relaxes mean independence given \(W\).

Suppose the coarsened covariate \(W\) has support \(\mathcal W\subseteq\mathbb R\). For \(d\in\{0,1\}\), assume \(\mathrm E[|Y(d)|]<\infty\) and \(0<e(w)<1\) for all \(w\in\mathcal W\), where \(e(w):=\Pr(D=1\mid W=w)\). Define the observed mean and the average unobserved potential outcome in cell \(w\) by
\begin{align*}
m_d(w)&:=\mathrm E[Y\mid D=d,W=w]=\mathrm E[Y(d)\mid D=d,W=w],\\
m_d^{\mathrm{cf}}(w)&:=\mathrm E[Y(d)\mid D=1-d,W=w].
\end{align*}
The second equality follows from the observed outcome equation \eqref{eq:observed_outcome}. The average \(m_d^{\mathrm{cf}}(w)\) is unobserved because \(Y(d)\) is not observed for units with \(D=1-d\).

For simplicity, I omit additional observed covariates. The restriction below extends to such covariates \(V\) by applying it separately at each value of \(V\). The average unobserved potential outcome for units with \(W=w\) and \(V=v\) is then bounded by the observed means for units with the same value of \(V\) and nearby values of \(W\). Section~\ref{sec:supp_additional_covariates} of the supplement develops this extension and discusses the case in which \(V\) contains continuous components.

\begin{definition}[$c$-mean dependence]
\label{def:c_mean_dependence}
Let \(d\in\{0,1\}\), \(c\geq0\), and \(w\in\mathcal W\). \(Y(d)\) is \(c\)-mean dependent with \(D\) given \(W=w\) if
\begin{align}
m_d^{\mathrm{cf}}(w)\in\left[\underline m_{d,c}(w),\overline m_{d,c}(w)\right],
\label{eq:c_mean_dependence_cell}
\end{align}
where
\[
\underline m_{d,c}(w):=\inf_{w'\in\mathcal N_c(w)}m_d(w'),
\qquad
\overline m_{d,c}(w):=\sup_{w'\in\mathcal N_c(w)}m_d(w'),
\]
and \(\mathcal N_c(w):=\{w'\in\mathcal W\mid |w'-w|\leq c\}\). If this condition holds for all \(w\in\mathcal W\), \(Y(d)\) is \(c\)-mean dependent with \(D\) given \(W\).
\end{definition}

Here, \(\mathcal N_c(w)\) is the \(c\)-neighborhood of \(w\) in the support \(\mathcal W\). For a vector-valued coarsened covariate, this neighborhood can be defined using the sup-norm in place of the absolute value. The condition requires the average unobserved potential outcome \(m_d^{\mathrm{cf}}(w)\) to lie between the smallest and largest observed means \(m_d(w')\) over cells \(w'\) in this neighborhood.

Thus, \(c\)-mean dependence relaxes exact mean independence given \(W\) by replacing equality with a local range restriction. The parameter \(c\) determines the size of the neighborhood and hence the range of observed means used to restrict the average unobserved potential outcome. When \(c=0\), the neighborhood contains only \(w\), so the condition reduces to mean independence given \(W\). As \(c\) increases, the neighborhood expands and the restriction becomes weaker. At the other extreme, when \(c\) is large enough to include all cells in each neighborhood, \(c\)-mean dependence requires only that average unobserved potential outcomes lie within the overall range of corresponding observed cell means. The role of \(c\) parallels the sensitivity parameters in \citet{masten_identification_2018} and \citet{masten_assessing_2024}. Here, \(c\) measures how far across the coarsened support observed means may be used to restrict an average unobserved potential outcome.

The next proposition characterizes what \(c\)-mean dependence implies for the ATT alone and for the ATT and the ATE jointly.

\begin{proposition}[Identified sets under $c$-mean dependence]
\label{prop:c_mean_dependence_bounds}
Suppose the expectations below are finite, and let $c\geq0$.
\begin{enumerate}[label=\textup{(\roman*)}]
\item If $Y\left(0\right)$ is $c$-mean dependent with $D$ given $W$, then
\begin{align*}
\tau_{\mathrm{ATT}}
\in
\left[
\underline\tau_{\mathrm{ATT},c},
\overline\tau_{\mathrm{ATT},c}
\right],
\end{align*}
where
\begin{align*}
\underline\tau_{\mathrm{ATT},c}
&:=
\mathrm{E}\left[Y\mid D=1\right]
-
\mathrm{E}\left[\overline m_{0,c}\left(W\right)\mid D=1\right],
\\
\overline\tau_{\mathrm{ATT},c}
&:=
\mathrm{E}\left[Y\mid D=1\right]
-
\mathrm{E}\left[\underline m_{0,c}\left(W\right)\mid D=1\right].
\end{align*}
\item If $Y\left(0\right)$ and $Y\left(1\right)$ are $c$-mean dependent with $D$ given $W$, then
\begin{align*}
\left(\tau_{\mathrm{ATT}},\tau_{\mathrm{ATE}}\right)
\in
\mathcal I_c,
\end{align*}
where
\[
\mathcal I_c
:=
\left\{
\left(t_1,t_2\right)\in\mathbb R^2
\;\middle|\;
\begin{aligned}
&\underline\tau_{\mathrm{ATT},c}
\leq
t_1
\leq
\overline\tau_{\mathrm{ATT},c},
\\
&\underline\tau_{\mathrm{ATE},c}+\pi_1\left(t_1-\underline\tau_{\mathrm{ATT},c}\right)
\leq
t_2
\leq
\overline\tau_{\mathrm{ATE},c}-\pi_1\left(\overline\tau_{\mathrm{ATT},c}-t_1\right)
\end{aligned}
\right\},
\]
\(\pi_1:=\mathrm{Pr}\left(D=1\right)\), and
\begin{equation}
\label{eq:c_mean_ate_bounds}
\begin{aligned}
\underline\tau_{\mathrm{ATE},c}
&:=
\mathrm{E}\left[
e\left(W\right)
\bigl(
m_{1}\left(W\right)
-
\overline m_{0,c}\left(W\right)
\bigr)
+
\left(1-e\left(W\right)\right)
\bigl(
\underline m_{1,c}\left(W\right)
-
m_{0}\left(W\right)
\bigr)
\right],
\\
\overline\tau_{\mathrm{ATE},c}
&:=
\mathrm{E}\left[
e\left(W\right)
\bigl(
m_{1}\left(W\right)
-
\underline m_{0,c}\left(W\right)
\bigr)
+
\left(1-e\left(W\right)\right)
\bigl(
\overline m_{1,c}\left(W\right)
-
m_{0}\left(W\right)
\bigr)
\right].
\end{aligned}
\end{equation}
\end{enumerate}
Moreover, the interval in \textup{(i)} and the set \(\mathcal I_c\) in \textup{(ii)} are, respectively, the sharp identified sets for \(\tau_{\mathrm{ATT}}\) and \((\tau_{\mathrm{ATT}},\tau_{\mathrm{ATE}})\): every value in the interval, and
every pair in \(\mathcal I_c\), is attained by some data-generating process
that satisfies the maintained assumptions and yields the observed distribution
of \(\left(Y,D,W\right)\).
\end{proposition}

The joint set \(\mathcal I_c\) yields sharp bounds for each effect separately. Its projection onto the first coordinate is the ATT interval in part~\textup{(i)}, and its projection onto the second coordinate is the interval \(\left[\underline\tau_{\mathrm{ATE},c},\overline\tau_{\mathrm{ATE},c}\right]\) defined in \eqref{eq:c_mean_ate_bounds}. Unlike the ATT bounds, the ATE bounds require both \(Y(0)\) and \(Y(1)\) to be \(c\)-mean dependent. At \(c=0\), both intervals shrink to single points, so the ATT and the ATE are point identified, as under mean independence given \(W\). The intervals widen weakly as \(c\) increases.

The joint set \(\mathcal I_c\) is strictly smaller than the Cartesian product \(\left[\underline\tau_{\mathrm{ATT},c},\overline\tau_{\mathrm{ATT},c}\right]\times\left[\underline\tau_{\mathrm{ATE},c},\overline\tau_{\mathrm{ATE},c}\right]\) of the intervals in part~\textup{(i)} and \eqref{eq:c_mean_ate_bounds}, unless the ATT is point identified. The reason is that the ATE is a weighted average of the ATT and the average treatment effect for untreated units, with weights \(\pi_1\) and \(1-\pi_1\). Hence, the ATE attains its upper bound only if the ATT attains its upper bound, and likewise for the lower bounds. The joint set therefore excludes pairs that the Cartesian product contains, such as the lower bound of the ATT together with the upper bound of the ATE. Section~\ref{sec:estimation_c_mean_dependence} develops estimation and inference for the ATT and ATE bounds separately.

\subsection{Interpreting and Choosing \texorpdfstring{\(c\)}{c}}
\label{sec:choosing_c}

The choice of \(c\) determines which neighboring cells are used to restrict each average unobserved potential outcome. Substantive or institutional knowledge about the comparability of these cells can guide this choice. Alternatively, researchers can vary \(c\) to assess the sensitivity of their conclusions.

In the school choice example, suppose \(W\) denotes ordered family income categories. Consider the mean test score that children attending private schools would attain if they attended public schools instead. At \(c=0\), this mean in each income category equals the observed mean for public school children in the same category. At \(c=1\), it may lie within the range of observed public school means in that category and the adjacent categories. At \(c=2\), the range also includes means from categories two steps away.

The next result adapts the monotonicity argument in Proposition 1 of \citet{manski_inference_2002} to justify \(c=1\) when \(X\) is grouped into consecutive intervals. Suppose \(W\) has support \(\mathcal W=\{1,\ldots,K\}\), where \(K\geq3\), and is generated by a partition of \(\mathcal X\) into
\[
A_k:=\mathcal X\cap[a_k,a_{k+1}),
\qquad
W=k\Longleftrightarrow X\in A_k,
\qquad k=1,\ldots,K,
\]
where \(a_1<\cdots<a_{K+1}\) are the category cutoffs. Thus, \(W\) indicates which interval contains the fine covariate \(X\).

\begin{proposition}[Monotonicity and \(1\)-mean dependence]
\label{prop:one_step_c_mean_dependence}
Let \(d\in\{0,1\}\). Suppose Assumption~\ref{ass:overlap} holds, together with Assumption~\ref{ass:mean_independence_y0} if \(d=0\) or Assumption~\ref{ass:mean_independence_y1} if \(d=1\), and \(\mathrm E[|Y(d)|]<\infty\). Suppose \(\mu_d\) is monotone. Then \(Y(d)\) is \(1\)-mean dependent with \(D\) given \(W\) if and only if \(m_d^{\mathrm{cf}}(1)\) and \(m_d^{\mathrm{cf}}(K)\) lie between \(m_d(1)\) and \(m_d(K)\).
\end{proposition}

Proposition~\ref{prop:one_step_c_mean_dependence} motivates \(c=1\) as a simple benchmark for assessing sensitivity to covariate coarsening. Under its assumptions, the average unobserved potential outcome \(m_d^{\mathrm{cf}}(k)\) in an interior cell \(k\) lies between the observed means \(m_d(k-1)\) and \(m_d(k+1)\), as the proof shows. Thus, comparisons with the neighboring cell on either side suffice to bound \(m_d^{\mathrm{cf}}(k)\), even when mean independence fails within cell \(k\). An endpoint cell has a neighboring cell on one side only, and the condition in the proposition extends this justification to the endpoint cells. It requires the range of observed means to cover the average unobserved potential outcomes in these two cells. When \(\mu_d\) is nondecreasing, it reduces to \(m_d^{\mathrm{cf}}(1)\geq m_d(1)\) and \(m_d^{\mathrm{cf}}(K)\leq m_d(K)\).

Outcomes measured before treatment can provide an empirical benchmark for \(c\). In the school choice example, private school attendance cannot affect test scores measured before children enroll, provided that families do not change their behavior in anticipation of enrollment. Its effect on these earlier scores is then zero. If mean independence held given the income category, comparing the earlier scores of private and public school children within each category would recover this zero effect. A nonzero difference therefore indicates a departure from mean independence. Since the effect on these earlier scores is zero, the bounds on it should contain zero, and the smallest \(c\) at which the data do not reject this necessary condition can serve as a benchmark for test scores after enrollment. Section~\ref{sec:application} applies this approach to earnings before application to the military.

When an additional finely measured covariate \(V\) is observed, its coarsening can provide another empirical benchmark. Conditional on \(W\) and treatment status, compare the observed means given \(V\) with those obtained after grouping \(V\) into ordered categories. Let \(U:=r(V)\) denote the category of \(V\). For \(d\in\{0,1\}\), define
\[
m_d^V(w,v):=\mathrm E[Y\mid D=d,W=w,V=v],
\qquad
m_d^U(w,u):=\mathrm E[Y\mid D=d,W=w,U=u].
\]
For each value \(v\), consider the smallest neighborhood of \(u=r(v)\) such that \(m_d^V(w,v)\) lies between the smallest and largest coarse means \(m_d^U(w,u')\) over \(u'\) in that neighborhood. The size of this neighborhood, when it exists, provides a reference for interpreting \(c\). This exercise is analogous to the observed covariate benchmark of \citet{masten_identification_2018}, which examines how adding an observed covariate changes the conditional probability of treatment. Here, the exercise instead examines how coarsening an observed covariate changes conditional mean functions.

Beyond these benchmarks, researchers can report how the bounds change over a range of \(c\), in the spirit of the sensitivity analysis of \citet{masten_identification_2018}. The robustness of a particular conclusion can also be summarized by its breakdown point, following \citet{masten_inference_2020}. For example, if \(\underline\tau_{\mathrm{ATT},0}\geq0\), the breakdown point \(c^*:=\sup\{c\geq0\mid\underline\tau_{\mathrm{ATT},c}\geq0\}\) summarizes robustness of the conclusion that the ATT is nonnegative, with \(c^*=\infty\) if the conclusion holds at every value of \(c\).

\section{Estimation and Inference}
\label{sec:estimation_c_mean_dependence}

The preceding section expresses the ATT and ATE bounds as functions of observed cell probabilities and conditional means. This section defines their plug-in estimators and derives the joint convergence of the lower and upper bound estimators for each effect. It then constructs confidence bands that cover both bound functions simultaneously over the sensitivity values under consideration. Let $\left\{\left(Y_i,D_i,W_i\right)\right\}_{i=1}^N$ denote the observed sample. Let $\mathcal C\subseteq\left[0,\infty\right)$ be a nonempty set of sensitivity values at which the bound functions are evaluated.

Replacing the population quantities in
Proposition~\ref{prop:c_mean_dependence_bounds} by their empirical
counterparts gives the following estimators. Here and below, whenever a denominator in a sample estimator, bootstrap analogue, or generic plug-in functional is zero, the corresponding ratio is set equal to zero.\footnote{Under
Assumptions~\ref{ass:c_mean_est_sampling}--\ref{ass:c_mean_est_regularity}
below, the probability that any denominator appearing in the sample estimators
or bootstrap analogues is zero converges to zero.}
For any sequence of scalar random variables \(Z_1,\ldots,Z_N\), write
\begin{equation}
\label{eq:empirical_mean}
\widehat{\mathrm{E}}\left[Z_i\right]
:=
\frac{1}{N}\sum_{i=1}^N Z_i,
\qquad
\widehat{\mathrm{E}}\left[Z_i\mid D_i=1\right]
:=
\frac{
\sum_{i=1}^N Z_i\mathbb{I}\left\{D_i=1\right\}
}{
\sum_{i=1}^N \mathbb{I}\left\{D_i=1\right\}
}.
\end{equation}
For $d\in\left\{0,1\right\}$, $w\in\mathcal W$, and $c\in\mathcal C$, define
\begin{align*}
\widehat e\left(w\right)
&:=
\frac{
\widehat{\mathrm{E}}\left[\mathbb{I}\left\{D_i=1,W_i=w\right\}\right]
}{
\widehat{\mathrm{E}}\left[\mathbb{I}\left\{W_i=w\right\}\right]
},
\qquad
\widehat m_{d}\left(w\right)
:=
\frac{
\widehat{\mathrm{E}}\left[Y_i\mathbb{I}\left\{D_i=d,W_i=w\right\}\right]
}{
\widehat{\mathrm{E}}\left[\mathbb{I}\left\{D_i=d,W_i=w\right\}\right]
},
\\
\widehat{\underline m}_{d,c}\left(w\right)
&:=
\min_{w'\in\mathcal{N}_c\left(w\right)}
\widehat m_{d}\left(w'\right),
\qquad
\widehat{\overline m}_{d,c}\left(w\right)
:=
\max_{w'\in\mathcal{N}_c\left(w\right)}
\widehat m_{d}\left(w'\right).
\end{align*}
The plug-in estimators of the ATE bound functions are
\[
\begin{aligned}
\widehat{\underline\tau}_{\mathrm{ATE},c}
&:=
\widehat{\mathrm{E}}\left[
\widehat e\left(W_i\right)
\left[
\widehat m_{1}\left(W_i\right)
-
\widehat{\overline m}_{0,c}\left(W_i\right)
\right]
+
\left[
1-\widehat e\left(W_i\right)
\right]
\left[
\widehat{\underline m}_{1,c}\left(W_i\right)
-
\widehat m_{0}\left(W_i\right)
\right]
\right],
\\
\widehat{\overline\tau}_{\mathrm{ATE},c}
&:=
\widehat{\mathrm{E}}\left[
\widehat e\left(W_i\right)
\left[
\widehat m_{1}\left(W_i\right)
-
\widehat{\underline m}_{0,c}\left(W_i\right)
\right]
+
\left[
1-\widehat e\left(W_i\right)
\right]
\left[
\widehat{\overline m}_{1,c}\left(W_i\right)
-
\widehat m_{0}\left(W_i\right)
\right]
\right],
\end{aligned}
\]
and those of the ATT bound functions are
\[
\begin{aligned}
\widehat{\underline\tau}_{\mathrm{ATT},c}
&:=
\widehat{\mathrm{E}}\left[Y_i\mid D_i=1\right]
-
\widehat{\mathrm{E}}\left[\widehat{\overline m}_{0,c}\left(W_i\right)
\mid D_i=1\right],
\\
\widehat{\overline\tau}_{\mathrm{ATT},c}
&:=
\widehat{\mathrm{E}}\left[Y_i\mid D_i=1\right]
-
\widehat{\mathrm{E}}\left[\widehat{\underline m}_{0,c}\left(W_i\right)
\mid D_i=1\right].
\end{aligned}
\]

Suppose the following sampling and support conditions hold.

\begin{assumption}[Sampling]
\label{ass:c_mean_est_sampling}
$\left\{\left(Y_i,D_i,W_i\right)\right\}_{i=1}^N$ are i.i.d.\ copies of
$\left(Y,D,W\right)$.
\end{assumption}

\begin{assumption}[Finite support]
\label{ass:c_mean_est_support}
The coarsened covariate $W$ has finite support $\mathcal W$.
\end{assumption}

\begin{assumption}[Regularity]
\label{ass:c_mean_est_regularity}
For each $d\in\left\{0,1\right\}$ and $w\in\mathcal W$,
$
\mathrm{Pr}\left(D=d,W=w\right)>0.
$
Moreover, $\mathrm{E}\left[Y^2\right]<\infty$.
\end{assumption}

The bound estimators depend on the data only through finitely many cell probabilities and cell means. To derive their asymptotic behavior, I write them as a map of these quantities and apply the functional delta method. For \(d\in\left\{0,1\right\}\) and \(w\in\mathcal W\), define the cell probabilities \(p_d(w):=\Pr\left(D=d,W=w\right)\) and their estimators \(\widehat p_d(w):=\widehat{\mathrm{E}}\left[\mathbb{I}\left\{D_i=d,W_i=w\right\}\right]\). Collect these probabilities together with the cell means \(m_d(w)\) and their estimators \(\widehat m_d(w)\) into
\begin{equation}
\label{eq:c_mean_first_stage_components}
\begin{aligned}
\vartheta
&:=
\left(
\left(p_d(w)\right)_{d\in\left\{0,1\right\},\,w\in\mathcal W},
\left(m_d(w)\right)_{d\in\left\{0,1\right\},\,w\in\mathcal W}
\right)\in\mathbb R^{4K},
\\
\widehat\vartheta
&:=
\left(
\left(\widehat p_d(w)\right)_{d\in\left\{0,1\right\},\,w\in\mathcal W},
\left(\widehat m_d(w)\right)_{d\in\left\{0,1\right\},\,w\in\mathcal W}
\right),
\end{aligned}
\end{equation}
where \(K:=|\mathcal W|\).
For any vector \(\eta\in\mathbb R^{4K}\), write \(\eta^p_d(w)\) and \(\eta^m_d(w)\) for its coordinates in the positions of \(p_d(w)\) and \(m_d(w)\) in \(\vartheta\). For \(d\in\left\{0,1\right\}\), \(w\in\mathcal W\), and \(c\in\mathcal C\), define
\[
\underline m_{d,c}\left(w;\eta\right):=\min_{w'\in\mathcal N_c(w)}\eta^m_d(w'),
\qquad
\overline m_{d,c}\left(w;\eta\right):=\max_{w'\in\mathcal N_c(w)}\eta^m_d(w').
\]
For \(\rho\in\left\{\mathrm{ATE},\mathrm{ATT}\right\}\), define
\(\phi_{\rho}:\mathbb R^{4K}\to\ell^\infty\left(\mathcal C,\mathbb R^2\right)\)
by \(\left[\phi_{\rho}\left(\eta\right)\right](c):=\left(\left[\phi_{\rho,L}\left(\eta\right)\right](c),\left[\phi_{\rho,U}\left(\eta\right)\right](c)\right)'\),\footnote{Let
\(\mathscr{A}\) be an arbitrary set and \(\mathscr{B}\) a Banach space. Then
\(\ell^{\infty}\left(\mathscr{A},\mathscr{B}\right)\) denotes the space of all
bounded functions \(f:\mathscr{A}\to\mathscr{B}\) such that
\(\sup_{a\in\mathscr{A}}\left\|f\left(a\right)\right\|_{\mathscr{B}}<\infty\),
equipped with the sup-norm
\(\left\|f\right\|_{\infty}:=
\sup_{a\in\mathscr{A}}\left\|f\left(a\right)\right\|_{\mathscr{B}}\) (see,
e.g., \citealp{van_der_vaart_weak_2023}).} where
\begin{equation}
\label{eq:c_mean_phi_components}
\begin{aligned}
\left[\phi_{\mathrm{ATE},L}\left(\eta\right)\right](c)
&:=
\sum_{w\in\mathcal W}
\left[
\eta^p_1(w)
\left(\eta^m_1(w)-\overline m_{0,c}\left(w;\eta\right)\right)
+
\eta^p_0(w)
\left(\underline m_{1,c}\left(w;\eta\right)-\eta^m_0(w)\right)
\right],
\\
\left[\phi_{\mathrm{ATE},U}\left(\eta\right)\right](c)
&:=
\sum_{w\in\mathcal W}
\left[
\eta^p_1(w)
\left(\eta^m_1(w)-\underline m_{0,c}\left(w;\eta\right)\right)
+
\eta^p_0(w)
\left(\overline m_{1,c}\left(w;\eta\right)-\eta^m_0(w)\right)
\right],
\\
\left[\phi_{\mathrm{ATT},L}\left(\eta\right)\right](c)
&:=
\frac{
\sum_{w\in\mathcal W}
\left[
\eta^p_1(w)
\left(\eta^m_1(w)-\overline m_{0,c}\left(w;\eta\right)\right)
\right]
}{
\sum_{w\in\mathcal W} \eta^p_1(w)
},
\\
\left[\phi_{\mathrm{ATT},U}\left(\eta\right)\right](c)
&:=
\frac{
\sum_{w\in\mathcal W}
\left[
\eta^p_1(w)
\left(\eta^m_1(w)-\underline m_{0,c}\left(w;\eta\right)\right)
\right]
}{
\sum_{w\in\mathcal W} \eta^p_1(w)
}.
\end{aligned}
\end{equation}
By construction, evaluating \(\phi_\rho\) at \(\vartheta\) and at \(\widehat\vartheta\) gives the bound functions and their estimators:
\begin{equation}
\label{eq:c_mean_phi_plugin_representation}
\begin{pmatrix}
\underline\tau_{\rho,\cdot}
\\
\overline\tau_{\rho,\cdot}
\end{pmatrix}
=
\phi_{\rho}\left(\vartheta\right),
\qquad
\begin{pmatrix}
\widehat{\underline\tau}_{\rho,\cdot}
\\
\widehat{\overline\tau}_{\rho,\cdot}
\end{pmatrix}
=
\phi_{\rho}\left(\widehat\vartheta\right),
\qquad
\rho\in\left\{\mathrm{ATE},\mathrm{ATT}\right\}.
\end{equation}

The next proposition establishes weak convergence of the bound estimators uniformly over
$c\in\mathcal C$.

\begin{proposition}[Convergence of ATE and ATT bound estimators]
\label{prop:c_mean_dependence_estimation}
Suppose Assumptions~\ref{ass:c_mean_est_sampling}--\ref{ass:c_mean_est_regularity}
hold, and let
$\mathcal C\subseteq\left[0,\infty\right)$ be nonempty. Then

\begin{enumerate}[label=\textup{(\roman*)}]
\item The ATE bound process satisfies
\[
\sqrt N
\begin{pmatrix}
\widehat{\underline\tau}_{\mathrm{ATE},\cdot}
-
\underline\tau_{\mathrm{ATE},\cdot}
\\
\widehat{\overline\tau}_{\mathrm{ATE},\cdot}
-
\overline\tau_{\mathrm{ATE},\cdot}
\end{pmatrix}
\rightsquigarrow
G_{\mathrm{ATE}}
\quad
\text{in }
\ell^\infty\left(\mathcal C,\mathbb R^2\right).
\]

\item The ATT bound process satisfies
\[
\sqrt N
\begin{pmatrix}
\widehat{\underline\tau}_{\mathrm{ATT},\cdot}
-
\underline\tau_{\mathrm{ATT},\cdot}
\\
\widehat{\overline\tau}_{\mathrm{ATT},\cdot}
-
\overline\tau_{\mathrm{ATT},\cdot}
\end{pmatrix}
\rightsquigarrow
G_{\mathrm{ATT}}
\quad
\text{in }
\ell^\infty\left(\mathcal C,\mathbb R^2\right).
\]
\end{enumerate}
In both statements,
\begin{equation}
\label{eq:c_mean_limit_processes}
G_{\rho}
:=
\phi_{\rho,\vartheta}'\left(Z\right),
\qquad
\rho\in\left\{\mathrm{ATE},\mathrm{ATT}\right\},
\end{equation}
where \(Z\) and \(\phi_{\rho,\vartheta}'\) are given in
Lemmas~\ref{lem:c_mean_first_stage}--\ref{lem:c_mean_bound_maps} of the
supplement, respectively.
\end{proposition}

Proposition~\ref{prop:c_mean_dependence_estimation} gives the joint
asymptotic behavior of the estimated lower and upper bound functions. Although
the first-stage limit \(Z\) is Gaussian, the resulting limit processes need not
be Gaussian because the maps \(\phi_{\mathrm{ATE}}\) and \(\phi_{\mathrm{ATT}}\)
are only Hadamard directionally differentiable. These maps take finite minima
and maxima of conditional means over \(\mathcal N_c\left(w\right)\), and the
corresponding directional derivatives are generally nonlinear when the relevant
minimum or maximum is not unique.

Let
\(\left\{\left(Y_i^*,D_i^*,W_i^*\right)\right\}_{i=1}^N\) denote a
generic nonparametric i.i.d.\ bootstrap sample drawn with replacement from the
original sample. Define \(\widehat p_d^*(w)\) and \(\widehat m_d^*(w)\) as \(\widehat p_d(w)\) and \(\widehat m_d(w)\), with the bootstrap sample in place of the original sample, and let
\[
\widehat\vartheta^*:=
\left(
\left(\widehat p_d^*(w)\right)_{d\in\left\{0,1\right\},\,w\in\mathcal W},
\left(\widehat m_d^*(w)\right)_{d\in\left\{0,1\right\},\,w\in\mathcal W}
\right).
\]
Since \(\phi_{\rho}\) is only Hadamard directionally differentiable, the
standard nonparametric bootstrap applied directly to the bound estimators need not consistently estimate the limit distribution in Proposition~\ref{prop:c_mean_dependence_estimation} \citep[see, e.g.,][]{fang_inference_2019}. Therefore, I follow \citet{fang_inference_2019} and
\citet{masten_inference_2020} and use a bootstrap procedure based on the
numerical directional derivative of
\citet{hongNumericalDeltaMethod2018}.
For the bound functional \(\phi_\rho\) defined in \eqref{eq:c_mean_phi_components}, define its numerical directional derivative, for \(r\in\mathbb R^{4K}\), by
\[
\widehat{\phi}_{\rho}'\left(r\right)
:=
\frac{
\phi_{\rho}\left(\widehat\vartheta+\epsilon_N r\right)
-
\phi_{\rho}\left(\widehat\vartheta\right)
}{
\epsilon_N
},
\]
where \(\epsilon_N\to0\) and \(\sqrt N\epsilon_N\to\infty\). I then approximate the limiting distribution of the bound process for \(\rho\) by
the distribution of
\[
\begin{pmatrix}
T_{\rho,L}^*
\\
T_{\rho,U}^*
\end{pmatrix}
:=
\widehat{\phi}_{\rho}'\left(
\sqrt N\left(\widehat\vartheta^*-\widehat\vartheta\right)
\right)
\in
\ell^\infty\left(\mathcal C,\mathbb R^2\right).
\]
Under the maintained conditions, Theorem~3.1 of
\citet{hongNumericalDeltaMethod2018} implies that this bootstrap process
consistently estimates the limiting distribution in
Proposition~\ref{prop:c_mean_dependence_estimation}.

Let \(\sigma:\mathcal C\to\left(0,\infty\right)\) be a known weighting
function such that
\(0<\inf_{c\in\mathcal C}\sigma\left(c\right)\leq
\sup_{c\in\mathcal C}\sigma\left(c\right)<\infty\). Define the bootstrap
statistic and its conditional \(\left(1-\alpha\right)\)-quantile by
\begin{align*}
M_{\rho}^*
&:=
\max\left\{
\sup_{c\in\mathcal C}
\frac{T_{\rho,L}^*\left(c\right)}{\sigma\left(c\right)},
\sup_{c\in\mathcal C}
\left(
-\frac{T_{\rho,U}^*\left(c\right)}{\sigma\left(c\right)}
\right)
\right\},
\\
\widehat z_{\rho}\left(1-\alpha\right)
&:=
\inf\left\{
z\in\mathbb R
\mid
\Pr\left(
M_{\rho}^*\leq z
\mid
\left\{\left(Y_i,D_i,W_i\right)\right\}_{i=1}^N
\right)
\geq 1-\alpha
\right\}.
\end{align*}
As outlined in \citet{freyberger_uniform_2018}, many choices of \(\sigma\)
yield valid uniform confidence bands. In the application, I use the simple
choice \(\sigma\left(c\right)=1\), which produces an equal-width confidence band. For a
given \(\sigma\), the confidence band is
\begin{align}
\mathrm{CB}_{\rho}\left(c;1-\alpha\right)
:=
\left[
\widehat{\underline\tau}_{\rho,c}
-
\frac{\sigma\left(c\right)\widehat z_{\rho}\left(1-\alpha\right)}{\sqrt N},
\,
\widehat{\overline\tau}_{\rho,c}
+
\frac{\sigma\left(c\right)\widehat z_{\rho}\left(1-\alpha\right)}{\sqrt N}
\right],
\label{eq:c_mean_confidence_band}
\end{align}
for \(c\in\mathcal C\). Since \(\mathcal W\) is finite, each neighborhood \(\mathcal N_c(w)\) is one of finitely many subsets of \(\mathcal W\). Hence, the bound and bootstrap processes take only finitely many distinct values as functions of \(c\). When \(\sigma\) is constant, as in the application, the suprema over \(\mathcal C\) in \(M_\rho^*\) are maxima over finitely many values, even when \(\mathcal C\) is an interval.

\begin{proposition}[Bootstrap validity of ATE and ATT confidence bands]
\label{prop:c_mean_dependence_bootstrap}
Suppose Assumptions~\ref{ass:c_mean_est_sampling}--\ref{ass:c_mean_est_regularity}
hold, and let \(\mathcal C\subseteq\left[0,\infty\right)\) be nonempty. Let \(\epsilon_N\) be a deterministic sequence with \(\epsilon_N\to0\) and \(\sqrt N\epsilon_N\to\infty\). For
\(\rho\in\left\{\mathrm{ATE},\mathrm{ATT}\right\}\) and
\(\alpha\in\left(0,1\right)\), define
\begin{align*}
M_{\rho}
&:=
\max\left\{
\sup_{c\in\mathcal C}
\frac{G_{\rho,L}\left(c\right)}{\sigma\left(c\right)},
\sup_{c\in\mathcal C}
\left(
-\frac{G_{\rho,U}\left(c\right)}{\sigma\left(c\right)}
\right)
\right\},
\\
z_{\rho}\left(1-\alpha\right)
&:=
\inf\left\{
z\in\mathbb R
\mid
\Pr\left(M_{\rho}\leq z\right)\geq 1-\alpha
\right\},
\end{align*}
where
\(G_{\rho}:=\left(G_{\rho,L},G_{\rho,U}\right)'\) is the
weak limit in Proposition~\ref{prop:c_mean_dependence_estimation}. Assume that
the distribution function of \(M_{\rho}\) is continuous and strictly increasing
in a neighborhood of \(z_{\rho}\left(1-\alpha\right)\). Then
\[
\lim_{N\to\infty}
\Pr\left(
\left[
\underline\tau_{\rho,c},
\overline\tau_{\rho,c}
\right]
\subseteq
\mathrm{CB}_{\rho}\left(c;1-\alpha\right)
\ \text{for all } c\in\mathcal C
\right)
=
1-\alpha.
\]
\end{proposition}

Because the distribution function of \(M_{\rho}\) is continuous and strictly increasing near \(z_{\rho}\left(1-\alpha\right)\), the bootstrap critical value \(\widehat z_{\rho}\left(1-\alpha\right)\) consistently estimates \(z_{\rho}\left(1-\alpha\right)\), by Corollary~3.2 of \citet{fang_santos_inference_2015}. The band covers the lower and upper bound functions simultaneously over \(c\in\mathcal C\), with probability approaching \(1-\alpha\). This allows researchers to assess the robustness of a conclusion across a range of sensitivity values while accounting for sampling uncertainty. When interest centers on a particular relaxation of mean independence, taking \(\mathcal C=\{c\}\) yields a confidence interval at that value of \(c\). In contrast, when the bounds at several values of \(c\) are considered jointly, the band is needed. Finally, since \(\widehat z_{\rho}(1-\alpha)\) can be negative, the band in \eqref{eq:c_mean_confidence_band} need not contain the estimated bounds. To ensure that it does, I use \(\max\{0,\widehat z_{\rho}(1-\alpha)\}\) in place of \(\widehat z_{\rho}(1-\alpha)\) in \eqref{eq:c_mean_confidence_band} \citep[see, e.g., equation~(24) of][]{masten_inference_2020}. Because \(\max\{0,\widehat z_{\rho}(1-\alpha)\}\geq\widehat z_{\rho}(1-\alpha)\), the resulting band contains the band in \eqref{eq:c_mean_confidence_band}. Hence, its asymptotic coverage probability is at least \(1-\alpha\). 

\section{Application}
\label{sec:application}

\subsection{Background and Data}

Voluntary military service offers young adults an alternative to civilian employment. Since individuals choose whether to apply and the military screens those who do, assessing its effects on subsequent earnings requires accounting for both sources of selection. In a classic study of the earnings effects of voluntary military service, \citet{angrist_estimating_1998} addresses these two sources of selection by comparing applicants who enlisted with applicants who did not, while adjusting for characteristics used in screening, including schooling and Armed Forces Qualification Test (AFQT) scores. Restricting the comparison to applicants makes the groups more comparable in their interest in military service. Adjusting for the screening characteristics addresses differences in which applicants are accepted.

The data released by the Social Security Administration report average earnings separately for veterans and nonveterans in cells defined by race, year of application, AFQT score group, schooling level at the time of application, and year of birth. Applicants in the same AFQT score group can nevertheless have different exact scores. For instance, applicants with higher exact scores may have stronger verbal and numerical skills and higher expected earnings even without military service. These applicants may be more likely to be accepted by the military, but they may also have better civilian opportunities and be more likely to decline enlistment. Veterans and nonveterans with the same application year, AFQT score group, schooling level, and birth year may therefore have different mean earnings without military service. I revisit Angrist's estimates and use \(c\)-mean dependence to examine which of his conclusions about the earnings effects of military service survive such differences. The resulting bounds depend on the data only through these average earnings and the population counts of veterans and nonveterans in each cell, which the released data also report.

I follow the sample restrictions and matching specification used in Table II, Panel A, of \citet{angrist_estimating_1998}. The sample consists of men aged 17 to 22 at application who applied to the military between 1979 and 1982, had completed at least ninth grade, and had AFQT scores in categories III and IV, excluding college graduates.\footnote{AFQT scores are percentile ranks relative to a reference population. As reported in Table III of \citet{angrist_estimating_1998}, the five groups are IIIa (50--64), IIIb (31--49), IVa (21--30), IVb (16--20), and IVc (10--15). The scores use the corrected 1944 scale described in his footnote 21.} The earnings data cover a stratified random sample of 128,968 white and 175,262 nonwhite applicants, drawn from populations of 849,983 and 403,686 applicants. The released data omit the veterans or nonveterans of a cell when fewer than 25 of them are in the population. Tables I and III of \citet{angrist_estimating_1998} report descriptive statistics and enlistment probabilities by AFQT score group. The analysis is conducted separately for white and nonwhite applicants. Within each group, veterans are compared with nonveterans who share the same application year, schooling level, birth year, and AFQT score group. These cell-level earnings differences are averaged using the distribution of veterans across cells that contain both veterans and nonveterans, yielding the matching estimate of the ATT. For these comparisons, I focus on annual Social Security taxable earnings in 1982, 1986, and 1991, measured in 1991 dollars. For veterans, earnings in 1982 largely reflect military pay, while earnings in later years increasingly reflect civilian earnings \citep[see, e.g.,][footnote~25]{angrist_estimating_1998}.

Let $V$ collect application year, schooling level, and birth year. Let $W$ denote the AFQT score group, with $\mathcal W=\{1,\ldots,5\}$ in increasing score order. I impose $c$-mean dependence given $W$ and $V$, as in Definition~\ref{def:c_mean_additional_covariates} of the supplement. However, some cells contain only veterans or only nonveterans, and the overlap condition \(0<\Pr(D=1\mid W=w,V=v)<1\) assumed there fails in these cells. Following \citet{angrist_estimating_1998}, I restrict the target population to the veterans in cells that also contain nonveterans, as described in Section~\ref{sec:s4_identification} of the supplement. This restriction excludes a negligible share of veterans, 0.01 percent for white applicants and 0.27 percent for nonwhite applicants.\footnote{The excluded veterans are in one of the 467 cells with veterans for white applicants and in 12 of the 441 such cells for nonwhite applicants.}

Since the data report earnings only as means, standard deviations, and sample counts for veterans and nonveterans in each cell, I use the inference procedure for grouped data in Section~\ref{sec:s4_grouped_inference} of the supplement for all confidence intervals and bands reported below. I select the numerical step size $\epsilon_N$ using the bootstrap selection rule proposed in Appendix B.4 of \citet{masten_inference_2020}, as implemented in Section~\ref{sec:bootstrap_step_selection} of the supplement. This rule selects the candidate step size whose estimated coverage under a fitted data distribution is closest to the nominal confidence level. Section~\ref{sec:bootstrap_step_selection} of the supplement shows that the resulting confidence intervals and bands remain valid at the selected step size.

\subsection{Placebo Tests with Pre-application Earnings}
\label{sec:application_placebo}

After enlistment, the earnings veterans would have received without military service are not observed, and mean independence given \(W\) and \(V\) cannot be tested directly with post-application earnings. Earnings before application can instead serve as a placebo outcome, since skills that raise earnings without military service likely also raised earnings before application, and these earnings are measured before anyone served. I assume that there are no anticipation effects, that is, military service does not affect earnings before application. The treatment effect on these earnings is then zero for every applicant. For a given \(c\), if \(c\)-mean dependence also holds for pre-application earnings, the bounds on the ATT for these earnings must contain zero. I test this implication with the confidence interval for these bounds and reject \(c\)-mean dependence for pre-application earnings when the interval excludes zero.\footnote{Since \(c\)-mean dependence implies \(c'\)-mean dependence for every \(c'\geq c\), testing \(c=0,1,\ldots\) in turn for a given comparison and stopping at the first value that is not rejected ensures that the probability of rejecting any true restriction for that comparison is asymptotically at most the nominal level, with no adjustment for multiple testing, by the argument in Example 9.2.2 of \citet{lehmann_testing_2022}.} At \(c=0\), the bounds reduce to the matching estimates, and the placebo test becomes the specification check discussed by \citet{angrist_estimating_1998}.

I carry out the test both separately for each application year and for comparisons pooled over application years. Pooled estimates are more precise, but departures from \(c\)-mean dependence in different cells and application years can have opposite signs and offset each other, which lowers the power of a pooled test against such departures. Let \(A\) denote the year of application, with support \(\mathcal A=\{1979,1980,1981,1982\}\), and \(Y_t\) earnings in calendar year \(t\). Write \(Y_t(1)\) and \(Y_t(0)\) for its potential outcomes if the applicant enters the military within two years following application (\(D=1\)) and otherwise (\(D=0\)). For \(a\in\mathcal A\), write \(\mathrm{ATT}(a,t):=\mathrm E[Y_t(1)-Y_t(0)\mid D=1,A=a]\) for the effect on earnings in year \(t\) of veterans who applied in year \(a\). Angrist pools the comparisons over application years holding the calendar year fixed, so that the pooled earnings come from the same year and the same labor market conditions, but the time before application differs across application cohorts. Earnings in 1978, for example, are measured one year before application for 1979 applicants but four years before application for 1982 applicants. Differences that occur at a fixed time before application, such as the dip in earnings just before application that Angrist observes for some cohorts, therefore fall in different calendar years for different cohorts. Pooling by calendar year averages such a difference in one cohort with the differences of the other cohorts at other times before application, which may make the placebo test less likely to detect such a difference. To complement his comparisons, I also pool them over application years holding the number \(k\) of years before application fixed. Specifically, with \(\omega_a:=\Pr(A=a\mid D=1)\), Angrist's pooled comparison for calendar year \(t\) and mine for \(k=1,2,3\) target
\[
\sum_{a\in\mathcal A}\omega_a\,\mathrm{ATT}(a,t)
\qquad\text{and}\qquad
\sum_{a\in\mathcal A}\omega_a\,\mathrm{ATT}(a,a-k),
\]
respectively, where the weights and effects refer to the veterans included in the comparison.\footnote{Under the confidentiality edit described in Appendix B.III of \citet{angrist_estimating_1998}, the released data do not report mean earnings for cells with very low total earnings or few workers with positive earnings. The comparisons use only cells whose earnings are reported, which cover at least 97 percent of veterans in every group and application year for \(k\leq3\).} By iterated expectations, the first target is \(\tau_{\mathrm{ATT}}\) of Section~\ref{sec:coarsening_setup} for these veterans when the outcome is earnings in calendar year \(t\), and the second is \(\tau_{\mathrm{ATT}}\) when the outcome is earnings \(k\) years before application. The second aligns cohorts by the number of years before application, as in event studies with staggered treatment timing \citep[see, e.g.,][]{callaway_difference--differences_2021}. For \(k=1\), for example, it combines the 1978 earnings of 1979 applicants with the 1981 earnings of 1982 applicants. A dip in the year before application in one cohort is then averaged with the differences of the other cohorts in their own year before application.

\begin{figure}[htbp]
\centering
\includegraphics[width=\textwidth]{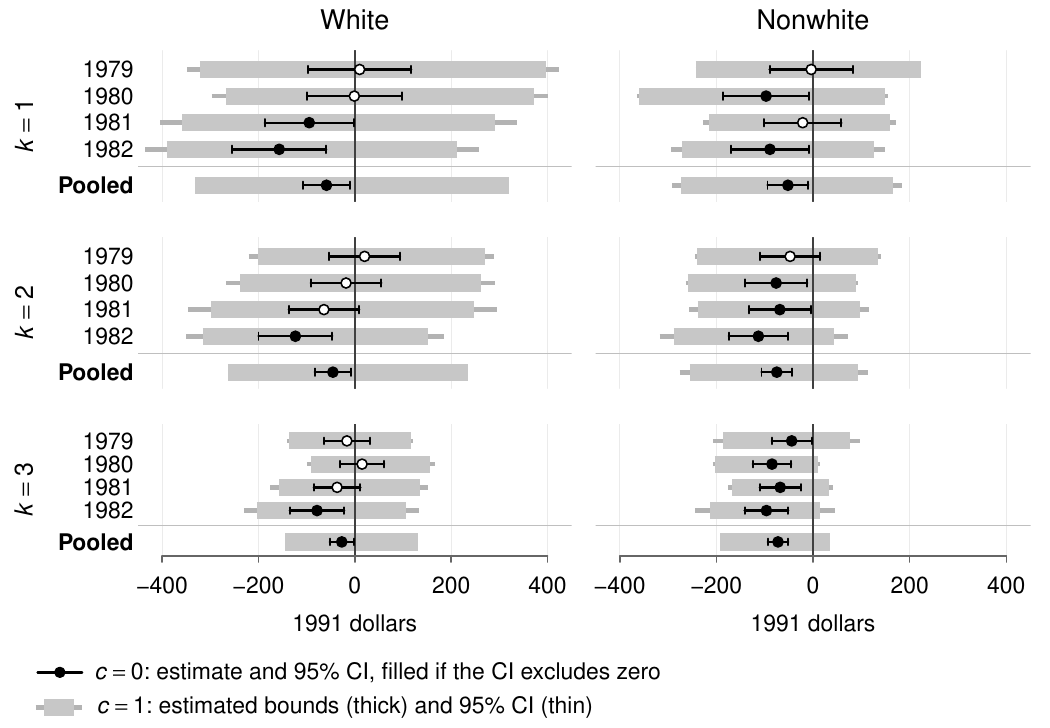}
\caption{Earnings Differences before Application}
\label{fig:pre_application_earnings}
\fnote{\textit{Notes:} The figure shows the placebo comparisons for earnings in year \(a-k\) of applicants who applied in year \(a\in\mathcal A\), measured in 1991 dollars. The left panels show white applicants and the right panels nonwhite applicants, and from top to bottom the panels show \(k=1,2,3\). In each panel, the bottom row pools the application years in \(\mathcal A\). Table~\ref{tab:placebo_eventtime} of the supplement reports the estimates and intervals, including those for \(k=4\) and \(5\).}
\end{figure}

Figure~\ref{fig:pre_application_earnings} shows that, before application, applicants who later enlisted tended to earn less than nonveterans in the same cell. Almost all of the matching estimates are negative. At \(c=0\), the confidence intervals exclude zero in more than half of the comparisons for individual application years, mostly for nonwhite applicants, and every estimate whose interval excludes zero is negative. When application years are pooled at the same \(k\), the intervals exclude zero for both groups at every \(k\). The comparisons pooled by calendar year, reported in Table~\ref{tab:placebo_calendar} of the supplement, show a similar but weaker pattern. At \(c=1\), by contrast, every confidence interval contains zero. The matching estimates for individual application years correspond to those in Angrist's Figure 3, which shows them without confidence intervals.

The pre-application earnings differences at \(c=0\) are modest in dollar terms, as Angrist notes, but they are systematic.\footnote{\citet{angrist_estimating_1998} describes the pre-application differences by veteran status in his Figure 2 and Table II as modest and the matching estimates in his Figure 3 as close to zero until the year of application.} The placebo test thus rejects mean independence given \(W\) and \(V\) for pre-application earnings but does not reject \(1\)-mean dependence. If mean independence held given the exact AFQT score and \(V\), these differences would reflect coarsening bias. Angrist also observes a modest dip in earnings just before application for some cohorts and relates it to transitory shocks, such as those that lower earnings before entry into training programs. Such shocks would also violate mean independence given \(W\) and \(V\).

Following Section~\ref{sec:choosing_c}, these comparisons also provide a benchmark for \(c\). Since \(c=1\) is the smallest value at which they do not reject \(c\)-mean dependence, it serves as the benchmark for post-application earnings.

\subsection{Results}
\label{sec:application_results}

Since the pre-application comparisons reject mean independence, I examine how the post-application effects suggested by the matching estimates change as \(c\) increases. At \(c=0\), veterans' mean earnings without military service equal the observed mean for nonveterans in the same cell, which yields Angrist's matching estimates (Table II, Panel A). A larger \(c\) allows veterans' mean earnings without military service to lie anywhere between the lowest and highest observed means for nonveterans in AFQT score groups up to \(c\) steps away, holding application year, schooling level, and birth year fixed. For \(c\geq4\), these groups include every AFQT score group with the same application year, schooling level, and birth year, and the bounds no longer change. Table~\ref{tab:angrist_sensitivity} reports the matching estimates at \(c=0\), the estimated ATT bounds for \(c\geq1\), and a 95 percent confidence band that covers the bounds at all \(c\in\mathcal C=\{0,1,2,3,4\}\) simultaneously.

\begin{table}[!ht]
\centering
\caption{Sensitivity of Estimated Earnings Effects to Relaxing Mean Independence}
\label{tab:angrist_sensitivity}
\small
\renewcommand{\arraystretch}{1.2}
\setlength{\tabcolsep}{5pt}
\begin{tabular}{@{}lcccccc@{}}
\toprule
Year & & \(c=0\) & \(c=1\) & \(c=2\) & \(c=3\) & \(c\geq4\) \\
\midrule
\multicolumn{7}{@{}l}{\textit{Panel A. White applicants}} \\
\addlinespace[3pt]
\multirow{2}{*}{1982} & Bounds & \(783\) & \([364,\,1{,}361]\) & \([234,\,1{,}889]\) & \([191,\,2{,}380]\) & \([183,\,2{,}633]\) \\
 & 95\% CB & \([712,\,854]\) & \([293,\,1{,}432]\) & \([163,\,1{,}960]\) & \([120,\,2{,}451]\) & \([112,\,2{,}704]\) \\
\arrayrulecolor{black!25}\cmidrule(l{2pt}r{2pt}){1-7}\arrayrulecolor{black}
\multirow{2}{*}{1986} & Bounds & \(-557\) & \([-1{,}216,\,376]\) & \([-1{,}437,\,1{,}239]\) & \([-1{,}516,\,2{,}007]\) & \([-1{,}533,\,2{,}380]\) \\
 & 95\% CB & \([-673,\,-442]\) & \([-1{,}332,\,492]\) & \([-1{,}552,\,1{,}354]\) & \([-1{,}632,\,2{,}123]\) & \([-1{,}648,\,2{,}496]\) \\
\arrayrulecolor{black!25}\cmidrule(l{2pt}r{2pt}){1-7}\arrayrulecolor{black}
\multirow{2}{*}{1991} & Bounds & \(30\) & \([-1{,}048,\,1{,}504]\) & \([-1{,}422,\,2{,}645]\) & \([-1{,}566,\,3{,}636]\) & \([-1{,}603,\,4{,}130]\) \\
 & 95\% CB & \([-117,\,177]\) & \([-1{,}195,\,1{,}651]\) & \([-1{,}569,\,2{,}792]\) & \([-1{,}713,\,3{,}783]\) & \([-1{,}750,\,4{,}277]\) \\
\midrule
\multicolumn{7}{@{}l}{\textit{Panel B. Nonwhite applicants}} \\
\addlinespace[3pt]
\multirow{2}{*}{1982} & Bounds & \(2{,}186\) & \([1{,}774,\,2{,}604]\) & \([1{,}573,\,2{,}808]\) & \([1{,}489,\,2{,}927]\) & \([1{,}454,\,2{,}962]\) \\
 & 95\% CB & \([2{,}123,\,2{,}249]\) & \([1{,}711,\,2{,}667]\) & \([1{,}510,\,2{,}870]\) & \([1{,}426,\,2{,}989]\) & \([1{,}392,\,3{,}025]\) \\
\arrayrulecolor{black!25}\cmidrule(l{2pt}r{2pt}){1-7}\arrayrulecolor{black}
\multirow{2}{*}{1986} & Bounds & \(721\) & \([-53,\,1{,}459]\) & \([-476,\,1{,}898]\) & \([-684,\,2{,}159]\) & \([-751,\,2{,}239]\) \\
 & 95\% CB & \([620,\,821]\) & \([-153,\,1{,}559]\) & \([-577,\,1{,}999]\) & \([-784,\,2{,}260]\) & \([-851,\,2{,}339]\) \\
\arrayrulecolor{black!25}\cmidrule(l{2pt}r{2pt}){1-7}\arrayrulecolor{black}
\multirow{2}{*}{1991} & Bounds & \(1{,}026\) & \([-256,\,2{,}255]\) & \([-1{,}040,\,2{,}965]\) & \([-1{,}445,\,3{,}338]\) & \([-1{,}574,\,3{,}450]\) \\
 & 95\% CB & \([893,\,1{,}160]\) & \([-390,\,2{,}389]\) & \([-1{,}173,\,3{,}099]\) & \([-1{,}579,\,3{,}472]\) & \([-1{,}708,\,3{,}584]\) \\
\bottomrule
\end{tabular}
\fnote{\textit{Notes:} For each group and year, the first row reports the matching estimate at \(c=0\) and the estimated ATT bounds for \(c\geq1\). The second row reports the 95 percent confidence band (CB), which covers the bounds at all values of \(c\) simultaneously (at \(c=0\), essentially the pointwise interval). Earnings are in 1991 dollars, rounded to the nearest dollar. For \(c\geq4\), the neighboring cells include every AFQT score group with the same application year, schooling level, and birth year, and larger values of \(c\) give the same results. The bootstrap uses 19{,}999 Gaussian draws, as described in Section~\ref{sec:s4_grouped_inference} of the supplement, with \(\sigma(c)=1\) and $\epsilon_N=\kappa\sqrt{\log N/N}$, where $\kappa\in\{0.1,0.15,0.25,0.35,0.5,0.7,1,1.4,2\}$ is chosen by the selection rule in Section~\ref{sec:bootstrap_step_selection} of the supplement. Table~\ref{tab:post_application_radii} of the supplement reports pointwise confidence intervals.}
\end{table}

As in \citet{angrist_estimating_1998}, the matching estimates for 1982 are positive for both groups, at \$783 for white applicants and \$2,186 for nonwhite applicants. As \(c\) increases, the estimated lower bounds decline at a decreasing rate, to \$364 and \$1,774 at the benchmark \(c=1\) and to \$183 and \$1,454 at \(c=4\). At the benchmark \(c=1\), veterans' mean earnings without military service may be as high as those of nonveterans in their own and the adjacent AFQT score groups. At \(c=4\), they may be as high as those of nonveterans in any AFQT score group with the same application year, schooling level, and birth year. The confidence band remains above zero at every \(c\). Hence, military service raised earnings in 1982 even under the weakest restriction.

\begin{figure}[htbp]
\centering
\includegraphics[width=\textwidth]{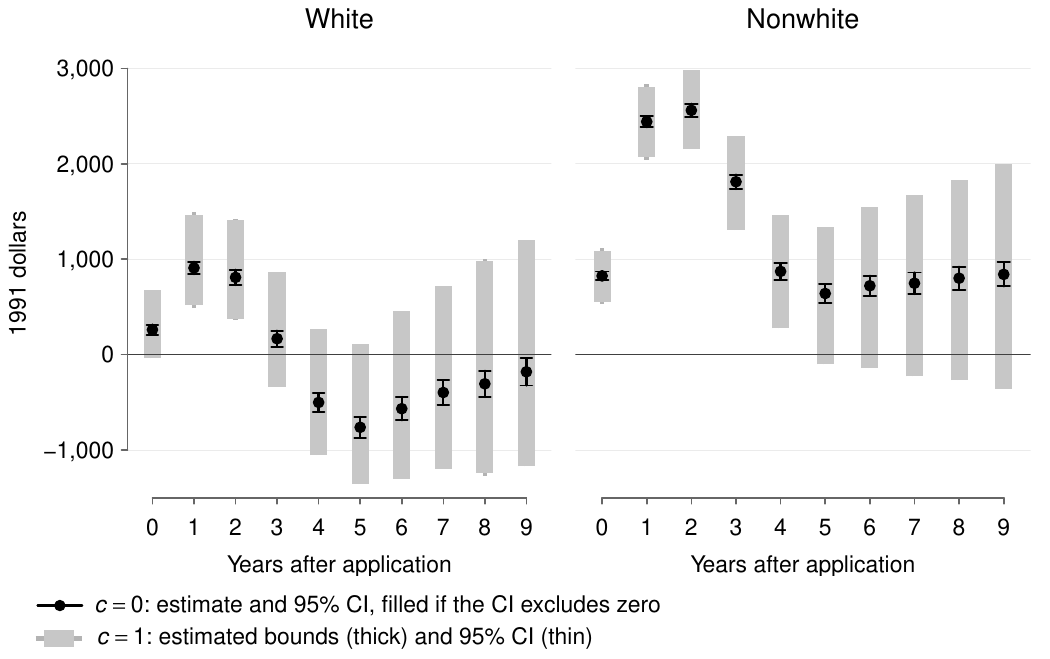}
\caption{Earnings Effects by Years after Application}
\label{fig:post_application_earnings}
\fnote{\textit{Notes:} The figure shows, for each number \(e\) of years after application, the matching estimates at \(c=0\) and the estimated ATT bounds at \(c=1\) for earnings in year \(a+e\), pooled over the application years \(a\in\mathcal A\), measured in 1991 dollars. Year 0 denotes the year of application. The left panel shows white applicants and the right panel nonwhite applicants. Table~\ref{tab:post_application_eventtime} of the supplement reports the estimates and intervals.}
\end{figure}

The results for 1986 and 1991 are more sensitive to the restriction on veterans' mean earnings without military service. At \(c=0\), the confidence band lies below zero for white applicants in 1986 and above zero for nonwhite applicants in 1986 and 1991, and it contains zero for white applicants in 1991. At every \(c\geq1\), including the benchmark \(c=1\), the estimated bounds and the confidence bands contain zero in all four cases. Thus, for later years, when earnings increasingly reflect civilian rather than military pay, the signs found at \(c=0\) rely on mean independence, which the pre-application comparisons reject for earlier earnings. Once veterans' mean earnings without military service may lie anywhere between those of nonveterans in their own and the adjacent AFQT score groups, the data no longer determine whether military service raised or lowered civilian earnings.

Table~\ref{tab:angrist_sensitivity} shows that the effects in 1982 are positive at every \(c\), while the signs of those in 1986 and 1991 are not determined at the benchmark \(c=1\). Because it pools application years at a fixed calendar year, as \citet{angrist_estimating_1998} does, it does not show how long after application the effects remain positive. In 1986, for example, the 1979 applicants are seven years past application, while the 1982 applicants are four years past application. Figure~\ref{fig:post_application_earnings} instead pools them at a fixed number \(e\) of years after application, which targets \(\sum_{a\in\mathcal A}\omega_a\,\mathrm{ATT}(a,a+e)\), the ATT for earnings \(e\) years after application, as for the pre-application comparisons in Section~\ref{sec:application_placebo}. At the benchmark \(c=1\), the confidence intervals exclude zero only in years 1 and 2 after application for white applicants and in years 0 to 4 for nonwhite applicants, where year 0 is the year of application. These years roughly cover the typical length of service, a median of three to four years according to \citet[footnote~25]{angrist_estimating_1998}.

\section{Conclusion}
\label{sec:conclusion}

This paper studies how discretizing covariates changes treatment effect
comparisons. The analysis shows that replacing a finely measured covariate by a
coarsened version is not merely a loss of precision. It can change the estimand,
and the direction and magnitude of the change depend on how treated and untreated
units in the same cell differ in mean outcomes without treatment and in average
treatment effects. The decomposition results make these two channels explicit and
clarify why seemingly innocuous discretization choices can matter for ATT and
ATE comparisons.

The sensitivity analysis provides a way to report what remains credible when
exact mean independence after coarsening is not imposed. The
\(c\)-mean-dependence restriction indexes the allowed departures from mean
independence given the coarsened covariate, and the estimation results provide uniform
inference for the resulting sharp bounds over a range of sensitivity values. Outcomes measured before treatment can be used to test mean independence and provide an empirical benchmark for \(c\).

The military service application illustrates how researchers can assess treatment effects when covariates are observed only in groups. For earnings before application, the placebo tests reject mean independence given the reported covariates in various comparisons for both white and nonwhite applicants, but they reject \(1\)-mean dependence in none. Military service raised earnings in 1982, when veterans' earnings largely reflect military pay, even under the weakest restriction considered. Whether it raised or lowered later civilian earnings depends on what veterans would have earned without service relative to nonveterans in the same AFQT score group. At the benchmark \(c=1\), the data do not determine the sign of this effect.

\clearpage
\spacingset{1}
\phantomsection
\addcontentsline{toc}{section}{References}
\bibliographystyle{chicago}
\bibliography{references/L1}

\ManuscriptSupplementSetup
\etocdepthtag.toc{supplement}
\makeatletter
\let\ManuscriptBibContext\the@ipfilectr
\ManuscriptSupplementFrontMatter
\global\let\the@ipfilectr\ManuscriptBibContext
\makeatother

\section{Additional Results for Section~\ref{sec:coarsening_setup}}

\subsection{Conditional Coarsening Bias}
\label{sec:supp_conditional_coarsening_bias}

Proposition~\ref{prop:coarsen_bias_decomposition} describes the bias in comparisons averaged across cells. Researchers may also be interested in treatment effects within a reported category. The same observed difference \(\theta(w)\) can then be compared with the average effect for treated units in that category or with the average effect for all units in it. For \(w\in\mathcal{W}\), define the conditional ATT and conditional ATE by
\begin{align*}
\tau_{\mathrm{ATT}}\left(w\right)
&:=
\mathrm{E}\left[
Y\left(1\right)-Y\left(0\right)
\mid D=1,W=w
\right],
\\
\tau_{\mathrm{ATE}}\left(w\right)
&:=
\mathrm{E}\left[
Y\left(1\right)-Y\left(0\right)
\mid W=w
\right].
\end{align*}
Recall the within-cell difference
\(\Delta_Z\left(w\right):=
\mathrm{E}\left[Z\mid D=1,W=w\right]
-
\mathrm{E}\left[Z\mid D=0,W=w\right]\)
from Section~\ref{sec:coarsening_setup}. The observed difference is \(\theta(w)=\Delta_Y(w)\). The first identity below holds under the assumptions of Proposition~\ref{prop:coarsen_bias_decomposition}\textup{(i)}, and the second under those of part~\textup{(ii)}:
\begin{align*}
\theta\left(w\right)
-
\tau_{\mathrm{ATT}}\left(w\right)
&=
\Delta_{\mu_0\left(X\right)}\left(w\right),
\\
\theta\left(w\right)
-
\tau_{\mathrm{ATE}}\left(w\right)
&=
\left(1-e\left(w\right)\right)
\Delta_{\tau\left(X\right)}\left(w\right)
+
\Delta_{\mu_0\left(X\right)}\left(w\right).
\end{align*}
For the conditional ATT, the bias is the difference between the two groups' average untreated potential outcomes within the cell. For the conditional ATE, the difference between their average treatment effects also enters, with weight \(1-e(w)\). Averaging the first identity over the distribution of \(W\) among treated units and the second over the population distribution of \(W\) recovers \(\widetilde B_0\) and \(B_\tau+B_0\) in Proposition~\ref{prop:coarsen_bias_decomposition}, respectively. The conditional formulas therefore show where the aggregate biases arise and how they can differ across reported categories.

\subsection{Quantile Coarsening Bias}
\label{sec:supp_quantile_coarsening_bias}

The ATT and ATE comparisons in Proposition~\ref{prop:coarsen_bias_decomposition} concern average outcomes. To study how coarsening affects comparisons at different parts of the outcome distribution, consider the quantile treatment effect on the treated (QTT) and the quantile treatment effect (QTE). These targets depend on the distributions of potential outcomes, not only on their conditional means \citep[see, e.g.,][]{frolich_propensity_2007}.

To identify these distributions using the fine covariate \(X\), maintain Assumption~\ref{ass:overlap} and strengthen the relevant mean independence conditions. For the QTT, replace Assumption~\ref{ass:mean_independence_y0} by
\[
Y\left(0\right)\perp D\mid X,
\]
while for the QTE impose
\[
Y\left(d\right)\perp D\mid X
\qquad
\text{for all }d\in\left\{0,1\right\}.
\]

Define the distributions whose quantiles enter the fine and coarsened comparisons. For a distribution function \(F\), let \(F^{-1}\) denote its generalized inverse. For \(d\in\left\{0,1\right\}\), let
\begin{align*}
F_d\left(y\right)
&:=
\mathrm{Pr}\left(Y\left(d\right)\leq y\right),
&
F_{d\mid D=1}\left(y\right)
&:=
\mathrm{Pr}\left(Y\left(d\right)\leq y\mid D=1\right),
\\
G_d\left(y\right)
&:=
\mathrm{E}\left[
\mathrm{Pr}\left(Y\leq y\mid D=d,W\right)
\right],
&
G_{0\mid D=1}\left(y\right)
&:=
\mathrm{E}\left[
\mathrm{Pr}\left(Y\leq y\mid D=0,W\right)
\mid D=1
\right].
\end{align*}
Here, \(F_d\) is the population distribution of \(Y(d)\), and \(F_{d\mid D=1}\) is its distribution among treated units. The coarsened distribution \(G_d\) averages the observed distribution function for treatment group \(d\) over the population distribution of \(W\). For the QTT, \(G_{0\mid D=1}\) instead uses the distribution of \(W\) among treated units, as in the ATT comparison of Section~\ref{sec:coarsening_setup}. Under the corresponding independence and overlap conditions, using \(X\) in place of \(W\) in these averages recovers the relevant potential outcome distributions.

Each coarsened quantile is obtained by averaging conditional distribution functions and then taking the generalized inverse. For \(q\in\left(0,1\right)\), define the QTT and QTE and their coarsened counterparts by
\begin{align*}
\tau_{\mathrm{QTT}}\left(q\right)
&:=
F_{1\mid D=1}^{-1}\left(q\right)
-
F_{0\mid D=1}^{-1}\left(q\right),
&
\tau_{\mathrm{QTE}}\left(q\right)
&:=
F_1^{-1}\left(q\right)
-
F_0^{-1}\left(q\right),
\\
\theta_{\mathrm{QTT}}\left(q\right)
&:=
F_{1\mid D=1}^{-1}\left(q\right)
-
\left(G_{0\mid D=1}\right)^{-1}\left(q\right),
&
\theta_{\mathrm{QTE}}\left(q\right)
&:=
\left(G_1\right)^{-1}\left(q\right)
-
\left(G_0\right)^{-1}\left(q\right).
\end{align*}
The resulting quantile coarsening biases are
\begin{align*}
\theta_{\mathrm{QTT}}\left(q\right)
-
\tau_{\mathrm{QTT}}\left(q\right)
&=
F_{0\mid D=1}^{-1}\left(q\right)
-
\left(G_{0\mid D=1}\right)^{-1}\left(q\right),
\\
\theta_{\mathrm{QTE}}\left(q\right)
-
\tau_{\mathrm{QTE}}\left(q\right)
&=
\left(
\left(G_1\right)^{-1}\left(q\right)
-
F_1^{-1}\left(q\right)
\right)
-
\left(
\left(G_0\right)^{-1}\left(q\right)
-
F_0^{-1}\left(q\right)
\right).
\end{align*}
For the QTT, the quantile of treated outcomes appears in both comparisons and cancels from the bias. The bias therefore depends only on the difference between the true and coarsened quantiles of untreated potential outcomes among treated units. For the QTE, the differences for both potential outcome distributions enter. Thus, coarsening can bias distributional comparisons as well as average ones. Because generalized inversion is nonlinear, these quantile biases need not admit the additive channel decomposition in Proposition~\ref{prop:coarsen_bias_decomposition}.

\subsection{Difference-in-Differences}
\label{sec:supp_did_coarsening_bias}

The same question arises in DiD designs: does conditional parallel trends remain valid when the covariates on which it holds are observed only in groups? Section~\ref{sec:coarsening_setup} studies this question for outcome levels. Here, the relevant object is the trend in the untreated potential outcome. I first derive the DiD counterpart of the ATT decomposition in Proposition~\ref{prop:coarsen_bias_decomposition}. I then use pre-treatment differences to restrict the post-treatment coarsening bias.

Let \(t\in\left\{-L,\ldots,0,1\right\}\), where \(L\geq0\), period \(1\) is the post-treatment period, and periods \(t\leq0\) are pre-treatment periods. Let \(D\in\left\{0,1\right\}\) indicate membership in the group treated in period \(1\). Write \(X\) for a pre-treatment covariate and \(W=g\left(X\right)\) for its coarsened version. For each \(t\in\left\{-L,\ldots,0,1\right\}\) and \(d\in\left\{0,1\right\}\), let \(Y_t\left(d\right)\) denote the potential outcome in period \(t\) under treatment status \(d\). The observed outcome is
\begin{align*}
Y_t
=
D Y_t\left(1\right)
+
\left(1-D\right)Y_t\left(0\right).
\end{align*}
The target is the ATT in period \(1\):
\begin{align*}
\tau_{\mathrm{ATT}}^{\mathrm{DiD}}
:=
\mathrm{E}\left[
Y_1\left(1\right)-Y_1\left(0\right)
\mid D=1
\right].
\end{align*}
For \(t\in\left\{-L+1,\ldots,1\right\}\), define
\[
\Delta_tY\left(d\right)
:=
Y_t\left(d\right)-Y_{t-1}\left(d\right),
\qquad
\Delta_tY
:=
Y_t-Y_{t-1}.
\]

\begin{assumption}[Conditional parallel trends for DiD]
\label{ass:supp_did_conditional_parallel_trends}
\[
\mathrm{E}\left[\Delta_1Y\left(0\right)\mid X,D\right]
=
\mathrm{E}\left[\Delta_1Y\left(0\right)\mid X\right]\qquad\text{a.s.}
\]
\end{assumption}

\begin{assumption}[No anticipation]
\label{ass:supp_did_no_anticipation}
For each \(t\leq0\),
\[
Y_t\left(1\right)=Y_t\left(0\right)
\qquad
\text{a.s.}
\]
\end{assumption}

\begin{assumption}[Overlap for DiD]
\label{ass:supp_did_overlap}
There exists \(\eta>0\) such that
\[
\eta
\leq
\mathrm{Pr}\left(D=1\mid X\right)
\leq
1-\eta
\qquad
\text{a.s.}
\]
\end{assumption}

Under conditional parallel trends given \(X\), the average trend for untreated units at each value of \(X\) equals the average trend that treated units with that value would have experienced without treatment. Together with no anticipation and overlap, this identifies the ATT as
\begin{align*}
\tau_{\mathrm{ATT}}^{\mathrm{DiD}}
=
\mathrm{E}\left[
\Delta_1Y
-
\mathrm{E}\left[\Delta_1Y\mid D=0,X\right]
\mid D=1
\right].
\end{align*}
When only \(W\) is observed, the comparison instead conditions on the reported cells. For \(t\in\left\{-L+1,\ldots,1\right\}\) and \(w\in\mathcal{W}\), define the coarsened DiD comparison by
\begin{align*}
\delta_t\left(w\right)
:=
\mathrm{E}\left[\Delta_tY\mid D=1,W=w\right]
-
\mathrm{E}\left[\Delta_tY\mid D=0,W=w\right].
\end{align*}
The corresponding coarsened ATT estimand in period \(1\) is
\begin{align*}
\theta_{\mathrm{ATT}}^{\mathrm{DiD}}
:=
\mathrm{E}\left[
\delta_1\left(W\right)
\mid D=1
\right].
\end{align*}

\begin{proposition}[Coarsening bias in DiD]
\label{prop:supp_coarsen_did_att_bias}
Suppose Assumptions~\ref{ass:supp_did_conditional_parallel_trends}--\ref{ass:supp_did_overlap} hold, and
suppose the expectations below are finite. Then
\begin{align*}
\theta_{\mathrm{ATT}}^{\mathrm{DiD}}
=
\tau_{\mathrm{ATT}}^{\mathrm{DiD}}
+
B_{\mathrm{ATT}}^{\mathrm{DiD}},
\end{align*}
where
\begin{align*}
B_{\mathrm{ATT}}^{\mathrm{DiD}}
:=
\mathrm{E}\left[
\mathrm{E}\left[
\Delta_1Y\left(0\right)
\mid D=1,W
\right]
-
\mathrm{E}\left[
\Delta_1Y\left(0\right)
\mid D=0,W
\right]
\mid D=1
\right].
\end{align*}
\end{proposition}

Conditional parallel trends requires the two groups to have the same average trend in \(Y(0)\) at each value of \(X\). Within a cell of \(W\), however, the groups can have different distributions of \(X\). Averaging the common trend at each value of \(X\) over these different distributions can therefore produce different average trends for the two groups. Proposition~\ref{prop:supp_coarsen_did_att_bias} gives the resulting bias, averaged over the distribution of \(W\) among treated units. It is the counterpart of the composition bias in Proposition~\ref{prop:coarsen_bias_decomposition}\textup{(i)}, with trends in \(Y(0)\) in place of its levels.

When \(L\geq1\), pre-treatment outcomes identify the corresponding differences for earlier periods. For each \(t=-L+1,\ldots,0\), define
\begin{align*}
\Gamma_t
:=
\mathrm{E}\left[\delta_t\left(W\right)\mid D=1\right].
\end{align*}
Under no anticipation, \(\Delta_tY=\Delta_tY\left(0\right)\) for \(t\leq0\). Hence, \(\Gamma_t\) is identified from the observed distribution. It measures the difference between the two groups' average trends in \(Y(0)\), using the same cells of \(W\) and the same distribution of \(W\) among treated units as \(B_{\mathrm{ATT}}^{\mathrm{DiD}}\). Let \(\Gamma_{\mathrm{pre}}:=\left(\Gamma_{-L+1},\ldots,\Gamma_0\right)\).

To use these pre-treatment differences to bound the treatment effect, I restrict the post-treatment coarsening bias relative to \(\Gamma_{\mathrm{pre}}\). The functions \(\underline b\) and \(\overline b\) specify its allowed range, with \(\underline b\left(\Gamma_{\mathrm{pre}}\right)\leq\overline b\left(\Gamma_{\mathrm{pre}}\right)\). The following assumption makes this link explicit. It parallels bounded variation restrictions in DiD sensitivity analysis \citep{manski_how_2018,rambachan_more_2023}.

\begin{assumption}[Bounded variation of coarsening bias]
\label{ass:supp_did_bounded_bias_variation}
Suppose \(L\geq1\). The post-treatment coarsening bias satisfies
\begin{align*}
B_{\mathrm{ATT}}^{\mathrm{DiD}}
\in
\left[
\underline b\left(\Gamma_{\mathrm{pre}}\right),
\overline b\left(\Gamma_{\mathrm{pre}}\right)
\right].
\end{align*}
\end{assumption}

Three choices illustrate different restrictions on how the post-treatment bias can depart from the pre-treatment differences. Exact bias stability equates the post-treatment bias with the most recent pre-treatment difference:
\[
\underline b\left(\Gamma_{\mathrm{pre}}\right)
=
\overline b\left(\Gamma_{\mathrm{pre}}\right)
:=
\Gamma_0.
\]
A range restriction allows the post-treatment bias to lie between the smallest and largest pre-treatment differences, widened by \(r\geq0\) on either side:
\begin{align*}
\underline b\left(\Gamma_{\mathrm{pre}}\right)
&:=
\min_{t\in\left\{-L+1,\ldots,0\right\}}
\Gamma_t
-
r,
\\
\overline b\left(\Gamma_{\mathrm{pre}}\right)
&:=
\max_{t\in\left\{-L+1,\ldots,0\right\}}
\Gamma_t
+
r.
\end{align*}
When \(L\geq2\), a bounded change restriction instead centers the allowed range at \(\Gamma_0\). It bounds the departure from this comparison by \(M\geq0\) times the largest change between consecutive pre-treatment differences:
\begin{align*}
\underline b\left(\Gamma_{\mathrm{pre}}\right)
&:=
\Gamma_0
-
M
\max_{t\in\left\{-L+2,\ldots,0\right\}}
\left|
\Gamma_t-\Gamma_{t-1}
\right|,
\\
\overline b\left(\Gamma_{\mathrm{pre}}\right)
&:=
\Gamma_0
+
M
\max_{t\in\left\{-L+2,\ldots,0\right\}}
\left|
\Gamma_t-\Gamma_{t-1}
\right|.
\end{align*}

\begin{corollary}[Bounds under bounded bias variation]
\label{cor:supp_did_bounded_bias_variation}
Suppose the assumptions of
Proposition~\ref{prop:supp_coarsen_did_att_bias} hold, and suppose
Assumption~\ref{ass:supp_did_bounded_bias_variation} holds. Then
\begin{align*}
\tau_{\mathrm{ATT}}^{\mathrm{DiD}}
\in
\left[
\theta_{\mathrm{ATT}}^{\mathrm{DiD}}
-
\overline b\left(\Gamma_{\mathrm{pre}}\right),
\,
\theta_{\mathrm{ATT}}^{\mathrm{DiD}}
-
\underline b\left(\Gamma_{\mathrm{pre}}\right)
\right].
\end{align*}
\end{corollary}

Corollary~\ref{cor:supp_did_bounded_bias_variation} translates the allowed range of post-treatment coarsening bias into bounds on the ATT. The upper bound on the bias gives the lower bound on the effect, and the lower bound on the bias gives the upper bound on the effect. Under exact bias stability, the interval collapses to the corrected DiD estimand
\[
\theta_{\mathrm{ATT}}^{\mathrm{DiD}}-\Gamma_0.
\]
The other restrictions allow the bias to change while still using the pre-treatment differences to bound the effect. This complements the sensitivity analysis in Section~\ref{sec:c_mean_dependence}, where observed means in neighboring cells restrict average unobserved potential outcomes. In the DiD design, pre-treatment differences instead restrict the post-treatment coarsening bias.

\section{Proofs for Section~\ref{sec:coarsening_setup}}
\begin{proof}[Proof of Proposition~\ref{prop:coarsen_bias_decomposition}]
Since $W=g\left(X\right)$, the coarsened covariate is measurable with
respect to $X$. By Assumption~\ref{ass:overlap},
\begin{align*}
\eta
\leq
e\left(W\right)
=
\mathrm{E}\left[e^*\left(X\right)\mid W\right]
\leq
1-\eta
\qquad
\text{a.s.},
\end{align*}
which implies that
$\mathrm{Pr}\left(D=d,W=w\right)>0$ for all
$d\in\left\{0,1\right\}$ and $w\in\mathcal{W}$. Hence, all
cellwise conditional means given $D=d$ and $W=w$ that appear below are
well defined.
Recall that \(\theta(w)=\Delta_Y(w)\), where
\(\Delta_Z(w):=\mathrm{E}\left[Z\mid D=1,W=w\right]
-\mathrm{E}\left[Z\mid D=0,W=w\right]\).
For \textup{(i)}, equation~\eqref{eq:observed_outcome} and
Assumption~\ref{ass:mean_independence_y0} imply
\begin{align*}
\mathrm{E}\left[Y\mid D=0,W\right]
&=
\mathrm{E}\left[Y\left(0\right)\mid D=0,W\right] \\
&=
\mathrm{E}\left[
\mathrm{E}\left[Y\left(0\right)\mid X,D=0,W\right]
\mid D=0,W
\right] \\
&=
\mathrm{E}\left[
\mathrm{E}\left[Y\left(0\right)\mid X,D=0\right]
\mid D=0,W
\right] \\
&=
\mathrm{E}\left[
\mathrm{E}\left[Y\left(0\right)\mid X\right]
\mid D=0,W
\right] \\
&=
\mathrm{E}\left[\mu_0\left(X\right)\mid D=0,W\right].
\end{align*}
Assumption~\ref{ass:mean_independence_y0} also implies
\begin{align*}
\mathrm{E}\left[Y\left(0\right)\mid D=1\right]
=
\mathrm{E}\left[
\mathrm{E}\left[\mu_0\left(X\right)\mid D=1,W\right]
\mid D=1
\right].
\end{align*}
First, consider the ATT. Using $\theta\left(W\right)
=
\mathrm{E}\left[Y\mid D=1,W\right]
-
\mathrm{E}\left[Y\mid D=0,W\right]$, this gives
\begin{align*}
\theta_{\mathrm{ATT}}
&=
\mathrm{E}\left[\theta\left(W\right)\mid D=1\right] \\
&=
\mathrm{E}\left[Y\mid D=1\right]
-
\mathrm{E}\left[
\mathrm{E}\left[\mu_0\left(X\right)\mid D=0,W\right]
\mid D=1
\right] \\
&=
\tau_{\mathrm{ATT}}
+
\mathrm{E}\left[
\mathrm{E}\left[\mu_0\left(X\right)\mid D=1,W\right]
-
\mathrm{E}\left[\mu_0\left(X\right)\mid D=0,W\right]
\mid D=1
\right] \\
&=
\tau_{\mathrm{ATT}}
+
\widetilde B_0.
\end{align*}
For \textup{(ii)}, suppose, in addition, that Assumption~\ref{ass:mean_independence_y1}
holds. The same steps give
\begin{align*}
\theta\left(W\right)
&=
\mathrm{E}\left[Y\mid D=1,W\right]
-
\mathrm{E}\left[Y\mid D=0,W\right] \\
&=
\mathrm{E}\left[\mu_1\left(X\right)\mid D=1,W\right]
-
\mathrm{E}\left[\mu_0\left(X\right)\mid D=0,W\right] \\
&=
\mathrm{E}\left[\tau\left(X\right)\mid D=1,W\right]
+
\mathrm{E}\left[\mu_0\left(X\right)\mid D=1,W\right]
-
\mathrm{E}\left[\mu_0\left(X\right)\mid D=0,W\right].
\end{align*}
For the ATE, the law of iterated expectations gives
\begin{align*}
\mathrm{E}\left[\tau\left(X\right)\mid W\right]
&=
e\left(W\right)
\mathrm{E}\left[\tau\left(X\right)\mid D=1,W\right] +
\left(1-e\left(W\right)\right)
\mathrm{E}\left[\tau\left(X\right)\mid D=0,W\right].
\end{align*}
Therefore,
\begin{align*}
\theta\left(W\right)
&=
\mathrm{E}\left[\tau\left(X\right)\mid D=1,W\right]
+
\mathrm{E}\left[\mu_0\left(X\right)\mid D=1,W\right]
-
\mathrm{E}\left[\mu_0\left(X\right)\mid D=0,W\right] \\
&=
\mathrm{E}\left[\tau\left(X\right)\mid W\right] \\
&\quad+
\left(1-e\left(W\right)\right)
\left(
\mathrm{E}\left[\tau\left(X\right)\mid D=1,W\right]
-
\mathrm{E}\left[\tau\left(X\right)\mid D=0,W\right]
\right) \\
&\quad+
\mathrm{E}\left[\mu_0\left(X\right)\mid D=1,W\right]
-
\mathrm{E}\left[\mu_0\left(X\right)\mid D=0,W\right].
\end{align*}
Taking expectations and using
$\tau_{\mathrm{ATE}}=\mathrm{E}\left[\tau\left(X\right)\right]
=\mathrm{E}\left[\mathrm{E}\left[\tau\left(X\right)\mid W\right]\right]$
gives
\begin{align*}
\theta_{\mathrm{ATE}}
&=
\mathrm{E}\left[\theta\left(W\right)\right] =
\tau_{\mathrm{ATE}}
+
B_{\tau}
+
B_0.
\end{align*}
This proves both decompositions.
\end{proof}

\begin{proof}[Proof of Corollary~\ref{cor:coarsen_harmless}]
For \textup{(i)}, suppose that
$\tau\left(X\right)=r_\tau\left(W\right)$ a.s.\ for some measurable function
$r_\tau$. Then
\begin{align*}
\mathrm{E}\left[\tau\left(X\right)\mid D=1,W\right]
&=
r_\tau\left(W\right)
=
\mathrm{E}\left[\tau\left(X\right)\mid D=0,W\right]
\qquad
\text{a.s.}
\end{align*}
Hence, by the definition of $B_{\tau}$, \(B_{\tau}=0\).

For \textup{(ii)}, suppose that
$\mu_0\left(X\right)=r_0\left(W\right)$ a.s.\ for some measurable function
$r_0$. Then
\begin{align*}
\mathrm{E}\left[\mu_0\left(X\right)\mid D=1,W\right]
&=
r_0\left(W\right)
=
\mathrm{E}\left[\mu_0\left(X\right)\mid D=0,W\right]
\qquad
\text{a.s.}
\end{align*}
Hence, by the definitions of $B_0$ and $\widetilde B_0$,
\(B_0=\widetilde B_0=0\).

Finally, for \textup{(iii)}, suppose that
$e^*\left(X\right)=e\left(W\right)$ a.s. Let $f\left(X\right)$ be either
$\tau\left(X\right)$ or $\mu_0\left(X\right)$. By Assumption~\ref{ass:overlap},
$e\left(W\right)$ and $1-e\left(W\right)$ are bounded away from zero
a.s. Hence,
\begin{align*}
\mathrm{E}\left[f\left(X\right)\mid D=1,W\right]
&=
\frac{
\mathrm{E}\left[D f\left(X\right)\mid W\right]
}{
e\left(W\right)
} \\
&=
\frac{
\mathrm{E}\left[
\mathrm{E}\left[D f\left(X\right)\mid X,W\right]
\mid W
\right]
}{
e\left(W\right)
} \\
&=
\frac{
\mathrm{E}\left[e^*\left(X\right)f\left(X\right)\mid W\right]
}{
e\left(W\right)
} \\
&=
\mathrm{E}\left[f\left(X\right)\mid W\right]
\quad
\text{a.s.}
\end{align*}
The same argument with \(1-D\) in place of \(D\) yields
\(\mathrm{E}\left[f\left(X\right)\mid D=0,W\right]
=\mathrm{E}\left[f\left(X\right)\mid W\right]\) a.s. Hence,
\begin{align*}
\mathrm{E}\left[f\left(X\right)\mid D=1,W\right]
=
\mathrm{E}\left[f\left(X\right)\mid D=0,W\right]
\qquad
\text{a.s.}
\end{align*}
Taking $f\left(X\right)=\tau\left(X\right)$ gives $B_{\tau}=0$, and taking
$f\left(X\right)=\mu_0\left(X\right)$ gives
$B_0=\widetilde B_0=0$.
\end{proof}

\begin{lemma}[Covariance inequality for monotone transforms]
\label{lem:monotone_covariance}
Let $X$ be a real-valued random variable with
$\mathrm{E}\left[\left|X\right|\right]<\infty$ and
$f:\mathbb{R}\to\mathbb{R}$ a bounded measurable function that is weakly
increasing on $\operatorname{supp}\left(X\right)$. Then
\begin{align*}
\mathrm{Cov}\left(X,f\left(X\right)\right)
=
\mathrm{E}\left[X f\left(X\right)\right]
-
\mathrm{E}\left[X\right]\mathrm{E}\left[f\left(X\right)\right]
\geq 0.
\end{align*}
If $f\left(X\right)$ is not a.s.\ constant, then the last inequality is strict.
\end{lemma}

\begin{proof}[Proof of Lemma~\ref{lem:monotone_covariance}]
Let $X'$ be an independent copy of $X$. Since $f$ is weakly increasing on
$\operatorname{supp}\left(X\right)$,
\begin{align}
\left(X-X'\right)
\left(f\left(X\right)-f\left(X'\right)\right)
\geq 0
\qquad
\text{a.s.}
\label{eq:monotone_product_nonnegative}
\end{align}
Taking expectations yields
\begin{align*}
0
&\leq
\mathrm{E}\left[
\left(X-X'\right)
\left(f\left(X\right)-f\left(X'\right)\right)
\right]\\
&=
\mathrm{E}\left[Xf\left(X\right)\right]
+
\mathrm{E}\left[X'f\left(X'\right)\right]
-
\mathrm{E}\left[Xf\left(X'\right)\right]
-
\mathrm{E}\left[X'f\left(X\right)\right]\\
&=
2\left(
\mathrm{E}\left[Xf\left(X\right)\right]
-
\mathrm{E}\left[X\right]\mathrm{E}\left[f\left(X\right)\right]
\right),
\end{align*}
where the last equality uses that $X'$ has the same distribution as $X$ and
that $X$ and $X'$ are independent. Consequently,
\begin{align*}
\mathrm{E}\left[Xf\left(X\right)\right]
\geq
\mathrm{E}\left[X\right]\mathrm{E}\left[f\left(X\right)\right].
\end{align*}

For strictness, suppose that $f\left(X\right)$ is not a.s.\ constant. Then
there exists $c\in\mathbb{R}$ such that
$\mathrm{Pr}\left(f\left(X\right)\leq c\right)>0$ and
$\mathrm{Pr}\left(f\left(X\right)>c\right)>0$. By independence of $X$ and
$X'$, and since $X'$ has the same distribution as $X$,
\begin{align*}
\mathrm{Pr}\left(f\left(X\right)\leq c<f\left(X'\right)\right)
&=
\mathrm{Pr}\left(f\left(X\right)\leq c\right)
\mathrm{Pr}\left(f\left(X'\right)>c\right)\\
&=
\mathrm{Pr}\left(f\left(X\right)\leq c\right)
\mathrm{Pr}\left(f\left(X\right)>c\right)
>0.
\end{align*}
Thus, the event
$\left\{f\left(X\right)\leq c<f\left(X'\right)\right\}$ has positive
probability. For $x,x'\in\operatorname{supp}\left(X\right)$, weak
monotonicity of $f$ implies that
$f\left(x\right)<f\left(x'\right)$ only if $x<x'$. Hence, on this event,
$X<X'$ and $f\left(X\right)<f\left(X'\right)$, which implies
\begin{align*}
\left(X-X'\right)
\left(f\left(X\right)-f\left(X'\right)\right)>0
\end{align*}
except on a null set. Together with
\eqref{eq:monotone_product_nonnegative}, this implies
\begin{align*}
\mathrm{E}\left[
\left(X-X'\right)
\left(f\left(X\right)-f\left(X'\right)\right)
\right]
>0.
\end{align*}
Repeating the algebra above gives
\begin{align*}
\mathrm{E}\left[Xf\left(X\right)\right]
>
\mathrm{E}\left[X\right]\mathrm{E}\left[f\left(X\right)\right].
\end{align*}
\end{proof}

\begin{proof}[Proof of Corollary~\ref{cor:positive_coarsening_bias}]
For \textup{(i)}, define, for each $w\in\mathcal{W}$,
\begin{align*}
\Delta_{\tau}\left(w\right)
:=
\mathrm{E}\left[\tau\left(X\right)\mid D=1,W=w\right]
-
\mathrm{E}\left[\tau\left(X\right)\mid D=0,W=w\right].
\end{align*}
By Assumption~\ref{ass:overlap}, $0<e\left(w\right)<1$. By the definition
of $e_{\tau}$ and iterated expectations,
\begin{align*}
e\left(w\right)
&=
\mathrm{Pr}\left(D=1\mid W=w\right)=
\mathrm{E}\left[
\mathrm{E}\left[D\mid \tau\left(X\right),W=w\right]
\mid W=w
\right]\\
&=
\mathrm{E}\left[
e_{\tau}\left(\tau\left(X\right),w\right)
\mid W=w
\right],
\end{align*}
and
\begin{align*}
\mathrm{E}\left[\tau\left(X\right)D\mid W=w\right]
&=
\mathrm{E}\left[
\mathrm{E}\left[
\tau\left(X\right)D
\mid \tau\left(X\right),W=w
\right]
\mid W=w
\right]\\
&=
\mathrm{E}\left[
\tau\left(X\right)
\mathrm{E}\left[D\mid \tau\left(X\right),W=w\right]
\mid W=w
\right]\\
&=
\mathrm{E}\left[
\tau\left(X\right)
e_{\tau}\left(\tau\left(X\right),w\right)
\mid W=w
\right].
\end{align*}
Similarly,
\begin{align*}
\mathrm{E}\left[\tau\left(X\right)\left(1-D\right)\mid W=w\right]
=\mathrm{E}\left[
\tau\left(X\right)
\left(1-e_{\tau}\left(\tau\left(X\right),w\right)\right)
\mid W=w
\right].
\end{align*}
Hence,
\begin{align*}
\mathrm{E}\left[\tau\left(X\right)\mid D=1,W=w\right]
&=
\frac{
\mathrm{E}\left[\tau\left(X\right)D\mid W=w\right]
}{
\mathrm{Pr}\left(D=1\mid W=w\right)
}=
\frac{
\mathrm{E}\left[
\tau\left(X\right)
e_{\tau}\left(\tau\left(X\right),w\right)
\mid W=w
\right]
}{
e\left(w\right)
},
\\
\mathrm{E}\left[\tau\left(X\right)\mid D=0,W=w\right]
&=
\frac{
\mathrm{E}\left[\tau\left(X\right)\left(1-D\right)\mid W=w\right]
}{
\mathrm{Pr}\left(D=0\mid W=w\right)
}=
\frac{
\mathrm{E}\left[
\tau\left(X\right)
\left(1-e_{\tau}\left(\tau\left(X\right),w\right)\right)
\mid W=w
\right]
}{
1-e\left(w\right)
}.
\end{align*}
It follows that
\begin{align*}
\Delta_{\tau}\left(w\right)
&=
\frac{
\mathrm{E}\left[
\tau\left(X\right)
e_{\tau}\left(\tau\left(X\right),w\right)
\mid W=w
\right]
}{
e\left(w\right)
}
-
\frac{
\mathrm{E}\left[
\tau\left(X\right)
\left(1-e_{\tau}\left(\tau\left(X\right),w\right)\right)
\mid W=w
\right]
}{
1-e\left(w\right)
}\\
&=
\frac{
\mathrm{E}\left[
\tau\left(X\right)
e_{\tau}\left(\tau\left(X\right),w\right)
\mid W=w
\right]
-
e\left(w\right)
\mathrm{E}\left[\tau\left(X\right)\mid W=w\right]
}{
e\left(w\right)\left(1-e\left(w\right)\right)
}\\
&=
\frac{
\mathrm{E}\left[
\tau\left(X\right)
e_{\tau}\left(\tau\left(X\right),w\right)
\mid W=w
\right]
-
\mathrm{E}\left[\tau\left(X\right)\mid W=w\right]
\mathrm{E}\left[
e_{\tau}\left(\tau\left(X\right),w\right)
\mid W=w
\right]
}{
e\left(w\right)\left(1-e\left(w\right)\right)
}.
\end{align*}

Under the conditional distribution given $W=w$,
Lemma~\ref{lem:monotone_covariance} applied to the random variable
$\tau\left(X\right)$ and the function
\(y\mapsto e_{\tau}\left(y,w\right)\) shows that the numerator in the last
display is nonnegative for all $w\in\mathcal{W}$.
Thus,
\begin{align*}
\Delta_{\tau}\left(w\right)\geq 0
\qquad
\text{for all } w\in\mathcal{W}.
\end{align*}
Choose a cell $w_0\in\mathcal{W}$ satisfying the nondegeneracy condition in \textup{(i)}. The strict inequality in
Lemma~\ref{lem:monotone_covariance} applies because
$e_{\tau}\left(\tau\left(X\right),w_0\right)$ is not a.s.\ constant
conditional on $W=w_0$. Hence,
\begin{align*}
\Delta_{\tau}\left(w_0\right)>0.
\end{align*}
Since \(w_0\in\mathcal{W}\) has positive probability and
\(1-e\left(w_0\right)>0\), the definition of \(B_{\tau}\) gives
\begin{align*}
B_{\tau}
&=
\mathrm{E}\left[
\left(1-e\left(W\right)\right)
\Delta_{\tau}\left(W\right)
\right]
>0.
\end{align*}

For \textup{(ii)}, define, for each
\(w\in\mathcal{W}\),
\begin{align*}
\Delta_0\left(w\right)
:=
\mathrm{E}\left[\mu_0\left(X\right)\mid D=1,W=w\right]
-
\mathrm{E}\left[\mu_0\left(X\right)\mid D=0,W=w\right].
\end{align*}
Repeating the preceding argument with $\mu_0\left(X\right)$ in place of
$\tau\left(X\right)$ and \(e_0\) in place of \(e_{\tau}\) gives
\(\Delta_0\left(w\right)\geq 0\) for all
\(w\in\mathcal{W}\). Choose a cell \(w_1\in\mathcal{W}\) satisfying the nondegeneracy condition in \textup{(ii)}. The strict inequality in
Lemma~\ref{lem:monotone_covariance} gives
\(\Delta_0\left(w_1\right)>0\). Since \(w_1\in\mathcal{W}\) has
positive probability, the definition of \(B_0\) gives
\begin{align*}
B_0
&=
\mathrm{E}\left[\Delta_0\left(W\right)\right]
>0.
\end{align*}

Finally, Assumption~\ref{ass:overlap} implies \(e\left(w_1\right)>0\), which yields
\(\mathrm{Pr}\left(W=w_1\mid D=1\right)>0\). Hence,
\begin{align*}
\widetilde B_0
&=
\mathrm{E}\left[\Delta_0\left(W\right)\mid D=1\right]
>0.
\end{align*}
\end{proof}

\begin{proof}[Proof of Proposition~\ref{prop:supp_coarsen_did_att_bias}]
Assumption~\ref{ass:supp_did_no_anticipation} gives
\[
\tau_{\mathrm{ATT}}^{\mathrm{DiD}}
=
\mathrm{E}\left[
\Delta_1Y\left(1\right)-\Delta_1Y\left(0\right)
\mid D=1
\right].
\]
The definition of \(\theta_{\mathrm{ATT}}^{\mathrm{DiD}}\) and the observed
outcome equation imply
\begin{align*}
\theta_{\mathrm{ATT}}^{\mathrm{DiD}}
&=
\mathrm{E}\left[
\mathrm{E}\left[\Delta_1Y\mid D=1,W\right]
-
\mathrm{E}\left[\Delta_1Y\mid D=0,W\right]
\mid D=1
\right]
\\
&=
\mathrm{E}\left[
\mathrm{E}\left[\Delta_1Y\left(1\right)\mid D=1,W\right]
-
\mathrm{E}\left[\Delta_1Y\left(0\right)\mid D=0,W\right]
\mid D=1
\right].
\end{align*}
Adding and subtracting
\(\mathrm{E}\left[\Delta_1Y\left(0\right)\mid D=1,W\right]\) inside the
outer expectation yields
\begin{align*}
\theta_{\mathrm{ATT}}^{\mathrm{DiD}}
&=
\mathrm{E}\left[
\mathrm{E}\left[
\Delta_1Y\left(1\right)-\Delta_1Y\left(0\right)
\mid D=1,W
\right]
\mid D=1
\right]
\\
&\quad+
\mathrm{E}\left[
\mathrm{E}\left[
\Delta_1Y\left(0\right)
\mid D=1,W
\right]
-
\mathrm{E}\left[
\Delta_1Y\left(0\right)
\mid D=0,W
\right]
\mid D=1
\right]
\\
&=
\tau_{\mathrm{ATT}}^{\mathrm{DiD}}
+
B_{\mathrm{ATT}}^{\mathrm{DiD}}.
\end{align*}
\end{proof}

\begin{proof}[Proof of Corollary~\ref{cor:supp_did_bounded_bias_variation}]
By Proposition~\ref{prop:supp_coarsen_did_att_bias},
\[
\tau_{\mathrm{ATT}}^{\mathrm{DiD}}
=
\theta_{\mathrm{ATT}}^{\mathrm{DiD}}
-
B_{\mathrm{ATT}}^{\mathrm{DiD}}.
\]
Assumption~\ref{ass:supp_did_bounded_bias_variation} restricts
\(B_{\mathrm{ATT}}^{\mathrm{DiD}}\) to the interval
\[
\left[
\underline b\left(\Gamma_{\mathrm{pre}}\right),
\overline b\left(\Gamma_{\mathrm{pre}}\right)
\right].
\]
Therefore, the affine map
\(b\mapsto \theta_{\mathrm{ATT}}^{\mathrm{DiD}}-b\) maps the
admissible bias interval into
\[
\left[
\theta_{\mathrm{ATT}}^{\mathrm{DiD}}
-
\overline b\left(\Gamma_{\mathrm{pre}}\right),
\,
\theta_{\mathrm{ATT}}^{\mathrm{DiD}}
-
\underline b\left(\Gamma_{\mathrm{pre}}\right)
\right],
\]
which gives the stated bounds.
\end{proof}

\section{Proofs for Section~\ref{sec:c_mean_dependence}}

As in the main text, \(W\) denotes the conditioning variable.
Recall that \(m_d(w):=\mathrm{E}\left[Y\mid D=d,W=w\right]\) and
\(m_d^{\mathrm{cf}}(w):=\mathrm{E}\left[Y(d)\mid D=1-d,W=w\right]\).
Also, as in Definition~\ref{def:c_mean_dependence}, for \(c\geq0\),
\(\underline m_{d,c}(w):=\inf_{u\in\mathcal{N}_c(w)}m_d(u)\) and
\(\overline m_{d,c}(w):=\sup_{u\in\mathcal{N}_c(w)}m_d(u)\).

\begin{proof}[Proof of Proposition~\ref{prop:c_mean_dependence_bounds}]
\underline{\textit{Validity.}}
For the ATT, if $Y\left(0\right)$ is $c$-mean dependent with \(D\) given
\(W\), then
\begin{align*}
\underline m_{0,c}\left(w\right)
\leq
m^{\mathrm{cf}}_{0}\left(w\right)
\leq
\overline m_{0,c}\left(w\right)
\qquad
\text{for all } w\in\mathcal{W}.
\end{align*}
Taking conditional expectations given $D=1$ gives
\begin{align*}
\mathrm{E}\left[
\underline m_{0,c}\left(W\right)
\mid D=1
\right]
\leq
\mathrm{E}\left[
m^{\mathrm{cf}}_{0}\left(W\right)
\mid D=1
\right]
\leq
\mathrm{E}\left[
\overline m_{0,c}\left(W\right)
\mid D=1
\right].
\end{align*}
Moreover, by equation~\eqref{eq:observed_outcome} and iterated expectations,
\[
\tau_{\mathrm{ATT}}
=
\mathrm{E}\left[Y\left(1\right)-Y\left(0\right)\mid D=1\right]
=
\mathrm{E}\left[Y\mid D=1\right]
-
\mathrm{E}\left[
m^{\mathrm{cf}}_{0}\left(W\right)
\mid D=1
\right].
\]
Combining the last two displays gives
\begin{align*}
\underline\tau_{\mathrm{ATT},c}
\leq
\tau_{\mathrm{ATT}}
\leq
\overline\tau_{\mathrm{ATT},c}.
\end{align*}

For part~\textup{(ii)}, suppose \(Y\left(1\right)\) and \(Y\left(0\right)\) are
\(c\)-mean dependent with \(D\) given \(W\). As in part~\textup{(i)},
\(\underline\tau_{\mathrm{ATT},c}\leq\tau_{\mathrm{ATT}}\leq\overline\tau_{\mathrm{ATT},c}\).
Write \(\underline\nu_c:=\underline\tau_{\mathrm{ATE},c}-\pi_1\underline\tau_{\mathrm{ATT},c}\) and \(\overline\nu_c:=\overline\tau_{\mathrm{ATE},c}-\pi_1\overline\tau_{\mathrm{ATT},c}\). By the definition of the joint set \(\mathcal I_c\) in part~\textup{(ii)},
\begin{align}
\label{eq:Ic_equivalence}
\left(t_1,t_2\right)\in\mathcal I_c
\quad\text{if and only if}\quad
t_1\in\left[\underline\tau_{\mathrm{ATT},c},\overline\tau_{\mathrm{ATT},c}\right]
\ \text{ and }\ 
t_2-\pi_1t_1\in\left[\underline\nu_c,\overline\nu_c\right].
\end{align}
Therefore, it remains to show that
\(\tau_{\mathrm{ATE}}-\pi_1\tau_{\mathrm{ATT}}\in\left[\underline\nu_c,\overline\nu_c\right]\).
By equation~\eqref{eq:observed_outcome} and iterated expectations,
\begin{align*}
\mathrm{E}\left[Y\left(1\right)\mid W\right]
&=
e\left(W\right)m_{1}\left(W\right)
+
\left(1-e\left(W\right)\right)
m^{\mathrm{cf}}_{1}\left(W\right),
\\
\mathrm{E}\left[Y\left(0\right)\mid W\right]
&=
e\left(W\right)m^{\mathrm{cf}}_{0}\left(W\right)
+
\left(1-e\left(W\right)\right)
m_{0}\left(W\right).
\end{align*}
Therefore,
\begin{align*}
\tau_{\mathrm{ATE}}
&=
\mathrm{E}\left[
e\left(W\right)
\bigl(
m_{1}\left(W\right)
-
m^{\mathrm{cf}}_{0}\left(W\right)
\bigr)
\right]
+
\mathrm{E}\left[
\left(1-e\left(W\right)\right)
\bigl(
m^{\mathrm{cf}}_{1}\left(W\right)
-
m_{0}\left(W\right)
\bigr)
\right].
\end{align*}
Note that, for any function \(f\) with \(\mathrm E[D|f(W)|]<\infty\),
\begin{align}
\label{eq:e_weighting}
\mathrm{E}\left[e\left(W\right)f\left(W\right)\right]=\mathrm{E}\left[Df\left(W\right)\right]=\pi_1\mathrm{E}\left[f\left(W\right)\mid D=1\right].
\end{align}
Applying \eqref{eq:e_weighting} with \(f=m_{1}-m^{\mathrm{cf}}_{0}\) gives
\[
\mathrm{E}\left[
e\left(W\right)
\bigl(
m_{1}\left(W\right)
-
m^{\mathrm{cf}}_{0}\left(W\right)
\bigr)
\right]
=
\pi_1\mathrm{E}\left[
m_{1}\left(W\right)
-
m^{\mathrm{cf}}_{0}\left(W\right)
\mid D=1
\right]
=
\pi_1\tau_{\mathrm{ATT}},
\]
where the last equality uses \(\mathrm{E}\left[Y\mid D=1\right]=\mathrm{E}\left[m_{1}\left(W\right)\mid D=1\right]\) and the expression for the ATT in part~\textup{(i)}. Hence,
\begin{align}
\label{eq:ate_att_identity}
\tau_{\mathrm{ATE}}-\pi_1\tau_{\mathrm{ATT}}
=
\mathrm{E}\left[
\left(1-e\left(W\right)\right)
\bigl(
m^{\mathrm{cf}}_{1}\left(W\right)
-
m_{0}\left(W\right)
\bigr)
\right].
\end{align}
Similarly, applying \eqref{eq:e_weighting} with \(f=m_{1}-\overline m_{0,c}\) and with \(f=m_{1}-\underline m_{0,c}\) gives
\[
\mathrm{E}\left[
e\left(W\right)
\bigl(
m_{1}\left(W\right)
-
\overline m_{0,c}\left(W\right)
\bigr)
\right]
=
\pi_1\underline\tau_{\mathrm{ATT},c},
\qquad
\mathrm{E}\left[
e\left(W\right)
\bigl(
m_{1}\left(W\right)
-
\underline m_{0,c}\left(W\right)
\bigr)
\right]
=
\pi_1\overline\tau_{\mathrm{ATT},c}.
\]
Subtracting these from \(\underline\tau_{\mathrm{ATE},c}\) and \(\overline\tau_{\mathrm{ATE},c}\) in \eqref{eq:c_mean_ate_bounds} gives
\begin{equation}
\label{eq:nu_expressions}
\begin{aligned}
\underline\nu_c
&=
\underline\tau_{\mathrm{ATE},c}-\pi_1\underline\tau_{\mathrm{ATT},c}
=
\mathrm{E}\left[
\left(1-e\left(W\right)\right)
\bigl(
\underline m_{1,c}\left(W\right)
-
m_{0}\left(W\right)
\bigr)
\right],
\\
\overline\nu_c
&=
\overline\tau_{\mathrm{ATE},c}-\pi_1\overline\tau_{\mathrm{ATT},c}
=
\mathrm{E}\left[
\left(1-e\left(W\right)\right)
\bigl(
\overline m_{1,c}\left(W\right)
-
m_{0}\left(W\right)
\bigr)
\right].
\end{aligned}
\end{equation}
Since \(1-e\left(W\right)\geq0\) and \(\underline m_{1,c}\left(W\right)\leq m^{\mathrm{cf}}_{1}\left(W\right)\leq\overline m_{1,c}\left(W\right)\), the expectation in \eqref{eq:ate_att_identity} lies between the two expectations in \eqref{eq:nu_expressions}. Hence, \(\underline\nu_c\leq\tau_{\mathrm{ATE}}-\pi_1\tau_{\mathrm{ATT}}\leq\overline\nu_c\).

\underline{\textit{Sharpness.}}
For the ATT, let
\(t\in
\left[
\underline\tau_{\mathrm{ATT},c},
\overline\tau_{\mathrm{ATT},c}
\right]\).
I construct below a joint law of the variables \(\left(Y\left(1\right),Y\left(0\right),Y,D,W\right)\) such that
\begin{enumerate}[label=\textup{(\alph*)}]
\item the marginal law of \(\left(Y,D,W\right)\) is the given observable law,
\item equation~\eqref{eq:observed_outcome} holds,
\item \(Y\left(0\right)\) is \(c\)-mean dependent with \(D\) given \(W\), and
\item the ATT equals \(t\).
\end{enumerate}
Since \(t\in\left[\underline\tau_{\mathrm{ATT},c},\overline\tau_{\mathrm{ATT},c}\right]\), there exists \(\lambda\in\left[0,1\right]\) such that \(t=\lambda\underline\tau_{\mathrm{ATT},c}+\left(1-\lambda\right)\overline\tau_{\mathrm{ATT},c}\). Define \(g_0:\mathcal{W}\to\mathbb{R}\) by
\[
g_0\left(w\right)
:=
\lambda\overline m_{0,c}\left(w\right)
+
\left(1-\lambda\right)\underline m_{0,c}\left(w\right),
\]
so that \(\underline m_{0,c}\left(w\right)\leq g_0\left(w\right)\leq\overline m_{0,c}\left(w\right)\) for all \(w\in\mathcal{W}\) and \(\mathrm{E}\left[Y\mid D=1\right]-\mathrm{E}\left[g_0\left(W\right)\mid D=1\right]=t\). Let \(g_1:\mathcal{W}\to\mathbb{R}\) be any integrable function. Define
\begin{align*}
Y\left(0\right)
&:=
\left(1-D\right)Y
+
D g_0\left(W\right),
\\
Y\left(1\right)
&:=
D Y
+
\left(1-D\right)g_1\left(W\right).
\end{align*}
Together with the given observable marginal law of \(\left(Y,D,W\right)\),
these definitions induce a joint law of
\(\left(Y\left(1\right),Y\left(0\right),Y,D,W\right)\) whose marginal law for
\(\left(Y,D,W\right)\) is unchanged by construction, which establishes \textup{(a)}. Since \(D\in\left\{0,1\right\}\), \(DY\left(1\right)+\left(1-D\right)Y\left(0\right)=DY+\left(1-D\right)Y=Y\), which establishes \textup{(b)}. Moreover, \(m^{\mathrm{cf}}_{0}\left(w\right)=\mathrm{E}\left[Y\left(0\right)\mid D=1,W=w\right]=g_0\left(w\right)\in\left[\underline m_{0,c}\left(w\right),\overline m_{0,c}\left(w\right)\right]\) for all \(w\in\mathcal{W}\). Hence, \textup{(c)} holds by Definition~\ref{def:c_mean_dependence}. Finally,
\[
\tau_{\mathrm{ATT}}
=
\mathrm{E}\left[Y\left(1\right)-Y\left(0\right)\mid D=1\right]
=
\mathrm{E}\left[Y\mid D=1\right]
-
\mathrm{E}\left[g_0\left(W\right)\mid D=1\right]
=
t,
\]
which establishes \textup{(d)}.

For part~\textup{(ii)}, let \(\left(t_1,t_2\right)\in\mathcal I_c\). By \eqref{eq:Ic_equivalence},
\(t_2-\pi_1t_1\in\left[\underline\nu_c,\overline\nu_c\right]\).
I construct below a joint law of the variables \(\left(Y\left(1\right),Y\left(0\right),Y,D,W\right)\) such that
\begin{enumerate}[label=\textup{(\alph*)}]
\item the marginal law of \(\left(Y,D,W\right)\) is the given observable law,
\item equation~\eqref{eq:observed_outcome} holds,
\item \(Y\left(0\right)\) and \(Y\left(1\right)\) are \(c\)-mean dependent with \(D\) given \(W\), and
\item the ATT equals \(t_1\) and the ATE equals \(t_2\).
\end{enumerate}
As in part~\textup{(i)}, there exist \(\lambda_0,\lambda_1\in\left[0,1\right]\) such that \(t_1=\lambda_0\underline\tau_{\mathrm{ATT},c}+\left(1-\lambda_0\right)\overline\tau_{\mathrm{ATT},c}\) and \(t_2-\pi_1t_1=\lambda_1\underline\nu_c+\left(1-\lambda_1\right)\overline\nu_c\). Define \(Y\left(0\right)\) and \(Y\left(1\right)\) as in part~\textup{(i)}, with \(\lambda_0\) in place of \(\lambda\) in \(g_0\) and with
\[
g_1\left(w\right)
:=
\lambda_1\underline m_{1,c}\left(w\right)
+
\left(1-\lambda_1\right)\overline m_{1,c}\left(w\right).
\]
Since \(\underline m_{1,c}\left(w\right)\leq g_1\left(w\right)\leq\overline m_{1,c}\left(w\right)\) for all \(w\in\mathcal{W}\), the same argument as in part~\textup{(i)} shows that \textup{(a)}--\textup{(c)} hold and that \(\tau_{\mathrm{ATT}}=t_1\). Since \eqref{eq:ate_att_identity} holds under \textup{(b)},
\begin{align*}
\tau_{\mathrm{ATE}}
=
\pi_1t_1
+
\mathrm{E}\left[
\left(1-e\left(W\right)\right)
\bigl(
g_1\left(W\right)
-
m_{0}\left(W\right)
\bigr)
\right]
=
\pi_1t_1
+
\lambda_1\underline\nu_c
+
\left(1-\lambda_1\right)\overline\nu_c
=
t_2,
\end{align*}
where the second equality follows from \eqref{eq:nu_expressions}. Hence, \textup{(d)} holds.
\end{proof}

\begin{proof}[Proof of Proposition~\ref{prop:one_step_c_mean_dependence}]
By Assumption~\ref{ass:mean_independence_y0} if \(d=0\) or Assumption~\ref{ass:mean_independence_y1} if \(d=1\), and \eqref{eq:observed_outcome},
\[
m_d(k)=\mathrm E[\mu_d(X)\mid D=d,W=k],
\qquad
m_d^{\mathrm{cf}}(k)=\mathrm E[\mu_d(X)\mid D=1-d,W=k].
\]
Assumption~\ref{ass:overlap} ensures that both treatment groups have positive probability in each cell. Without loss of generality, suppose \(\mu_d\) is nondecreasing. Since all fine values in \(A_k\) precede those in \(A_{k+1}\), \(\mu_d(x)\leq\mu_d(x')\) for all \(x\in A_k\) and \(x'\in A_{k+1}\). By the display above, \(m_d(k)\) and \(m_d^{\mathrm{cf}}(k)\) are means of \(\mu_d(X)\) over \(A_k\), while \(m_d(k+1)\) and \(m_d^{\mathrm{cf}}(k+1)\) are means of \(\mu_d(X)\) over \(A_{k+1}\). Hence,
\[
\max\{m_d(k),m_d^{\mathrm{cf}}(k)\}
\leq
\min\{m_d(k+1),m_d^{\mathrm{cf}}(k+1)\},
\qquad k=1,\ldots,K-1.
\]
In particular, \(m_d(k-1)\leq m_d^{\mathrm{cf}}(k)\leq m_d(k+1)\) for \(k=2,\ldots,K-1\), and \(m_d(k)\) is nondecreasing.

At the bottom cell, \(\mathcal N_1(1)=\{1,2\}\) and the observed means are nondecreasing. Hence, \(1\)-mean dependence requires \(m_d(1)\leq m_d^{\mathrm{cf}}(1)\leq m_d(2)\). The inequality above with \(k=1\) gives \(m_d^{\mathrm{cf}}(1)\leq m_d(2)\). At the bottom cell, \(1\)-mean dependence is therefore equivalent to \(m_d^{\mathrm{cf}}(1)\geq m_d(1)\). Similarly, at the top cell, \(1\)-mean dependence requires \(m_d(K-1)\leq m_d^{\mathrm{cf}}(K)\leq m_d(K)\), and the same inequality with \(k=K-1\) gives \(m_d(K-1)\leq m_d^{\mathrm{cf}}(K)\). At the top cell, it is therefore equivalent to \(m_d^{\mathrm{cf}}(K)\leq m_d(K)\). Since the inequalities for \(k=2,\ldots,K-1\) give \(1\)-mean dependence at every interior cell, \(1\)-mean dependence holds at all cells if and only if \(m_d^{\mathrm{cf}}(1)\geq m_d(1)\) and \(m_d^{\mathrm{cf}}(K)\leq m_d(K)\). Since \(m_d(k)\) is nondecreasing, \(m_d^{\mathrm{cf}}(1)\leq m_d(2)\leq m_d(K)\) and \(m_d^{\mathrm{cf}}(K)\geq m_d(K-1)\geq m_d(1)\). Hence, these two inequalities hold if and only if \(m_d^{\mathrm{cf}}(1)\) and \(m_d^{\mathrm{cf}}(K)\) lie between \(m_d(1)\) and \(m_d(K)\).
\end{proof}

\section{Proofs for Section~\ref{sec:estimation_c_mean_dependence}}
\label{sec:supp_c_mean_dependence_estimation}

Recall that
\(\vartheta:=((p_d(w))_{d\in\{0,1\},\,w\in\mathcal W},
(m_d(w))_{d\in\{0,1\},\,w\in\mathcal W})\) and that
\(\phi_\rho(\vartheta)=(\underline\tau_{\rho,\cdot},
\overline\tau_{\rho,\cdot})'\) for
\(\rho\in\{\mathrm{ATE},\mathrm{ATT}\}\), as in
\eqref{eq:c_mean_first_stage_components}--\eqref{eq:c_mean_phi_plugin_representation}.
For the first-stage argument, also write, for
\(d\in\left\{0,1\right\}\) and \(w\in\mathcal W\),
\[
s_d(w):=\mathrm{E}\left[
Y\mathbb{I}\left\{D=d,W=w\right\}
\right],
\qquad
\widehat s_d(w):=
\widehat{\mathrm{E}}\left[
Y_i\mathbb{I}\left\{D_i=d,W_i=w\right\}
\right],
\]
where \(\widehat{\mathrm{E}}\) is the empirical mean in \eqref{eq:empirical_mean}.
For notational ease in the first-stage arguments, let
\[
\zeta
:=
\left(
\left(p_d(w)\right)_{d\in\left\{0,1\right\},\,w\in\mathcal W},
\left(s_d(w)\right)_{d\in\left\{0,1\right\},\,w\in\mathcal W}
\right),
\qquad
\widehat\zeta
:=
\left(
\left(\widehat p_d(w)\right)_{d\in\left\{0,1\right\},\,w\in\mathcal W},
\left(\widehat s_d(w)\right)_{d\in\left\{0,1\right\},\,w\in\mathcal W}
\right).
\]
Then \(m_d(w)=s_d(w)/p_d(w)\), and
\(\widehat m_d(w)=\widehat s_d(w)/\widehat p_d(w)\) whenever
\(\widehat p_d(w)>0\). As in the main text, \(\widehat m_d(w)\) is set
equal to zero whenever the denominator is zero, an event that occurs with
probability approaching zero under
Assumptions~\ref{ass:c_mean_est_sampling}--\ref{ass:c_mean_est_regularity}.
Let \(\pi_1:=\Pr\left(D=1\right)=\sum_{w\in\mathcal W} p_1(w)\). For
\(d\in\left\{0,1\right\}\), \(w\in\mathcal W\), and
\(c\in\mathcal C\), note that
\[
\underline m_{d,c}(w)=\underline m_{d,c}\left(w;\vartheta\right),
\qquad
\overline m_{d,c}(w)=\overline m_{d,c}\left(w;\vartheta\right).
\]

\begin{lemma}[Convergence of first-stage estimators]
\label{lem:c_mean_first_stage}
Suppose Assumptions~\ref{ass:c_mean_est_sampling}--\ref{ass:c_mean_est_regularity}
hold. Then
\[
\sqrt N\left(\widehat\vartheta-\vartheta\right)
\rightsquigarrow
Z
\qquad
\text{in }\mathbb R^{4K},
\]
where \(Z:=\left(\left(Z^p_d(w)\right)_{d\in\left\{0,1\right\},\,w\in\mathcal W},
\left(Z^m_d(w)\right)_{d\in\left\{0,1\right\},\,w\in\mathcal W}\right)\)
is a mean-zero Gaussian vector with
covariance matrix characterized by, for
\(d,d'\in\left\{0,1\right\}\) and \(w,w'\in\mathcal W\),
\begin{align*}
\mathrm{E}\left[Z^p_d(w)Z^p_{d'}(w')\right]
&=
p_d(w)\mathbb{I}\left\{d=d',w=w'\right\}
-p_d(w)p_{d'}(w'),
\\
\mathrm{E}\left[Z^m_d(w)Z^m_{d'}(w')\right]
&=
\frac{\mathbb{I}\left\{d=d',w=w'\right\}}{p_d(w)}
\mathrm{E}\left[
\left(Y-m_d(w)\right)^2
\mid D=d,W=w
\right],
\\
\mathrm{E}\left[Z^p_d(w)Z^m_{d'}(w')\right]
&=
0.
\end{align*}
\end{lemma}

\begin{proof}[Proof of Lemma~\ref{lem:c_mean_first_stage}]
Under Assumptions~\ref{ass:c_mean_est_sampling}--\ref{ass:c_mean_est_support}, the coordinates of \(\widehat\zeta\) are
finite-dimensional averages of i.i.d.\ summands.
Assumption~\ref{ass:c_mean_est_regularity} gives finite second moments. The
multivariate CLT yields
\[
\sqrt N\left(\widehat\zeta-\zeta\right)
\rightsquigarrow
\widetilde Z
\qquad
\text{in }\mathbb R^{4K},
\]
where \(\widetilde Z\) is mean-zero Gaussian. Since
\(p_d(w)>0\) for all \(d\in\left\{0,1\right\}\) and \(w\in\mathcal W\) by
Assumption~\ref{ass:c_mean_est_regularity}, the map
\[
\left(
\left(q_d(w)\right)_{d\in\left\{0,1\right\},\,w\in\mathcal W},
\left(t_d(w)\right)_{d\in\left\{0,1\right\},\,w\in\mathcal W}
\right)
\mapsto
\left(
\left(q_d(w)\right)_{d\in\left\{0,1\right\},\,w\in\mathcal W},
\left(t_d(w)/q_d(w)\right)_{d\in\left\{0,1\right\},\,w\in\mathcal W}
\right)
\]
is continuously differentiable at
\(\zeta\). It maps \(\zeta\) to \(\vartheta\) and, with probability
approaching one, maps \(\widehat\zeta\) to \(\widehat\vartheta\). The delta
method then yields the asserted joint weak convergence. The estimators \(\widehat p_d(w)\) and \(\widehat m_d(w)\) are asymptotically linear with influence functions
\[
\mathbb{I}\left\{D=d,W=w\right\}-p_d(w)
\qquad\text{and}\qquad
\frac{\mathbb{I}\left\{D=d,W=w\right\}\left(Y-m_d(w)\right)}{p_d(w)},
\]
respectively, and the covariances of these influence functions give the covariance matrix in the statement.
\end{proof}

\begin{lemma}[Hadamard directional differentiability of ATE and ATT bound functionals]
\label{lem:c_mean_bound_maps}
Suppose Assumptions~\ref{ass:c_mean_est_support}--\ref{ass:c_mean_est_regularity}
hold. Then

\begin{enumerate}[label=\textup{(\roman*)}]
\item The ATE bound functional \(\phi_{\mathrm{ATE}}\) is Hadamard
directionally differentiable at \(\vartheta\). Its derivative
\(\phi_{\mathrm{ATE},\vartheta}'\) satisfies
\footnotesize\begin{align*}
&\left[\phi_{\mathrm{ATE},\vartheta}'(r)\right](c)
\\
&=
\begin{pmatrix}
\sum_{w\in\mathcal W}
\left[\begin{aligned}
&r^p_1(w)\left(m_1(w)-\overline m_{0,c}(w)\right)
+p_1(w)\left(r^m_1(w)-\max_{w'\in\overline\Psi_{0,c}(w)}r^m_0(w')\right)\\
&+r^p_0(w)\left(\underline m_{1,c}(w)-m_0(w)\right)
+p_0(w)\left(\min_{w'\in\underline\Psi_{1,c}(w)}r^m_1(w')-r^m_0(w)\right)
\end{aligned}\right]
\\
\sum_{w\in\mathcal W}
\left[\begin{aligned}
&r^p_1(w)\left(m_1(w)-\underline m_{0,c}(w)\right)
+p_1(w)\left(r^m_1(w)-\min_{w'\in\underline\Psi_{0,c}(w)}r^m_0(w')\right)\\
&+r^p_0(w)\left(\overline m_{1,c}(w)-m_0(w)\right)
+p_0(w)\left(\max_{w'\in\overline\Psi_{1,c}(w)}r^m_1(w')-r^m_0(w)\right)
\end{aligned}\right]
\end{pmatrix},
\end{align*}\normalsize
for all
\(r=\left(
\left(r^p_d(w)\right)_{d\in\left\{0,1\right\},\,w\in\mathcal W},
\left(r^m_d(w)\right)_{d\in\left\{0,1\right\},\,w\in\mathcal W}
\right)\in\mathbb R^{4K}\), with
\[
\underline\Psi_{d,c}(w):=\operatorname{arg\,min}_{w'\in\mathcal N_c(w)}m_d(w'),
\qquad
\overline\Psi_{d,c}(w):=\operatorname{arg\,max}_{w'\in\mathcal N_c(w)}m_d(w'),
\]
for \(d\in\left\{0,1\right\}\), \(w\in\mathcal W\), and \(c\in\mathcal C\).

\item The ATT bound functional \(\phi_{\mathrm{ATT}}\) is Hadamard
directionally differentiable at \(\vartheta\). Its derivative
\(\phi_{\mathrm{ATT},\vartheta}'\) satisfies
\footnotesize\begin{align*}
&\left[\phi_{\mathrm{ATT},\vartheta}'(r)\right](c)
\\
&=
\begin{pmatrix}
\dfrac{1}{\pi_1}
\sum_{w\in\mathcal W}
\left[\begin{aligned}
&r^p_1(w)\left(m_1(w)-\overline m_{0,c}(w)\right)\\
&+p_1(w)\left(r^m_1(w)-\max_{w'\in\overline\Psi_{0,c}(w)}r^m_0(w')\right)
\end{aligned}\right]
-\dfrac{\left[\phi_{\mathrm{ATT},L}\left(\vartheta\right)\right](c)}{\pi_1}
\sum_{w\in\mathcal W} r^p_1(w)
\\
\dfrac{1}{\pi_1}
\sum_{w\in\mathcal W}
\left[\begin{aligned}
&r^p_1(w)\left(m_1(w)-\underline m_{0,c}(w)\right)\\
&+p_1(w)\left(r^m_1(w)-\min_{w'\in\underline\Psi_{0,c}(w)}r^m_0(w')\right)
\end{aligned}\right]
-\dfrac{\left[\phi_{\mathrm{ATT},U}\left(\vartheta\right)\right](c)}{\pi_1}
\sum_{w\in\mathcal W} r^p_1(w)
\end{pmatrix},
\end{align*}\normalsize
for all
\(r=\left(
\left(r^p_d(w)\right)_{d\in\left\{0,1\right\},\,w\in\mathcal W},
\left(r^m_d(w)\right)_{d\in\left\{0,1\right\},\,w\in\mathcal W}
\right)\in\mathbb R^{4K}\).
\end{enumerate}
\end{lemma}

\begin{proof}[Proof of Lemma~\ref{lem:c_mean_bound_maps}]
Since \(p_1(w)>0\) for all \(w\in\mathcal W\) by
Assumption~\ref{ass:c_mean_est_regularity}, \(\pi_1>0\). Hence, the map
\((b,\pi)\mapsto b/\pi\) in the ATT maps is continuously differentiable at each
\((b,\pi_1)\). The maps in \eqref{eq:c_mean_phi_components} also involve
\(\underline m_{d,c}(w;\eta)\) and \(\overline m_{d,c}(w;\eta)\). For fixed \((d,w,c)\),
since \(\mathcal N_c(w)\) is finite,
Lemma~C4 of \citet{masten_inference_2020} gives Hadamard directional
differentiability of the map
\(v\mapsto\max_{w'\in\mathcal N_c(w)}v(w')\) at
\(\left(m_d(w')\right)_{w'\in\mathcal N_c(w)}\), with derivative
\(\max_{w'\in\overline\Psi_{d,c}(w)}r^m_d(w')\) in direction
\(\left(r^m_d(w')\right)_{w'\in\mathcal N_c(w)}\).
Applying the same result to \(-v\), together with \(\inf A=-\sup(-A)\), gives the derivative
\(\min_{w'\in\underline\Psi_{d,c}(w)}r^m_d(w')\) of the map
\(v\mapsto\min_{w'\in\mathcal N_c(w)}v(w')\).
Combining these maps with addition, multiplication, and division and applying
the chain rule in Lemma~C2 of \citet{masten_inference_2020} shows that, for
each \(c\in\mathcal C\), the maps
\(\eta\mapsto\left[\phi_{\mathrm{ATE}}\left(\eta\right)\right](c)\)
and
\(\eta\mapsto\left[\phi_{\mathrm{ATT}}\left(\eta\right)\right](c)\)
are Hadamard directionally differentiable at \(\vartheta\), with the derivatives
displayed in the statement.

To pass to the process indexed by \(\mathcal C\), note that each neighborhood \(\mathcal N_c\left(w\right)\) is one of finitely many subsets of the finite set \(\mathcal W\). Therefore, the collections of finite-dimensional maps \(\eta\mapsto\left[\phi_{\mathrm{ATE}}\left(\eta\right)\right](c)\) and \(\eta\mapsto\left[\phi_{\mathrm{ATT}}\left(\eta\right)\right](c)\), \(c\in\mathcal C\), are finite. The supremum over \(c\in\mathcal C\) is then a maximum over these finitely many maps, each of which is Hadamard directionally differentiable at \(\vartheta\). Hence, \(\phi_{\mathrm{ATE}}\) and \(\phi_{\mathrm{ATT}}\) are Hadamard directionally differentiable at \(\vartheta\) as maps into \(\ell^\infty\left(\mathcal C,\mathbb R^2\right)\), with the derivatives displayed in the statement.
\end{proof}

\begin{proof}[Proof of Proposition~\ref{prop:c_mean_dependence_estimation}]
The identities in \eqref{eq:c_mean_phi_plugin_representation} hold as elements of \(\ell^\infty\left(\mathcal C,\mathbb R^2\right)\).

By Lemma~\ref{lem:c_mean_first_stage},
\[
\sqrt N\left(\widehat\vartheta-\vartheta\right)
\rightsquigarrow
Z
\qquad
\text{in }\mathbb R^{4K}.
\]
By Lemma~\ref{lem:c_mean_bound_maps},
\(\phi_{\mathrm{ATE}}\) and \(\phi_{\mathrm{ATT}}\) are Hadamard
directionally differentiable at \(\vartheta\). Applying Theorem~2.1 of
\citet{fang_inference_2019}, with \(G_{\rho}\) as defined in
\eqref{eq:c_mean_limit_processes}, yields
\[
\sqrt N
\left[
\phi_{\mathrm{ATE}}\left(\widehat\vartheta\right)
-
\phi_{\mathrm{ATE}}\left(\vartheta\right)
\right]
\rightsquigarrow
G_{\mathrm{ATE}}
\quad
\text{in }
\ell^\infty\left(\mathcal C,\mathbb R^2\right),
\]
and
\[
\sqrt N
\left[
\phi_{\mathrm{ATT}}\left(\widehat\vartheta\right)
-
\phi_{\mathrm{ATT}}\left(\vartheta\right)
\right]
\rightsquigarrow
G_{\mathrm{ATT}}
\quad
\text{in }
\ell^\infty\left(\mathcal C,\mathbb R^2\right).
\]
\end{proof}

\begin{proof}[Proof of Proposition~\ref{prop:c_mean_dependence_bootstrap}]
Fix \(\rho\in\left\{\mathrm{ATE},\mathrm{ATT}\right\}\). Let
\(\widehat\zeta^*\) denote the bootstrap analogue of \(\widehat\zeta\).
The finite class of functions
\[
\left\{
\mathbb{I}\left\{D=d,W=w\right\},
Y\mathbb{I}\left\{D=d,W=w\right\}
\mid
d\in\left\{0,1\right\},\, w\in\mathcal W
\right\}
\]
has a square-integrable envelope under
Assumption~\ref{ass:c_mean_est_regularity}. By Theorem~3.7.1 of
\citet{van_der_vaart_weak_2023},
\[
\sqrt N\left(\widehat\zeta^*-\widehat\zeta\right)
\stackrel{d^*}{\rightsquigarrow}
\widetilde Z
\qquad
\text{in }\mathbb R^{4K}
\]
in probability, where \(\widetilde Z\) is the Gaussian limit of
\(\sqrt N\left(\widehat\zeta-\zeta\right)\).\footnote{For a bootstrap statistic \(T_N^*\) and a random element \(T\), write \(T_N^*\stackrel{d^*}{\rightsquigarrow}T\) in probability if the supremum \(\sup_{f\in\mathrm{BL}_1}\left|\mathrm{E}^*\left[f\left(T_N^*\right)\right]-\mathrm{E}\left[f\left(T\right)\right]\right|\) converges to zero in probability. Here, \(\mathrm{E}^*\) is the expectation under the bootstrap measure \(\Pr^*\) given the sample, and \(\mathrm{BL}_1\) is the set of functions bounded by one in absolute value and Lipschitz with constant at most one. See Section~3.7 of \citet{van_der_vaart_weak_2023}.}
The ratio map displayed in the proof of Lemma~\ref{lem:c_mean_first_stage} is continuously differentiable at \(\zeta\), because \(p_d(w)>0\) for all \(d\in\left\{0,1\right\}\) and \(w\in\mathcal W\) by Assumption~\ref{ass:c_mean_est_regularity}. Moreover, the sample
and bootstrap denominators are strictly positive with probability approaching
one. Therefore, Theorem~3.10.11 of \citet{van_der_vaart_weak_2023}, applied to this smooth ratio map, yields
\[
\sqrt N\left(\widehat\vartheta^*-\widehat\vartheta\right)
\stackrel{d^*}{\rightsquigarrow}
Z
\qquad
\text{in }\mathbb R^{4K}
\]
in probability, where \(Z\) is the Gaussian vector in
Lemma~\ref{lem:c_mean_first_stage}.

By Lemma~\ref{lem:c_mean_bound_maps}, \(\phi_{\rho}\) is Hadamard
directionally differentiable at \(\vartheta\), and
\(\phi_{\rho,\vartheta}'\left(Z\right)=G_{\rho}\). Theorem~3.1 of
\citet{hongNumericalDeltaMethod2018}, together with
\(\epsilon_N\to0\) and \(\sqrt N\epsilon_N\to\infty\), implies that
\[
\begin{pmatrix}
T_{\rho,L}^*
\\
T_{\rho,U}^*
\end{pmatrix}
=
\widehat\phi_{\rho}'\left(
\sqrt N\left(\widehat\vartheta^*-\widehat\vartheta\right)
\right)
\stackrel{d^*}{\rightsquigarrow}
G_{\rho}
\qquad
\text{in }\ell^\infty\left(\mathcal C,\mathbb R^2\right)
\]
in probability.

Define \(\mathcal M_{\rho}:\ell^\infty\left(\mathcal C,\mathbb R^2\right)
\to\mathbb R\) by
\[
\mathcal M_{\rho}\left(g\right)
:=
\max\left\{
\sup_{c\in\mathcal C}
\frac{g_L\left(c\right)}{\sigma\left(c\right)},
\sup_{c\in\mathcal C}
\left(
-\frac{g_U\left(c\right)}{\sigma\left(c\right)}
\right)
\right\}.
\]
Since \(\sigma\) is bounded away from zero and infinity, each term in
\(\mathcal M_{\rho}\) is a rescaled supremum functional. By Lemma~C5 of
\citet{masten_inference_2020}, the supremum functional is Lipschitz continuous
with respect to the sup-norm. The maximum of two Lipschitz functions is again
Lipschitz. Hence, by the continuous mapping theorem for Lipschitz maps in
Proposition~10.7 of \citet{kosorok_introduction_2008},
\[
M_{\rho}^*
=
\mathcal M_{\rho}\left(
\begin{pmatrix}
T_{\rho,L}^*
\\
T_{\rho,U}^*
\end{pmatrix}
\right)
\stackrel{d^*}{\rightsquigarrow}
\mathcal M_{\rho}\left(G_{\rho}\right)
=
M_{\rho}
\]
in probability.
Since the distribution function of \(M_{\rho}\) is continuous and strictly
increasing in a neighborhood of \(z_{\rho}\left(1-\alpha\right)\),
Corollary~3.2 of \citet{fang_santos_inference_2015} gives
\[
\widehat z_{\rho}\left(1-\alpha\right)
\xrightarrow{\Pr}
z_{\rho}\left(1-\alpha\right).
\]

By Proposition~\ref{prop:c_mean_dependence_estimation} and the same continuous
mapping argument,
\[
\mathcal M_{\rho}\left(
\sqrt N
\begin{pmatrix}
\widehat{\underline\tau}_{\rho,\cdot}
-
\underline\tau_{\rho,\cdot}
\\
\widehat{\overline\tau}_{\rho,\cdot}
-
\overline\tau_{\rho,\cdot}
\end{pmatrix}
\right)
\rightsquigarrow
M_{\rho}.
\]
The event
\[
\left[
\underline\tau_{\rho,c},
\overline\tau_{\rho,c}
\right]
\subseteq
\mathrm{CB}_{\rho}\left(c;1-\alpha\right)
\quad
\text{for all }c\in\mathcal C
\]
is equivalent to
\[
\mathcal M_{\rho}\left(
\sqrt N
\begin{pmatrix}
\widehat{\underline\tau}_{\rho,\cdot}
-
\underline\tau_{\rho,\cdot}
\\
\widehat{\overline\tau}_{\rho,\cdot}
-
\overline\tau_{\rho,\cdot}
\end{pmatrix}
\right)
\leq
\widehat z_{\rho}\left(1-\alpha\right).
\]
The convergence of the statistic, the consistency of the bootstrap critical
value, and the continuity of the distribution function of \(M_{\rho}\) at
\(z_{\rho}\left(1-\alpha\right)\) imply that the probability of this event
converges to
\(\Pr\left(M_{\rho}\leq z_{\rho}\left(1-\alpha\right)\right)=1-\alpha\).
\end{proof}

\section{Additional Covariates}
\label{sec:supp_additional_covariates}

\subsection{Identification}
\label{sec:s4_identification}

Suppose \(V\) is an additional observed covariate vector with support \(\mathcal V\), which may contain discrete and continuous components. For \(v\in\mathcal V\), let \(\mathcal W_v\) denote the support of \(W\) given \(V=v\). Assume \(\mathrm E[|Y(d)|]<\infty\) for \(d\in\{0,1\}\) and \(0<e(w,v)<1\) for almost every \((w,v)\) in the support of \((W,V)\), where \(e(w,v):=\Pr(D=1\mid W=w,V=v)\). Define the observed mean and the average unobserved potential outcome in cell \((w,v)\) by \(m_d(w,v):=\mathrm E[Y\mid D=d,W=w,V=v]\) and \(m_d^{\mathrm{cf}}(w,v):=\mathrm E[Y(d)\mid D=1-d,W=w,V=v]\).

\begin{definition}[$c$-mean dependence given additional covariates]
\label{def:c_mean_additional_covariates}
Let \(d\in\{0,1\}\), \(c\geq0\), and \((w,v)\) be in the support of \((W,V)\). \(Y(d)\) is \(c\)-mean dependent with \(D\) given \(W=w\) and \(V=v\) if
\[
m_d^{\mathrm{cf}}(w,v)\in\left[\underline m_{d,c}(w,v),\overline m_{d,c}(w,v)\right],
\]
where
\[
\underline m_{d,c}(w,v):=\inf_{w'\in\mathcal N_c(w,v)}m_d(w',v),
\qquad
\overline m_{d,c}(w,v):=\sup_{w'\in\mathcal N_c(w,v)}m_d(w',v),
\]
and \(\mathcal N_c(w,v):=\{w'\in\mathcal W_v\mid |w'-w|\leq c\}\). If this condition holds for almost every \((w,v)\) in the support of \((W,V)\), \(Y(d)\) is \(c\)-mean dependent with \(D\) given \(W\) and \(V\).
\end{definition}

Here, \(\mathcal N_c(w,v)\) is the \(c\)-neighborhood of \(w\) in \(\mathcal W_v\). The identified sets in Proposition~\ref{prop:c_mean_dependence_bounds} extend to this setting by replacing each function of \(W\) with the corresponding function of \((W,V)\).

\begin{corollary}[Identification with additional covariates]
\label{cor:additional_covariates_identification}
Suppose the conditions above hold, and let \(c\geq0\). Replace \(c\)-mean dependence given \(W\) by \(c\)-mean dependence given \(W\) and \(V\), the functions \(e(W)\), \(m_d(W)\), \(\underline m_{d,c}(W)\), and \(\overline m_{d,c}(W)\) by \(e(W,V)\), \(m_d(W,V)\), \(\underline m_{d,c}(W,V)\), and \(\overline m_{d,c}(W,V)\), and the observed distribution of \((Y,D,W)\) by that of \((Y,D,W,V)\). Assume that the expectations in Proposition~\ref{prop:c_mean_dependence_bounds} are finite with these replacements. Then the conclusions of Proposition~\ref{prop:c_mean_dependence_bounds} continue to hold.
\end{corollary}

When overlap fails in some cells, as in the application in Section~\ref{sec:application}, one can take \(\mathrm E[Y(1)-Y(0)\mid D=1,\,0<e(W,V)<1]\) and \(\mathrm E[Y(1)-Y(0)\mid 0<e(W,V)<1]\) as the ATT and ATE \citep[see, e.g.,][]{heckman_matching_1998,angrist_estimating_1998}. Corollaries~\ref{cor:additional_covariates_identification}--\ref{cor:additional_covariates_inference} continue to hold for these targets when Definition~\ref{def:c_mean_additional_covariates} is imposed at \((w,v)\) with \(0<e(w,v)<1\) and its infimum and supremum are taken over the neighboring cells in which \(m_d(w',v)\) exists.

The rest of this section studies estimation and inference for these bounds. When \(V\) has finite support, the bounds are plug-in functionals of finitely many cell probabilities and cell means, as in Section~\ref{sec:estimation_c_mean_dependence}, and the results there carry over directly. Section~\ref{sec:s4_discrete} also covers data that report the outcome only through cell-level summaries, as in the application in Section~\ref{sec:application}. Section~\ref{sec:s4_continuous} outlines estimation when \(V\) contains continuous components.

\subsection{Discrete Covariates and Grouped Data}
\label{sec:s4_discrete}
\label{sec:s4_grouped_inference}

\begin{corollary}[Estimation and inference with discrete additional covariates]
\label{cor:additional_covariates_inference}
Suppose \(V\) has finite support, \(\{(Y_i,D_i,W_i,V_i)\}_{i=1}^N\) are i.i.d.\ copies of \((Y,D,W,V)\), and \(\mathrm E[Y^2]<\infty\). Define the estimators in Section~\ref{sec:estimation_c_mean_dependence} with \((W_i,V_i)\) in place of \(W_i\) and \(\mathcal N_c(w,v)\) in place of \(\mathcal N_c(w)\). Then, under the remaining conditions of Propositions~\ref{prop:c_mean_dependence_estimation}--\ref{prop:c_mean_dependence_bootstrap}, their conclusions continue to hold for the bound functions in Corollary~\ref{cor:additional_covariates_identification}.
\end{corollary}

Corollary~\ref{cor:additional_covariates_inference} uses individual observations. However, administrative data sometimes report the outcome only through cell-level summaries, for example to satisfy confidentiality requirements. The Social Security earnings data in \citet{angrist_estimating_1998}, which I use in the application in Section~\ref{sec:application}, are of this form. Suppose instead that, for each cell \((w,v)\) and treatment status \(d\), the data report the outcome only through its sample count \(N_d(w,v)\), sample mean \(\widehat m_d(w,v)\), and sample variance \(\widehat v_d(w,v)\). The data also report population counts for each cell and treatment status. These counts, normalized to sum to one over \(d\) and \((w,v)\), give the known probabilities \(p_d(w,v):=\Pr(D=d,W=w,V=v)\), and I set \(\widehat p_d(w,v):=p_d(w,v)\). When overlap fails in some cells, I use the target population and neighborhoods described in Section~\ref{sec:s4_identification}. For inference, I use the following grouped sampling conditions.
\begin{assumption}[Grouped sampling]
\label{ass:s4_grouped_sampling}
For each cell \((d,w,v)\) with \(p_d(w,v)>0\), the reported mean \(\widehat m_d(w,v)\) and variance \(\widehat v_d(w,v)\) are based on an i.i.d.\ sample of \(N_d(w,v)\) observations from the distribution of \(Y\mid D=d,W=w,V=v\), which has finite variance, and the samples are independent across these cells. The counts \(N_d(w,v)\) are nonrandom and satisfy \(N_d(w,v)/N\to\lambda_d(w,v)>0\), where \(N\) is the total sample count. The cells and the probabilities \(p_d(w,v)\) do not change with \(N\).
\end{assumption}

Under Assumption~\ref{ass:s4_grouped_sampling}, larger samples add observations within each covariate group but do not lead to finer groups. \citet{angrist_estimating_1998} also treats matching weights as fixed and veteran and nonveteran earnings as independent (see his footnote~27).

To accommodate the potentially nonsmooth bound functionals, I adapt the numerical directional derivative bootstrap in Section~\ref{sec:estimation_c_mean_dependence} to these data. Since the data report average outcomes for each cell rather than individual outcomes, individual observations cannot be resampled. Instead, I construct the first-stage bootstrap draws from the reported means, variances, and sample counts, using the Gaussian approximation described in Section~2 of \citet{hongNumericalDeltaMethod2018}. Given the reported sample means and variances, independently draw \(Z_d^{m,*}(w,v)\sim N(0,N\widehat v_d(w,v)/N_d(w,v))\) for the cells in Assumption~\ref{ass:s4_grouped_sampling}. Since the application reports only the ATT, Proposition~\ref{prop:s4_grouped_inference} states the result for the ATT bounds.

\begin{proposition}[Inference from grouped data]
\label{prop:s4_grouped_inference}
Let \(\mathcal C\subseteq[0,\infty)\) be nonempty. Suppose Assumption~\ref{ass:s4_grouped_sampling} holds, \(\Pr(D=1,\,0<e(W,V)<1)>0\), and the data report every cell mean that enters the ATT bounds for \(c\in\mathcal C\). The conclusion of Proposition~\ref{prop:c_mean_dependence_estimation} about weak convergence holds for the ATT bounds with \(Z\) given in the proof of this proposition. Under the regularity conditions in Proposition~\ref{prop:c_mean_dependence_bootstrap}, its conclusion about the ATT confidence band also holds for the bootstrap procedure constructed above.
\end{proposition}

Unlike Propositions~\ref{prop:c_mean_dependence_estimation} and~\ref{prop:c_mean_dependence_bootstrap}, Proposition~\ref{prop:s4_grouped_inference} uses cell means from independent samples of fixed size, with known cell probabilities. It also replaces Assumption~\ref{ass:c_mean_est_regularity}, which requires every cell to contain both treated and untreated units, by two conditions that concern only the target population and the cells that enter the bounds. In the application, the target population omits 0.01 percent of white and 0.27 percent of nonwhite veterans. For post-application earnings, the data report every cell mean that enters the bounds. For pre-application earnings, the confidentiality edit masks some cell means, and Section~\ref{sec:supp_pre_application} describes which veterans the placebo comparisons include.

\subsection{Continuous Covariates}
\label{sec:s4_continuous}

Corollary~\ref{cor:additional_covariates_identification} also covers the case in which \(V\) contains continuous components. The bounds then average conditional bounds over the distribution of \(V\). Given \(V=v\), these conditional bounds depend on the conditional cell probabilities \(p_d(w\mid v):=\Pr(D=d,W=w\mid V=v)\) and on the minima and maxima of the conditional means \(m_d(w',v)\) over \(w'\in\mathcal N_c(w,v)\). Two estimation approaches are available. I describe them briefly and leave their formal asymptotic theory to future work.

The first approach follows a common strategy in partially identified models, in which identification is nonparametric while estimation uses flexible parametric or semiparametric first-stage estimators \citep[see, e.g.,][]{masten_assessing_2024}. One can estimate \(p_d(w\mid v)\) with a multinomial logit model and \(m_d(w,v)\) with linear regressions on transformations of \(v\) within each cell \((d,w)\). A more flexible alternative for \(m_d(w,v)\) is distribution regression \citep{chernozhukov_inference_2013}. For each \(d\in\{0,1\}\), \(w\in\mathcal W\), and threshold \(y\), it fits a binary response regression with dependent variable \(\mathbb I\{Y\le y\}\) among units with \(D=d\) and \(W=w\),
\[
\Pr(Y\le y\mid D=d,W=w,V=v)=\Lambda\left(B(v)'\theta_{d,w}(y)\right),
\]
where \(B(v)\) is a vector of transformations of \(v\), \(\theta_{d,w}(y)\) is a parameter vector, and \(\Lambda\) is a known link function such as logit or probit. Since the binary regressions are fitted separately at each threshold, \(\Lambda(B(v)'\widehat\theta_{d,w}(y))\) need not be nondecreasing in \(y\). Let \(\widehat F_{d,w}(\cdot\mid v)\) denote its nondecreasing rearrangement in \(y\) \citep{chernozhukov_inference_2013}, set equal to one at the largest threshold. Integrating this fitted conditional distribution function over \(y\) gives
\[
\widehat m_d(w,v):=\int y\,d\widehat F_{d,w}(y\mid v),
\]
which is a finite sum when the thresholds form a finite grid, such as the observed values of \(Y\). The fitted values are then plugged into the conditional bounds, which are averaged over the empirical distribution of \(V\), or over that of the treated units for the ATT. The resulting bound functionals are Hadamard directionally differentiable in the first-stage parameters, even when a minimum or maximum over neighboring cells is attained at several cells. Provided the first-stage estimators satisfy suitable regularity conditions, the functional delta method should yield \(\sqrt N\) convergence of the bound estimators, and the numerical directional derivative bootstrap in Section~\ref{sec:estimation_c_mean_dependence} should apply \citep{fang_inference_2019,hongNumericalDeltaMethod2018}.

The second approach uses double machine learning with orthogonal scores and cross-fitting \citep{chernozhukov_double_debiased_2018}, which allows flexible machine learning estimators of \(p_d(w\mid v)\) and \(m_d(w,v)\). In the present setting, each conditional bound is a weighted sum of conditional means and of their minima and maxima over finitely many neighboring cells, and each such minimum or maximum can be written as the value of a linear program. Hence, the bounds are averages over \(V\) of the values of conditional linear programs, the class of parameters studied in recent work by \citet{khandamiryan_semenova_2026}, and their method should be applicable. They show that their cross-fitted plug-in estimator is \(\sqrt N\)-consistent and asymptotically normal when the first-stage estimators converge in the sup-norm faster than \(N^{-1/4}\). However, their theory requires the cell that attains each minimum or maximum to be unique almost surely, together with a margin condition that bounds the probability of near ties. These conditions rule out, for example, neighboring cells whose conditional means coincide for a set of values of \(V\) with positive probability, which the first approach allows.

\subsection{Proofs}
\label{sec:s4_proofs}

\begin{proof}[Proof of Corollary~\ref{cor:additional_covariates_identification}]
Given \(V=v\), Definition~\ref{def:c_mean_additional_covariates} is the \(c\)-mean-dependence restriction in Proposition~\ref{prop:c_mean_dependence_bounds}, with \(W\) ranging over \(\mathcal W_v\) and with the neighborhoods \(\mathcal N_c(\cdot,v)\). The validity part of the proof of Proposition~\ref{prop:c_mean_dependence_bounds} then applies at each \((w,v)\), and taking conditional expectations over \((W,V)\) given \(D=1\) and given \(D=0\), as in that proof, yields the bounds. By the law of iterated expectations,
\begin{align*}
\tau_{\mathrm{ATT}}&=\mathrm E[Y\mid D=1]-\mathrm E[m_0^{\mathrm{cf}}(W,V)\mid D=1],\\
\tau_{\mathrm{ATE}}-\pi_1\tau_{\mathrm{ATT}}&=(1-\pi_1)\left(\mathrm E[m_1^{\mathrm{cf}}(W,V)\mid D=0]-\mathrm E[Y\mid D=0]\right),
\end{align*}
where \(\pi_1=\Pr(D=1)\). Hence, the joint set is contained in the set of the same form as \(\mathcal I_c\). For sharpness, choose \((\lambda_0,\lambda_1)\in[0,1]^2\) and define \(g_d(w,v):=\lambda_d\overline m_{d,c}(w,v)+(1-\lambda_d)\underline m_{d,c}(w,v)\). The construction in the proof of Proposition~\ref{prop:c_mean_dependence_bounds} applies with \(g_d(w,v)\) in place of \(m_d^{\mathrm{cf}}(w,v)\), and every point of this set is attained by some \((\lambda_0,\lambda_1)\in[0,1]^2\).
\end{proof}

\begin{proof}[Proof of Corollary~\ref{cor:additional_covariates_inference}]
Since \(V\) has finite support, the support of \((W,V)\) is finite. By the overlap condition, \(\Pr(D=d,W=w,V=v)>0\) for each \(d\in\{0,1\}\) and each \((w,v)\) in this support. Hence, Assumptions~\ref{ass:c_mean_est_sampling}--\ref{ass:c_mean_est_regularity} hold with \((W,V)\) in place of \(W\), and Lemma~\ref{lem:c_mean_first_stage} applies to the vector in \eqref{eq:c_mean_first_stage_components} with cell probabilities \(\Pr(D=d,W=w,V=v)\) and cell means \(m_d(w,v)\). Each neighborhood \(\mathcal N_c(w,v)\) contains \(w\) and is one of finitely many subsets of the finite set \(\mathcal W_v\). Hence, the bound functionals combine finitely many min/max maps with addition, multiplication, and division, as in Lemma~\ref{lem:c_mean_bound_maps}. The arguments in the proofs of Propositions~\ref{prop:c_mean_dependence_estimation}--\ref{prop:c_mean_dependence_bootstrap} apply with only notational changes. When overlap fails in some cells, the same argument applies to the probabilities and means of the cells with positive probability, which include every cell that enters the bounds, provided that the target population has positive probability.
\end{proof}

\begin{proof}[Proof of Proposition~\ref{prop:s4_grouped_inference}]
Let \(\vartheta\) and \(\widehat\vartheta\) collect the probabilities and means of the cells in Assumption~\ref{ass:s4_grouped_sampling}. Since \(\widehat p_d(w,v)=p_d(w,v)\) and \(N_d(w,v)/N\to\lambda_d(w,v)>0\) under this assumption, the central limit theorem applied to the outcomes in each cell and the independence across cells yield
\[
\sqrt N\left(\widehat\vartheta-\vartheta\right)
\rightsquigarrow Z,
\]
where \(Z:=(Z^p,Z^m)\), \(Z^p:=0\), and \(Z^m:=(Z^m_d(w,v))\) is a mean-zero Gaussian vector with covariance matrix given by
\[
\mathrm E\left[Z^m_d(w,v)Z^m_{d'}(w',v')\right]
=\frac{\mathbb I\{d=d',\,(w,v)=(w',v')\}}{\lambda_d(w,v)}
\operatorname{Var}\left(Y\mid D=d,W=w,V=v\right)
\]
for the cells in Assumption~\ref{ass:s4_grouped_sampling}. The argument in the proof of Lemma~\ref{lem:c_mean_bound_maps} applies to \(\phi_{\mathrm{ATT}}\) with the neighborhoods in Section~\ref{sec:s4_identification} and with the sums restricted to the target cells, since every neighborhood is a nonempty subset of a finite set of cells, and the treated share of the target population is positive. The proof of Proposition~\ref{prop:c_mean_dependence_estimation} then applies with the preceding first-stage convergence in place of the conclusion of Lemma~\ref{lem:c_mean_first_stage}, which yields \(G_{\mathrm{ATT}}=\phi'_{\mathrm{ATT},\vartheta}(Z)\).

For the bootstrap, set \(Z^{p,*}:=0\) and let \(Z^{m,*}\) collect the draws \(Z_d^{m,*}(w,v)\). Since the cell sample variances are consistent by the law of large numbers and \(N_d(w,v)/N\to\lambda_d(w,v)>0\), the variances of the independent Gaussian draws converge to those of \(Z^m\). Hence,
\[
\left(Z^{p,*},Z^{m,*}\right)
\stackrel{d^*}{\rightsquigarrow}Z
\]
in probability, given the reported sample means and variances. The bootstrap statistic for grouped data evaluates the numerical directional derivative \(\widehat\phi_{\mathrm{ATT}}'\) at \(\left(Z^{p,*},Z^{m,*}\right)\) instead of at \(\sqrt N(\widehat\vartheta^*-\widehat\vartheta)\). Under the regularity conditions of Proposition~\ref{prop:c_mean_dependence_bootstrap}, the numerical delta method and quantile arguments in the proof of that proposition then yield the conclusion about the ATT confidence band.
\end{proof}

\section{Bootstrap Selection of \texorpdfstring{$\epsilon_N$}{epsilon N}}
\label{sec:bootstrap_step_selection}

The bootstrap procedure in Section~\ref{sec:estimation_c_mean_dependence} computes the numerical directional derivative of \citet{hongNumericalDeltaMethod2018} with a step size $\epsilon_N$ that satisfies $\epsilon_N\to0$ and $\sqrt N\epsilon_N\to\infty$. I select $\epsilon_N$ by adapting the heuristic bootstrap selection rule in Appendix B.4 of \citet{masten_inference_2020} to the present confidence bands. Fix $\rho\in\{\mathrm{ATE},\mathrm{ATT}\}$, the set $\mathcal C$ of sensitivity values, and the weighting function $\sigma$ of the confidence band in \eqref{eq:c_mean_confidence_band}. Let $F$ denote the distribution of the observed variables $(Y,D,W)$, or of $(Y,D,W,V)$ when additional covariates are used. For a step size $\epsilon$, let $\mathrm{CP}_N(\epsilon;F)$ denote the probability that the band in \eqref{eq:c_mean_confidence_band}, computed from a sample of size $N$ from $F$ with critical value $\max\{0,\widehat z_\rho(1-\alpha)\}$, contains the lower and upper bound functions under $F$ at every $c\in\mathcal C$. I estimate this probability by replacing $F$ with an estimate $\widehat F$ and select the step size from a prespecified grid $\{\epsilon_1,\ldots,\epsilon_L\}$ by the following double bootstrap procedure.
\begin{enumerate}
\item Draw $M$ simulated samples from $\widehat F$.
\begin{itemize}
\item Under i.i.d.\ sampling, $\widehat F$ is the empirical distribution of the original data, and each simulated sample is a nonparametric bootstrap sample of size $N$.
\item Under the grouped sampling in Assumption~\ref{ass:s4_grouped_sampling}, $\widehat F$ has the cell probabilities $p_d(w,v)$ and the outcome distribution $N(\widehat m_d(w,v),\widehat v_d(w,v))$ in cell $(d,w,v)$, and each simulated sample is a parametric bootstrap sample with $N_d(w,v)$ draws in each cell.
\end{itemize}
\item For each step size $\epsilon_\ell$ in the grid:
\begin{enumerate}[label=(\alph*)]
\item Treat each simulated sample as the observed data and construct the confidence band in \eqref{eq:c_mean_confidence_band} with step size $\epsilon_\ell$ and critical value $\max\{0,\widehat z_\rho(1-\alpha)\}$, using the nonparametric bootstrap of Section~\ref{sec:estimation_c_mean_dependence} or the Gaussian bootstrap of Section~\ref{sec:s4_grouped_inference}, respectively. Check whether this band contains the bounds estimated from the original data at every $c\in\mathcal C$. These estimates are the true bounds when $\widehat F$ is treated as the true distribution.
\item Approximate $\mathrm{CP}_N(\epsilon_\ell;\widehat F)$ by the average of these coverage indicators over the $M$ simulated samples.
\end{enumerate}
\item Select the step size $\widehat\epsilon_N$ that solves
\[
\min_{\ell=1,\ldots,L}\left|\mathrm{CP}_N\left(\epsilon_\ell;\widehat F\right)-\left(1-\alpha\right)\right|,
\]
using the approximation in step~2(b), with ties broken in favor of the smallest step size.
\end{enumerate} In the application in Section~\ref{sec:application}, $M=400$, each confidence band in step~2 uses 499 bootstrap draws, and the grid consists of the $L=9$ step sizes $\epsilon_\ell=\kappa_\ell\sqrt{\log N/N}$, where $\kappa_1,\ldots,\kappa_9$ are $0.1$, $0.15$, $0.25$, $0.35$, $0.5$, $0.7$, $1$, $1.4$, and $2$. Each step size satisfies the conditions of Proposition~\ref{prop:c_mean_dependence_bootstrap}, namely $\epsilon_\ell\to0$ and $\sqrt N\epsilon_\ell\to\infty$ as $N\to\infty$.\footnote{Section~4 of \citet{masten_inference_2020} uses multiples of $N^{-1/2}$ from 1 to 10. In units of $N^{-1/2}$, the present grid ranges from about $0.34$ to $7$ at the sample sizes used in Section~\ref{sec:application}.} As a result, the band computed at the selected step size \(\widehat\epsilon_N\) has asymptotic coverage probability at least \(1-\alpha\). To see this, write \(\widehat z_{\rho}\left(1-\alpha;\epsilon\right)\) for the critical value \(\widehat z_{\rho}\left(1-\alpha\right)\) in \eqref{eq:c_mean_confidence_band} computed with step size \(\epsilon\). The critical values at the grid points satisfy
\[
\max_{1\leq\ell\leq L}\left|\widehat z_{\rho}\left(1-\alpha;\epsilon_\ell\right)-z_{\rho}\left(1-\alpha\right)\right|\xrightarrow{\Pr}0,
\]
because each of them converges in probability to \(z_{\rho}\left(1-\alpha\right)\) by the proof of Proposition~\ref{prop:c_mean_dependence_bootstrap}, or that of Proposition~\ref{prop:s4_grouped_inference} for grouped data, and the grid has a fixed number of elements. Since \(\widehat\epsilon_N\) belongs to the grid, \(\widehat z_{\rho}\left(1-\alpha;\widehat\epsilon_N\right)\) also converges in probability to \(z_{\rho}\left(1-\alpha\right)\), which yields the coverage conclusion as in these proofs.

\section{Additional Results for the Application}
\setcounter{table}{0}
\renewcommand{\thetable}{S.\arabic{table}}
\setcounter{figure}{0}
\renewcommand{\thefigure}{S.\arabic{figure}}
\renewcommand{\theHtable}{supp.\arabic{table}}

\subsection{Placebo Tests with Pre-application Earnings}
\label{sec:supp_pre_application}

Tables~\ref{tab:placebo_eventtime} and~\ref{tab:placebo_calendar} report the placebo tests of Section~\ref{sec:application_placebo} by the number of years before application and by calendar year, respectively. The calendar years are 1974 to 1978, the years that precede application for every cohort. At \(c=0\), the pooled estimates differ from those in Table II of \citet{angrist_estimating_1998} by at most \$2 after rounding, because the comparisons here also require reported earnings in the adjacent AFQT score groups, as described below. Without this requirement, they reproduce his estimates for every year from 1974 to 1978. In Table~\ref{tab:placebo_calendar}, the confidence interval at \(c=0\) excludes zero in 16 of the 40 comparisons by race, application year, and calendar year, with a negative estimate in every case. When application years are pooled within each calendar year, the interval excludes zero for nonwhite applicants in each year from 1975 to 1978, again with negative estimates, but in no year for white applicants. When application years are pooled by \(k\), the interval for white applicants excludes zero at every \(k\).

Following \citet{angrist_estimating_1998}, I exclude cells whose earnings are masked by the confidentiality edit. At a given \(c\), the bounds for a target cell use the earnings of nonveterans in the AFQT score groups within distance \(c\) of the cell. Because the confidence interval at \(c=1\) contains zero in every comparison, the sequential test of Section~\ref{sec:application_placebo} never goes beyond \(c=1\), and the placebo tests use only \(c=0\) and \(c=1\). A target cell is therefore included only if the released data report the mean earnings of its veterans and of the nonveterans in its own and the adjacent AFQT score groups, so that the matching estimate at \(c=0\) and the bounds at \(c=1\) concern the same veterans and differ only in the value of \(c\). The included cells contain at least 97 percent of the veterans in every group and application year for \(k\leq3\) and fewer in earlier years, especially for the youngest applicants.

\clearpage
\begingroup
\spacingset{1}
\small
\renewcommand{\arraystretch}{1.2}
\setlength{\tabcolsep}{5pt}
\begin{longtable}{@{}lccccc@{}}
\caption{Placebo Tests by Years before Application}\label{tab:placebo_eventtime}\\
\toprule
Cohort & \(k\) & \(c=0\) estimate & 95\% CI & \(c=1\) bounds & 95\% CI \\
\midrule
\endfirsthead
\caption[]{Placebo Tests by Years before Application (continued)}\\
\toprule
Cohort & \(k\) & \(c=0\) estimate & 95\% CI & \(c=1\) bounds & 95\% CI \\
\midrule
\endhead
\bottomrule
\endfoot
\bottomrule
\endlastfoot
\multicolumn{6}{@{}l}{\textit{Panel A. White applicants}} \\*
\addlinespace[3pt]
\multirow{5}{*}{1979} & 1 & \(10\) & \([-97,\,117]\) & \([-322,\,397]\) & \([-349,\,424]\) \\*
 & 2 & \(20\) & \([-53,\,94]\) & \([-202,\,270]\) & \([-221,\,289]\) \\*
 & 3 & \(-16\) & \([-63,\,31]\) & \([-137,\,117]\) & \([-141,\,122]\) \\*
 & 4 & \(-21\) & \([-54,\,12]\) & \([-94,\,61]\) & \([-101,\,67]\) \\*
 & 5 & \(-20\) & \([-54,\,14]\) & \([-92,\,64]\) & \([-104,\,76]\) \\
\arrayrulecolor{black!25}\cmidrule(l{2pt}r{2pt}){1-6}\arrayrulecolor{black}
\multirow{5}{*}{1980} & 1 & \(-1\) & \([-99,\,97]\) & \([-268,\,371]\) & \([-297,\,401]\) \\*
 & 2 & \(-18\) & \([-91,\,55]\) & \([-238,\,261]\) & \([-268,\,291]\) \\*
 & 3 & \(15\) & \([-31,\,61]\) & \([-90,\,157]\) & \([-99,\,166]\) \\*
 & 4 & \(-0\) & \([-29,\,28]\) & \([-77,\,91]\) & \([-87,\,101]\) \\*
 & 5 & \(4\) & \([-18,\,26]\) & \([-45,\,52]\) & \([-59,\,67]\) \\
\arrayrulecolor{black!25}\cmidrule(l{2pt}r{2pt}){1-6}\arrayrulecolor{black}
\multirow{5}{*}{1981} & 1 & \(-95\) & \([-187,\,-2]\) & \([-358,\,290]\) & \([-404,\,337]\) \\*
 & 2 & \(-64\) & \([-137,\,9]\) & \([-299,\,247]\) & \([-347,\,295]\) \\*
 & 3 & \(-37\) & \([-85,\,12]\) & \([-157,\,135]\) & \([-175,\,153]\) \\*
 & 4 & \(-44\) & \([-74,\,-15]\) & \([-111,\,43]\) & \([-122,\,54]\) \\*
 & 5 & \(-32\) & \([-55,\,-8]\) & \([-78,\,28]\) & \([-96,\,46]\) \\
\arrayrulecolor{black!25}\cmidrule(l{2pt}r{2pt}){1-6}\arrayrulecolor{black}
\multirow{5}{*}{1982} & 1 & \(-157\) & \([-254,\,-59]\) & \([-391,\,213]\) & \([-435,\,257]\) \\*
 & 2 & \(-123\) & \([-200,\,-47]\) & \([-316,\,152]\) & \([-350,\,185]\) \\*
 & 3 & \(-78\) & \([-135,\,-22]\) & \([-203,\,107]\) & \([-230,\,134]\) \\*
 & 4 & \(-52\) & \([-89,\,-15]\) & \([-137,\,59]\) & \([-164,\,86]\) \\*
 & 5 & \(-39\) & \([-64,\,-14]\) & \([-82,\,20]\) & \([-101,\,39]\) \\
\arrayrulecolor{black!25}\cmidrule(l{2pt}r{2pt}){1-6}\arrayrulecolor{black}
\multirow{5}{*}{Pooled} & 1 & \(-59\) & \([-108,\,-10]\) & \([-332,\,320]\) & \([-332,\,320]\) \\*
 & 2 & \(-45\) & \([-83,\,-8]\) & \([-264,\,235]\) & \([-264,\,235]\) \\*
 & 3 & \(-27\) & \([-52,\,-2]\) & \([-144,\,131]\) & \([-144,\,131]\) \\*
 & 4 & \(-29\) & \([-45,\,-13]\) & \([-104,\,64]\) & \([-104,\,64]\) \\*
 & 5 & \(-21\) & \([-34,\,-8]\) & \([-72,\,40]\) & \([-72,\,40]\) \\
\newpage
\multicolumn{6}{@{}l}{\textit{Panel B. Nonwhite applicants}} \\*
\addlinespace[3pt]
\multirow{5}{*}{1979} & 1 & \(-4\) & \([-90,\,82]\) & \([-243,\,223]\) & \([-243,\,223]\) \\*
 & 2 & \(-47\) & \([-109,\,14]\) & \([-239,\,135]\) & \([-244,\,140]\) \\*
 & 3 & \(-44\) & \([-85,\,-3]\) & \([-186,\,77]\) & \([-207,\,98]\) \\*
 & 4 & \(-22\) & \([-48,\,4]\) & \([-97,\,45]\) & \([-97,\,46]\) \\*
 & 5 & \(-0\) & \([-23,\,23]\) & \([-49,\,49]\) & \([-50,\,50]\) \\
\arrayrulecolor{black!25}\cmidrule(l{2pt}r{2pt}){1-6}\arrayrulecolor{black}
\multirow{5}{*}{1980} & 1 & \(-97\) & \([-185,\,-9]\) & \([-360,\,150]\) & \([-365,\,154]\) \\*
 & 2 & \(-76\) & \([-141,\,-12]\) & \([-260,\,90]\) & \([-263,\,94]\) \\*
 & 3 & \(-85\) & \([-125,\,-45]\) & \([-203,\,9]\) & \([-208,\,14]\) \\*
 & 4 & \(-42\) & \([-67,\,-16]\) & \([-110,\,23]\) & \([-114,\,27]\) \\*
 & 5 & \(-22\) & \([-44,\,-0]\) & \([-78,\,30]\) & \([-92,\,44]\) \\
\arrayrulecolor{black!25}\cmidrule(l{2pt}r{2pt}){1-6}\arrayrulecolor{black}
\multirow{5}{*}{1981} & 1 & \(-21\) & \([-102,\,59]\) & \([-215,\,159]\) & \([-228,\,172]\) \\*
 & 2 & \(-69\) & \([-133,\,-5]\) & \([-239,\,98]\) & \([-256,\,115]\) \\*
 & 3 & \(-68\) & \([-110,\,-26]\) & \([-167,\,33]\) & \([-176,\,42]\) \\*
 & 4 & \(-39\) & \([-64,\,-15]\) & \([-103,\,23]\) & \([-118,\,38]\) \\*
 & 5 & \(-19\) & \([-39,\,-0]\) & \([-50,\,11]\) & \([-53,\,14]\) \\
\arrayrulecolor{black!25}\cmidrule(l{2pt}r{2pt}){1-6}\arrayrulecolor{black}
\multirow{5}{*}{1982} & 1 & \(-89\) & \([-169,\,-9]\) & \([-270,\,126]\) & \([-293,\,149]\) \\*
 & 2 & \(-113\) & \([-174,\,-52]\) & \([-288,\,43]\) & \([-317,\,72]\) \\*
 & 3 & \(-96\) & \([-142,\,-51]\) & \([-213,\,14]\) & \([-245,\,46]\) \\*
 & 4 & \(-73\) & \([-105,\,-40]\) & \([-149,\,7]\) & \([-163,\,21]\) \\*
 & 5 & \(-59\) & \([-85,\,-32]\) & \([-122,\,-0]\) & \([-137,\,15]\) \\
\arrayrulecolor{black!25}\cmidrule(l{2pt}r{2pt}){1-6}\arrayrulecolor{black}
\multirow{5}{*}{Pooled} & 1 & \(-52\) & \([-94,\,-10]\) & \([-273,\,166]\) & \([-292,\,184]\) \\*
 & 2 & \(-75\) & \([-107,\,-43]\) & \([-256,\,93]\) & \([-276,\,114]\) \\*
 & 3 & \(-73\) & \([-94,\,-51]\) & \([-192,\,34]\) & \([-192,\,34]\) \\*
 & 4 & \(-43\) & \([-56,\,-30]\) & \([-113,\,25]\) & \([-113,\,25]\) \\*
 & 5 & \(-23\) & \([-34,\,-12]\) & \([-72,\,24]\) & \([-72,\,24]\) \\
\end{longtable}
\vspace{-0.5\baselineskip}
{\small\noindent\interlinepenalty=10000 \textit{Notes:} The table reports the placebo comparisons for earnings in year \(a-k\) of applicants who applied in year \(a\in\mathcal A\): the matching estimates at \(c=0\), the estimated ATT bounds at \(c=1\), and their 95 percent pointwise confidence intervals. Earnings are in 1991 dollars, rounded to the nearest dollar. A value shown as \(-0\) denotes a negative value that rounds to zero. When the bootstrap critical value is negative, it is replaced by zero, and the confidence interval equals the estimated bounds. The comparisons use only cells whose earnings are reported, as described in Section~\ref{sec:supp_pre_application}.\par}
\endgroup

\clearpage
\begingroup
\spacingset{1}
\small
\renewcommand{\arraystretch}{1.2}
\setlength{\tabcolsep}{5pt}
\begin{longtable}{@{}lccccc@{}}
\caption{Placebo Tests by Calendar Year}\label{tab:placebo_calendar}\\
\toprule
Cohort & Year & \(c=0\) estimate & 95\% CI & \(c=1\) bounds & 95\% CI \\
\midrule
\endfirsthead
\caption[]{Placebo Tests by Calendar Year (continued)}\\
\toprule
Cohort & Year & \(c=0\) estimate & 95\% CI & \(c=1\) bounds & 95\% CI \\
\midrule
\endhead
\bottomrule
\endfoot
\bottomrule
\endlastfoot
\multicolumn{6}{@{}l}{\textit{Panel A. White applicants}} \\*
\addlinespace[3pt]
\multirow{5}{*}{1979} & 1974 & \(-20\) & \([-54,\,14]\) & \([-92,\,64]\) & \([-104,\,76]\) \\*
 & 1975 & \(-21\) & \([-54,\,12]\) & \([-94,\,61]\) & \([-101,\,67]\) \\*
 & 1976 & \(-16\) & \([-63,\,31]\) & \([-137,\,117]\) & \([-141,\,122]\) \\*
 & 1977 & \(20\) & \([-53,\,94]\) & \([-202,\,270]\) & \([-221,\,289]\) \\*
 & 1978 & \(10\) & \([-97,\,117]\) & \([-322,\,397]\) & \([-349,\,424]\) \\
\arrayrulecolor{black!25}\cmidrule(l{2pt}r{2pt}){1-6}\arrayrulecolor{black}
\multirow{5}{*}{1980} & 1974 & \(-2\) & \([-25,\,22]\) & \([-47,\,43]\) & \([-59,\,55]\) \\*
 & 1975 & \(4\) & \([-18,\,26]\) & \([-45,\,52]\) & \([-59,\,67]\) \\*
 & 1976 & \(-0\) & \([-29,\,28]\) & \([-77,\,91]\) & \([-87,\,101]\) \\*
 & 1977 & \(15\) & \([-31,\,61]\) & \([-90,\,157]\) & \([-99,\,166]\) \\*
 & 1978 & \(-18\) & \([-91,\,55]\) & \([-238,\,261]\) & \([-268,\,291]\) \\
\arrayrulecolor{black!25}\cmidrule(l{2pt}r{2pt}){1-6}\arrayrulecolor{black}
\multirow{5}{*}{1981} & 1974 & \(-48\) & \([-91,\,-4]\) & \([-87,\,4]\) & \([-110,\,26]\) \\*
 & 1975 & \(-40\) & \([-70,\,-9]\) & \([-75,\,4]\) & \([-90,\,19]\) \\*
 & 1976 & \(-32\) & \([-55,\,-8]\) & \([-78,\,28]\) & \([-96,\,46]\) \\*
 & 1977 & \(-44\) & \([-74,\,-15]\) & \([-111,\,43]\) & \([-122,\,54]\) \\*
 & 1978 & \(-37\) & \([-85,\,12]\) & \([-157,\,135]\) & \([-175,\,153]\) \\
\arrayrulecolor{black!25}\cmidrule(l{2pt}r{2pt}){1-6}\arrayrulecolor{black}
\multirow{5}{*}{1982} & 1974 & \(16\) & \([-50,\,81]\) & \([-36,\,16]\) & \([-113,\,92]\) \\*
 & 1975 & \(-35\) & \([-73,\,4]\) & \([-48,\,-11]\) & \([-67,\,8]\) \\*
 & 1976 & \(-17\) & \([-38,\,5]\) & \([-45,\,18]\) & \([-58,\,31]\) \\*
 & 1977 & \(-39\) & \([-64,\,-14]\) & \([-82,\,20]\) & \([-101,\,39]\) \\*
 & 1978 & \(-52\) & \([-89,\,-15]\) & \([-137,\,59]\) & \([-164,\,86]\) \\
\arrayrulecolor{black!25}\cmidrule(l{2pt}r{2pt}){1-6}\arrayrulecolor{black}
\multirow{5}{*}{Pooled} & 1974 & \(-15\) & \([-35,\,4]\) & \([-73,\,49]\) & \([-73,\,50]\) \\*
 & 1975 & \(-14\) & \([-31,\,3]\) & \([-69,\,46]\) & \([-69,\,46]\) \\*
 & 1976 & \(-14\) & \([-33,\,4]\) & \([-92,\,77]\) & \([-92,\,77]\) \\*
 & 1977 & \(-9\) & \([-34,\,16]\) & \([-122,\,131]\) & \([-122,\,131]\) \\*
 & 1978 & \(-24\) & \([-60,\,12]\) & \([-214,\,216]\) & \([-214,\,216]\) \\
\newpage
\multicolumn{6}{@{}l}{\textit{Panel B. Nonwhite applicants}} \\*
\addlinespace[3pt]
\multirow{5}{*}{1979} & 1974 & \(-0\) & \([-23,\,23]\) & \([-49,\,49]\) & \([-50,\,50]\) \\*
 & 1975 & \(-22\) & \([-48,\,4]\) & \([-97,\,45]\) & \([-97,\,46]\) \\*
 & 1976 & \(-44\) & \([-85,\,-3]\) & \([-186,\,77]\) & \([-207,\,98]\) \\*
 & 1977 & \(-47\) & \([-109,\,14]\) & \([-239,\,135]\) & \([-244,\,140]\) \\*
 & 1978 & \(-4\) & \([-90,\,82]\) & \([-243,\,223]\) & \([-243,\,223]\) \\
\arrayrulecolor{black!25}\cmidrule(l{2pt}r{2pt}){1-6}\arrayrulecolor{black}
\multirow{5}{*}{1980} & 1974 & \(-5\) & \([-23,\,14]\) & \([-42,\,35]\) & \([-51,\,44]\) \\*
 & 1975 & \(-22\) & \([-44,\,-0]\) & \([-78,\,30]\) & \([-92,\,44]\) \\*
 & 1976 & \(-42\) & \([-67,\,-16]\) & \([-110,\,23]\) & \([-114,\,27]\) \\*
 & 1977 & \(-85\) & \([-125,\,-45]\) & \([-203,\,9]\) & \([-208,\,14]\) \\*
 & 1978 & \(-76\) & \([-141,\,-12]\) & \([-260,\,90]\) & \([-263,\,94]\) \\
\arrayrulecolor{black!25}\cmidrule(l{2pt}r{2pt}){1-6}\arrayrulecolor{black}
\multirow{5}{*}{1981} & 1974 & \(-4\) & \([-28,\,20]\) & \([-33,\,20]\) & \([-49,\,36]\) \\*
 & 1975 & \(-6\) & \([-19,\,8]\) & \([-30,\,15]\) & \([-36,\,21]\) \\*
 & 1976 & \(-19\) & \([-39,\,-0]\) & \([-50,\,11]\) & \([-53,\,14]\) \\*
 & 1977 & \(-39\) & \([-64,\,-15]\) & \([-103,\,23]\) & \([-118,\,38]\) \\*
 & 1978 & \(-68\) & \([-110,\,-26]\) & \([-167,\,33]\) & \([-176,\,42]\) \\
\arrayrulecolor{black!25}\cmidrule(l{2pt}r{2pt}){1-6}\arrayrulecolor{black}
\multirow{5}{*}{1982} & 1974 & \(-8\) & \([-80,\,63]\) & \([-8,\,27]\) & \([-54,\,73]\) \\*
 & 1975 & \(-21\) & \([-53,\,11]\) & \([-56,\,17]\) & \([-96,\,57]\) \\*
 & 1976 & \(-17\) & \([-42,\,8]\) & \([-58,\,18]\) & \([-97,\,57]\) \\*
 & 1977 & \(-59\) & \([-85,\,-32]\) & \([-122,\,-0]\) & \([-137,\,15]\) \\*
 & 1978 & \(-73\) & \([-105,\,-40]\) & \([-149,\,7]\) & \([-163,\,21]\) \\
\arrayrulecolor{black!25}\cmidrule(l{2pt}r{2pt}){1-6}\arrayrulecolor{black}
\multirow{5}{*}{Pooled} & 1974 & \(-2\) & \([-16,\,12]\) & \([-44,\,40]\) & \([-44,\,40]\) \\*
 & 1975 & \(-19\) & \([-32,\,-6]\) & \([-74,\,32]\) & \([-74,\,32]\) \\*
 & 1976 & \(-35\) & \([-52,\,-18]\) & \([-117,\,38]\) & \([-117,\,38]\) \\*
 & 1977 & \(-58\) & \([-81,\,-35]\) & \([-175,\,49]\) & \([-175,\,49]\) \\*
 & 1978 & \(-54\) & \([-86,\,-22]\) & \([-209,\,94]\) & \([-226,\,112]\) \\
\end{longtable}
\vspace{-0.5\baselineskip}
{\small\noindent\interlinepenalty=10000 \textit{Notes:} The table reports the placebo comparisons for earnings in each calendar year of applicants who applied in year \(a\in\mathcal A\): the matching estimates at \(c=0\), the estimated ATT bounds at \(c=1\), and their 95 percent pointwise confidence intervals. Earnings are in 1991 dollars, rounded to the nearest dollar. A value shown as \(-0\) denotes a negative value that rounds to zero. When the bootstrap critical value is negative, it is replaced by zero, and the confidence interval equals the estimated bounds. The comparisons use only cells whose earnings are reported, as described in Section~\ref{sec:supp_pre_application}.\par}
\endgroup

\subsection{Pointwise Confidence Intervals}
\label{sec:supp_post_application_radii}

\begin{figure}[htbp]
\centering
\includegraphics[width=\textwidth]{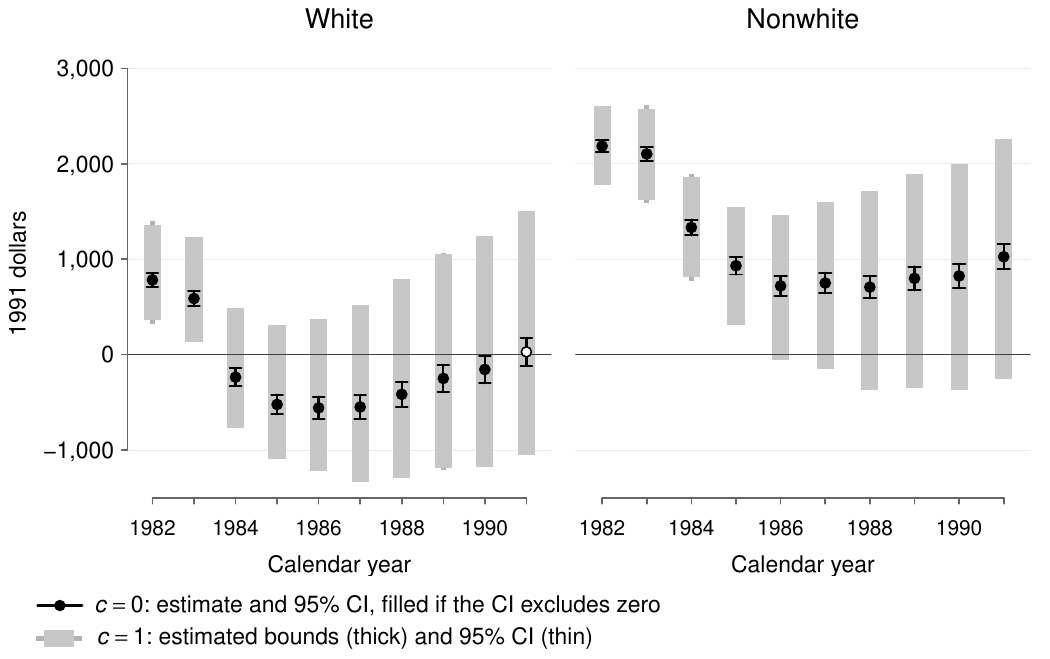}
\caption{Earnings Effects by Calendar Year}
\label{fig:post_application_calendar}
\fnote{\textit{Notes:} The figure shows, for each calendar year \(t\), the matching estimates at \(c=0\) and the estimated ATT bounds at \(c=1\) for earnings in year \(t\), pooled over the application years \(a\in\mathcal A\) as in Table~\ref{tab:angrist_sensitivity}, measured in 1991 dollars. The left panel shows white applicants and the right panel nonwhite applicants. Table~\ref{tab:post_application_radii} reports the values for 1982, 1986, and 1991.}
\end{figure}

Table~\ref{tab:post_application_radii} reports the estimated bounds in Table~\ref{tab:angrist_sensitivity} with 95 percent pointwise confidence intervals, that is, the intervals of Section~\ref{sec:estimation_c_mean_dependence} with \(\mathcal C=\{c\}\). Figure~\ref{fig:post_application_calendar} shows the same comparisons at \(c=0\) and \(c=1\) for every calendar year from 1982 to 1991. In both, the step size is selected separately for each group, year, and \(c\) by the rule in Section~\ref{sec:bootstrap_step_selection}.

\begin{table}[htbp]
\centering
\caption{Pointwise Confidence Intervals for Post-application Earnings}
\label{tab:post_application_radii}
\small
\renewcommand{\arraystretch}{1.2}
\setlength{\tabcolsep}{5pt}
\begin{tabular}{@{}lccc@{}}
\toprule
Year & \(c\) & Estimate or bounds & 95\% CI \\
\midrule
\multicolumn{4}{@{}l}{\textit{Panel A. White applicants}} \\*
\addlinespace[3pt]
\multirow{5}{*}{1982} & 0 & \(783\) & \([712,\,854]\) \\*
 & 1 & \([364,\,1{,}361]\) & \([321,\,1{,}404]\) \\*
 & 2 & \([234,\,1{,}889]\) & \([180,\,1{,}943]\) \\*
 & 3 & \([191,\,2{,}380]\) & \([191,\,2{,}380]\) \\*
 & 4 & \([183,\,2{,}633]\) & \([154,\,2{,}662]\) \\
\arrayrulecolor{black!25}\cmidrule(l{2pt}r{2pt}){1-4}\arrayrulecolor{black}
\multirow{5}{*}{1986} & 0 & \(-557\) & \([-674,\,-441]\) \\*
 & 1 & \([-1{,}216,\,376]\) & \([-1{,}216,\,376]\) \\*
 & 2 & \([-1{,}437,\,1{,}239]\) & \([-1{,}437,\,1{,}239]\) \\*
 & 3 & \([-1{,}516,\,2{,}007]\) & \([-1{,}516,\,2{,}007]\) \\*
 & 4 & \([-1{,}533,\,2{,}380]\) & \([-1{,}533,\,2{,}380]\) \\
\arrayrulecolor{black!25}\cmidrule(l{2pt}r{2pt}){1-4}\arrayrulecolor{black}
\multirow{5}{*}{1991} & 0 & \(30\) & \([-118,\,177]\) \\*
 & 1 & \([-1{,}048,\,1{,}504]\) & \([-1{,}048,\,1{,}504]\) \\*
 & 2 & \([-1{,}422,\,2{,}645]\) & \([-1{,}422,\,2{,}645]\) \\*
 & 3 & \([-1{,}566,\,3{,}636]\) & \([-1{,}566,\,3{,}636]\) \\*
 & 4 & \([-1{,}603,\,4{,}130]\) & \([-1{,}669,\,4{,}197]\) \\
\midrule
\multicolumn{4}{@{}l}{\textit{Panel B. Nonwhite applicants}} \\*
\addlinespace[3pt]
\multirow{5}{*}{1982} & 0 & \(2{,}186\) & \([2{,}124,\,2{,}248]\) \\*
 & 1 & \([1{,}774,\,2{,}604]\) & \([1{,}774,\,2{,}604]\) \\*
 & 2 & \([1{,}573,\,2{,}808]\) & \([1{,}543,\,2{,}838]\) \\*
 & 3 & \([1{,}489,\,2{,}927]\) & \([1{,}481,\,2{,}935]\) \\*
 & 4 & \([1{,}454,\,2{,}962]\) & \([1{,}454,\,2{,}962]\) \\
\arrayrulecolor{black!25}\cmidrule(l{2pt}r{2pt}){1-4}\arrayrulecolor{black}
\multirow{5}{*}{1986} & 0 & \(721\) & \([619,\,822]\) \\*
 & 1 & \([-53,\,1{,}459]\) & \([-53,\,1{,}459]\) \\*
 & 2 & \([-476,\,1{,}898]\) & \([-476,\,1{,}898]\) \\*
 & 3 & \([-684,\,2{,}159]\) & \([-684,\,2{,}159]\) \\*
 & 4 & \([-751,\,2{,}239]\) & \([-751,\,2{,}239]\) \\
\arrayrulecolor{black!25}\cmidrule(l{2pt}r{2pt}){1-4}\arrayrulecolor{black}
\multirow{5}{*}{1991} & 0 & \(1{,}026\) & \([895,\,1{,}157]\) \\*
 & 1 & \([-256,\,2{,}255]\) & \([-256,\,2{,}255]\) \\*
 & 2 & \([-1{,}040,\,2{,}965]\) & \([-1{,}083,\,3{,}009]\) \\*
 & 3 & \([-1{,}445,\,3{,}338]\) & \([-1{,}510,\,3{,}403]\) \\*
 & 4 & \([-1{,}574,\,3{,}450]\) & \([-1{,}650,\,3{,}526]\) \\
\bottomrule
\end{tabular}
\fnote{\textit{Notes:} For each group and year, the table reports the matching estimate at \(c=0\) and the estimated ATT bounds for \(c\geq1\), with 95 percent pointwise confidence intervals. Earnings are in 1991 dollars, rounded to the nearest dollar. When the bootstrap critical value is negative, it is replaced by zero, as described in Section~\ref{sec:estimation_c_mean_dependence}, and the confidence interval equals the estimated bounds. Table~\ref{tab:angrist_sensitivity} reports the confidence band that covers the bounds at all values of \(c\) simultaneously.}
\end{table}

\clearpage
\subsection{Estimates by Years after Application}
\label{sec:supp_post_application_eventtime}

Table~\ref{tab:post_application_eventtime} reports the estimates and intervals in Figure~\ref{fig:post_application_earnings}. As in Table~\ref{tab:post_application_radii}, the step size is selected separately for each group, number \(e\) of years after application, and \(c\) by the rule in Section~\ref{sec:bootstrap_step_selection}.

\begin{table}[htbp]
\centering
\caption{Estimates by Years after Application}
\label{tab:post_application_eventtime}
\small
\renewcommand{\arraystretch}{1.2}
\setlength{\tabcolsep}{5pt}
\begin{tabular}{@{}ccccc@{}}
\toprule
\(e\) & \(c=0\) estimate & 95\% CI & \(c=1\) bounds & 95\% CI \\
\midrule
\multicolumn{5}{@{}l}{\textit{Panel A. White applicants}} \\*
\addlinespace[3pt]
0 & \(262\) & \([210,\,314]\) & \([-38,\,676]\) & \([-38,\,676]\) \\*
1 & \(909\) & \([845,\,974]\) & \([519,\,1{,}464]\) & \([487,\,1{,}495]\) \\*
2 & \(810\) & \([731,\,889]\) & \([379,\,1{,}414]\) & \([368,\,1{,}425]\) \\*
3 & \(167\) & \([83,\,252]\) & \([-336,\,863]\) & \([-336,\,863]\) \\*
4 & \(-500\) & \([-597,\,-403]\) & \([-1{,}053,\,271]\) & \([-1{,}053,\,271]\) \\*
5 & \(-761\) & \([-869,\,-652]\) & \([-1{,}355,\,110]\) & \([-1{,}355,\,110]\) \\*
6 & \(-566\) & \([-688,\,-444]\) & \([-1{,}300,\,458]\) & \([-1{,}300,\,458]\) \\*
7 & \(-396\) & \([-525,\,-267]\) & \([-1{,}194,\,725]\) & \([-1{,}194,\,725]\) \\*
8 & \(-305\) & \([-443,\,-168]\) & \([-1{,}239,\,977]\) & \([-1{,}269,\,1{,}007]\) \\*
9 & \(-179\) & \([-322,\,-37]\) & \([-1{,}169,\,1{,}197]\) & \([-1{,}169,\,1{,}197]\) \\
\midrule
\multicolumn{5}{@{}l}{\textit{Panel B. Nonwhite applicants}} \\*
\addlinespace[3pt]
0 & \(826\) & \([781,\,871]\) & \([554,\,1{,}091]\) & \([529,\,1{,}116]\) \\*
1 & \(2{,}442\) & \([2{,}385,\,2{,}499]\) & \([2{,}073,\,2{,}809]\) & \([2{,}044,\,2{,}838]\) \\*
2 & \(2{,}561\) & \([2{,}494,\,2{,}628]\) & \([2{,}154,\,2{,}983]\) & \([2{,}154,\,2{,}983]\) \\*
3 & \(1{,}812\) & \([1{,}737,\,1{,}887]\) & \([1{,}308,\,2{,}292]\) & \([1{,}307,\,2{,}294]\) \\*
4 & \(873\) & \([787,\,959]\) & \([277,\,1{,}469]\) & \([277,\,1{,}469]\) \\*
5 & \(641\) & \([546,\,737]\) & \([-100,\,1{,}336]\) & \([-100,\,1{,}336]\) \\*
6 & \(722\) & \([618,\,827]\) & \([-141,\,1{,}550]\) & \([-141,\,1{,}550]\) \\*
7 & \(749\) & \([636,\,861]\) & \([-227,\,1{,}675]\) & \([-227,\,1{,}675]\) \\*
8 & \(802\) & \([681,\,922]\) & \([-265,\,1{,}827]\) & \([-265,\,1{,}827]\) \\*
9 & \(842\) & \([716,\,967]\) & \([-356,\,1{,}998]\) & \([-356,\,1{,}998]\) \\
\bottomrule
\end{tabular}
\fnote{\textit{Notes:} For each group and each number of years \(e\) after application, the table reports the matching estimate at \(c=0\) and the estimated ATT bounds at \(c=1\) for earnings in year \(a+e\), pooled over the application years \(a\in\mathcal A\), with 95 percent pointwise confidence intervals. Earnings are in 1991 dollars, rounded to the nearest dollar. When the bootstrap critical value is negative, it is replaced by zero, and the confidence interval equals the estimated bounds. Figure~\ref{fig:post_application_earnings} plots these estimates and intervals.}
\end{table}

\clearpage
\spacingset{1}
\phantomsection
\addcontentsline{toc}{section}{References}
\bibliographystyle{chicago}
\bibliography{references/L1}

\end{document}